\documentclass[screen,manuscript,acmsmall,nonacm]{acmart}
\setcopyright{none}

\newcommand*{\condition}[1]{\textsc{#1}\xspace}

\newcommand*{\Refl}{\condition{(1) Refl}}
\newcommand*{\ValueQuestRefl}{\condition{(2) ValueQuestRefl}}
\newcommand*{\ValueReflPrompt}{\condition{(3) ValueReflPrompt}}

\newcommand{\SubItem}[1]{
    {\setlength\itemindent{15pt} \item[-] #1}}

\newcommand*{\concept}[1]{\textit{\MakeUppercase #1}\xspace}

\newcommand*{\vds}{\concept{ValueDiscrepancyScore}}
\newcommand*{\knowledge}{\concept{KnowledgeRetentionScore}}
\newcommand*{\selA}{\concept{Selection1}}
\newcommand*{\selB}{\concept{Selection2}}
\newcommand*{\IUIPC}{\concept{PrivacyConcern}}

\AtBeginDocument{%
  \providecommand\BibTeX{{%
    \normalfont B\kern-0.5em{\scshape i\kern-0.25em b}\kern-0.8em\TeX}}}

\setcopyright{acmcopyright}
\copyrightyear{2025}
\acmYear{2025}
\acmDOI{XXXXXXX.XXXXXXX}

\usepackage{appendix}

\usepackage{colortbl}
\usepackage{xspace} 
\usepackage{tabularx}

\usepackage{xspace}
\newcommand*{\eg}{e.g.,\@\xspace}
\newcommand*{\ie}{i.e.,\@\xspace}

\usepackage{csquotes} 

\usepackage{color, colortbl}
\definecolor{Highlight}{RGB}{232,239,248}

\makeatletter
\patchcmd{\hyper@makecurrent}{%
 \ifx\Hy@param\Hy@chapterstring
 \let\Hy@param\Hy@chapapp
 \fi
}{%
 \iftoggle{inappendix}{
 \@checkappendixparam{chapter}%
 \@checkappendixparam{section}%
 \@checkappendixparam{subsection}%
 \@checkappendixparam{subsubsection}%
 \@checkappendixparam{paragraph}%
 \@checkappendixparam{subparagraph}%
 }{}%
}{}{\errmessage{failed to patch}}

\newcommand*{\@checkappendixparam}[1]{%
 \def\@checkappendixparamtmp{#1}%
 \ifx\Hy@param\@checkappendixparamtmp
 \let\Hy@param\Hy@appendixstring
 \fi
}
\makeatletter
\newtoggle{inappendix}
\togglefalse{inappendix}

\apptocmd{\appendix}{\toggletrue{inappendix}}{}{\errmessage{failed to patch}}
\apptocmd{\subappendices}{\toggletrue{inappendix}}{}{\errmessage{failed to patch}}

\begin{document}


\title[Designing Value-Centered Consent Interfaces]{Designing Value-Centered Consent Interfaces: A Mixed-Methods Approach to Support Patient Values in Data-Sharing Decisions}

\author{David Leimstädtner}
\email{david.leimstaedtner@fu-berlin.com}
\orcid{0000-0002-1445-3153}
\affiliation{%
\institution{Freie Universität Berlin}
\city{Berlin}
\country{Germany}
}

\author{Peter Sörries}
\email{peter.soerries@fu-berlin.de}
\orcid{0000-0003-0493-2895}
\affiliation{%
\institution{Freie Universität Berlin}
\city{Berlin}
\country{Germany}
}

\author{Claudia Müller-Birn}
\email{clmb@inf.fu-berlin.de}
\orcid{0000-0002-5143-1770}
\affiliation{%
\institution{Freie Universität Berlin}
\city{Berlin}
\country{Germany}}
 
\renewcommand{\shortauthors}{Leimstädtner et al.}

\begin{abstract}
In the digital health domain, ethical data collection practices are crucial for ensuring the availability of quality datasets that drive medical advancement. 
Data donation, allowing patients to share their medical data for secondary research purposes, presents a promising resource for such datasets. 
Yet, current consent user interfaces mediating data-sharing decisions are found to favor data collectors’ values over those of data subjects.
%
%
Seeking to establish value-centered data collection practices in digital health, we investigate the design of consent user interfaces that support end-users in making value-congruent health data-sharing decisions.
Focusing our research efforts on the situated context of health data donation at the psychosomatic unit of a university hospital, we demonstrate how a human-centered design can ground technology within the perspective of a vulnerable group.
We employed an exploratory sequential mixed-method approach consisting of five phases: (1) Participatory workshops elicit patient values, informing the (2) design of a proposed \textit{Value-Centered Consent Interface}. An (3) online experiment demonstrates our interface element’s effect, increasing value congruence in data-sharing decisions. Our proposed consent user interface design is then adapted to the research context through a (4) co-creation workshop with domain experts and (5) a user interface evaluation with patients.
Our work contributes to recent discourse in CSCW concerning ethical implications of new data practices within their socio-technological context by exploring patient values on medical data-sharing, introducing a novel consent interface leveraging reflection to support value-congruent decision-making, and providing a situated evaluation of the proposed consent user interface with patients.
\end{abstract}

\ccsdesc[500]{Human-centered computing~Empirical studies in collaborative and social computing}
\ccsdesc[300]{Security and privacy~Social aspects of security and privacy}
\ccsdesc[300]{Applied computing~Health care information systems}

\keywords{Health Data, Values, Decision Support, Consent Interfaces, Data Donation, Electronic Health Records}

\maketitle

\section{Introduction}
The emergence of large-scale knowledge infrastructures in digital health (e.g., the \textit{European Health Data Space}) increased the availability of patient data for secondary research purposes~\cite{marcus_european_2022}, promising data-driven medical innovation~\cite{scheuermanDatasetsHavePolitics2021,duckert_protecting_2022}.
Yet, these shifts in knowledge infrastructure have also exacerbated power differentials disadvantaging the values of patients, who make their medical data available for secondary research purposes~\cite{cajander_electronic_2019,kassamPatientPerspectivesPreferences2023}, requiring innovative approaches for ensuring core tenets of research ethics~\cite{tsengDataStewardshipClinical2024}.
To empower patients to sovereignly handle the rising amount of sensitive data disclosure inquiries~\cite{malhotra_internet_2004, appenzeller_cpiq_2021}, suitable modes of obtaining informed consent are required~\cite{wilcoxAIConsentFutures2023,kassamPatientPerspectivesPreferences2023,tsengDataStewardshipClinical2024}, safeguarding patients' informed and deliberate decision-making~\cite{maus_enhancing_2020}.
Consent for secondary research purposes is collected through a legally binding paper document, a \textit{consent form}, through which patients provide informed consent for their medical data to be used beyond their original purpose of collection~\cite{zenker_data_2021}.
The complexities of data use in digital health coupled with a multitude of regulatory requirements have resulted in burdensome consent documents that are demonstrated to be poorly understood~\cite{desutterImplementationElectronicInformed2020} and even lead their readers to ignore privacy policies altogether, i.e., so-called \textit{consent fatigue}~\cite{tschiderConsentMythImproving2018, fassl_stop_2021, utz_informed_2019}.
In this regard, current modes of obtaining informed consent fall short, as they present \textit{``incompatible and, to some extent, abusive privacy practices''}~\cite{tschiderConsentMythImproving2018}, failing to promote individual patient interests sufficiently.
Hence, while the ethical principles of respect for persons and individual autonomy (enshrined in the World Health Organization's 1964 \textit{Declaration of Helsinki} and derived legislature~\cite{kayeDynamicConsentPatient2015,worldmedicalassociationWorldMedicalAssociation2013,mcgraw_privacy_2021}) require patients' informed consent for the research use of their data, current consent interfaces\footnote{Throughout this article, the term \textit{interface} is used to refer to \textit{user interface}.} may impede individuals from effectively deriving deliberate decisions~\cite{tschiderConsentMythImproving2018,fassl_stop_2021}, as common design patterns were found to prioritize the values of data-collecting entities (see~\cite{utz_informed_2019, fassl_stop_2021,cajander_electronic_2019, wilcoxAIConsentFutures2023, kassamPatientPerspectivesPreferences2023}). 
These issues concerning the modes in which consent interfaces shape data collection practices raise ethical concerns about the use of data collected and the quality of the resulting datasets~\cite{wilcoxAIConsentFutures2023}. Specifically, this concerns the possibility of biased datasets and a failure to recognize what values are represented therein~\cite{scheuermanDatasetsHavePolitics2021}.
In this regard, researchers in Computer-Supported Cooperative Work (CSCW) and  Human-Computer Interaction (HCI) have called for efforts to advance a human-centered perspective on the role and treatment of data subjects~\cite{fassl_stop_2021,cajander_electronic_2019,wilcoxAIConsentFutures2023} to alleviate the misalignment between the values of \textit{``those who contribute data and those who use data''}~\cite{fassl_stop_2021}. 
This is further exacerbated in digital healthcare, given the substantial power differentials and information asymmetries inherent to the patient role ~\cite{tschiderConsentMythImproving2018}.
Specifically, contextualized approaches for improving the mechanisms for supporting data subjects' values in obtaining informed consent for the use of medical data are needed~\cite{tsengDataStewardshipClinical2024}.
Hence, supporting patients' decision-making also requires taking into account the situated individual interests and
values of data subjects to enable an ethical mode of consent provision in digital health applications~\cite{kassamPatientPerspectivesPreferences2023,eardleyExplanationAdoptionSupporting2023,tschiderConsentMythImproving2018}.
Recent work across CSCW~\cite{huh-yooItWildWild2020,wilcoxAIConsentFutures2023,eardleyExplanationAdoptionSupporting2023,scheuermanDatasetsHavePolitics2021} and HCI~\cite{wong_bringing_2019,eardleyExplanationAdoptionSupporting2023} noted a lack of studies applying participatory approaches to examine data-sharing processes within their sociotechnical contexts and from the perspective of those who make their data available. 
Thus, we posit that for approaching user interface design (e.g., for consent interfaces) grounded within a situated understanding of the values of its end-users (e.g., patients contributing health data), participatory approaches within human-centered design present a way toward designing technology representative of their values~\cite{birhanePowerPeopleOpportunities2022,wong_bringing_2019}. 

Our research explored the design space for supporting data subjects' values in health-data-sharing, asking: \textit{``How can human-centered approaches inform the design of consent interface that supports people in making data-sharing decisions consistent with their values in the medical context?''}
Methodically, we followed an exploratory sequential mixed-methods approach~\cite{creswell_designing_2017}, combining qualitative and quantitative methods. Therein, the initial qualitative phase, i.e., eliciting patient values through participatory workshops, served to ground the conceptualization of our \emph{Value-Centered Consent Interface}. 
The following quantitative phase evaluated the effects of our proposed interface elements through an online experiment, examining to what degree these succeed in supporting value-congruent decision-making. Next, we conducted a co-creation workshop with domain experts to contextualize a final high-fidelity prototype, which was subsequently evaluated with inpatients at a university hospital. 
Our research offers the following contributions to the CSCW community:
\begin{enumerate}
\item \textit{Mixed-Methods Approach to Value-Centered Interface Design:} We combine qualitative and quantitative methods within an exploratory sequential study and convey how this approach helped us to align our consent interface designs with patients' values and preferences regarding data-sharing. 

\item \textit{Value-Centered Reflection Prompt:} We propose a novel design element for a consent interface that fosters value-congruent decision-making. This \textit{Value-Centered Reflection Prompt} encourages data subjects to reflect on their values and consider them in the decision-making process.

\item \textit{Situated Application:} We carried out a domain-specific and contextualized adaptation to and evaluation of the proposed consent interface design in a university hospital data donation application context, further demonstrating the importance of sociotechnical contextualization for HCI research within digital health.
\end{enumerate}

This research effort consists of five subsequent phases, reported as follows: First, relevant literature regarding data-sharing decision-making, consent interface design, and supporting patient values through technology design is reviewed (see \autoref{sec:relatedwork}). Next, our participatory workshop for patient value elicitation and the resulting design requirements are discussed (see \autoref{sec:valueelecitationworkshops}) and consequently translated into a consent interface concept in the design phase (see \autoref{sec:design}). We then evaluate the effect of the proposed interface interventions in an online experiment (see \autoref{sec:experimentaleval}) before bringing it into the application context of health data donation at a German university hospital, where we conducted a co-creation workshop with domain experts (see \autoref{sec:experteval}) and user interface evaluations with patients (see \autoref{sec:usertests}).

\section{Related Work}
\label{sec:relatedwork}
In approaching the design space for consent interfaces based on patient values, we first (1) establish value congruence as a quality indicator for informed decisions, (2) review methodological approaches to leverage stakeholder values for technology design, and (3) examine how current consent interfaces shape how individuals derive data-sharing decisions. Finally, we (4) reflect on prior work concerning design strategies toward supporting individuals in making value-congruent decisions. 

\subsection{Value Congruence as Quality Indicator for Informed Decisions}
\label{sec:informeddecision}
%
%
The most comprehensive conceptualization of informed decision-making stems from the medical field, where it has been investigated extensively within the context of patient-oriented decision-support~\cite{mullen_measures_2006,ghanouni_common_2016}. Two elements are thereby considered fundamental: First, a decision must be based on relevant knowledge, and second, it must be congruent with the patient's values~\cite{witteman_design_2016, munro_choosing_2016}.
Value congruence describes this match between a selected choice and the decision-maker's values. Values are hereby understood to be an individual's attitudes and preferences toward the outcomes or attributes of the available options~\cite{munro_choosing_2016}.
Within HCI, Friedman et al.~\cite{friedman_survey_2017} argued for the importance of values in decision-making, defining them as \emph{``what is important to people in their lives, with a focus on ethics and morality''}.
Value congruence presents the key component of decision quality because even if there is good knowledge of the options available, if the option chosen does not align with the decision maker's values, it cannot constitute a quality decision~\cite{munro_choosing_2016}.
In light of the central role values play in determining the quality of a given decision, promoting value clarification, i.e., the process of determining what is important to an individual in the context of a given decision, is key to supporting their decision-making process~\cite{witteman_design_2016}.
Previous work has demonstrated how value clarification can serve to integrate patient values in a digital health context:
Lim et al.~\cite{lim_facilitating_2019} argued that patients' values may not align with the expectations of other clinical stakeholders, i.e., healthcare providers. To examine patient values, patients were given a worksheet with questions and value cards to contextualize, assess, and reflect on their health-related needs~\cite{lim_facilitating_2019}. The authors concluded that such an approach can support patients in gaining a deeper understanding of their values and help them reflect on their self-care, providing valuable directions for designing healthcare technologies that foster these values. 
Similarly, Berry et al.~\cite{berry_supporting_2021} showed how supporting patients in exploring and articulating their values creates an understanding, which can consequently serve as a valuable perspective for critiquing the current state of healthcare and envisioning possible ideal futures. Further, they highlight the use of value clarification methods to elicit patients’ values and preferences in healthcare contexts where patients’ personal values are not sufficiently communicated. 

In summary, value congruence, i.e., the alignment of decision outcomes and the data subject's personal values, presents a useful criterion when assessing decision quality in a medical data donation context. 
Consent interfaces that support end-users' decision-making should target the value congruence of decision outcomes as a quality indicator. 

\subsection{Designing for Patient Values in Digital Health}
\label{sec:valuesindesign}
Value sensitive design (VSD)~\cite{friedman_value_2019} is a theoretically grounded approach that focuses on the values of those who actively engage with technology or are affected by it. By investigating human values, VSD aims to inform and enhance technology in ways that align with the interests of its stakeholders—those who are directly or indirectly impacted by the technology.
Therein, VSD constitutes an iterative methodology that integrates \emph{conceptual}, \emph{empirical}, and \emph{technical} investigation: Initial conceptual investigations identify, define, and categorize fundamental values, which are refined within subsequent investigations. Empirical investigations can use manifold quantitative or qualitative methods to investigate individual behavior and values to ground the conceptual investigation within the socio-technical context affected by technology use. Technical investigations explore the relationship between realized or possible future technology designs, their implications within a broader context of use, and their impact on direct and indirect stakeholder values~\cite{friedman_value_2019}.
VSD remains non-prescriptive regarding the research methods employed within these investigations, prompting researchers and designers to adapt methods (or combinations thereof) according to the design process and project scope~\cite{friedman_survey_2017}. 

Along with broad adaption across HCI and CSCW since its introduction in the 1990s (see~\cite{friedman_survey_2017}), criticisms of VSD have been voiced, urging for VSD research efforts to employ empirical approaches toward value discovery~\cite{le_dantec_values_2009} as for more substantial alignment with participatory design (PD) to strengthen the voice of participants in research outcomes~\cite{borning_next_2012}.
PD employs democratic decision-making processes in socio-technical designs, seeking for researchers, designers, and people (affected by a design) to contribute equally to a design~\cite{muller_participatory_2007, schuler_participatory_1993, bannon_human_1995, norman_user_1986}. Promoting individuals' agency and autonomy in the design process through PD~\cite{ertner_five_2010,bratteteigUnpackingNotionParticipation2016b} can help overcome power imbalances and realize the values of vulnerable stakeholder groups, for example, patients, through meaningful technology conceptions~\cite{mcdonald_privacy_2020, dahl_facilitating_2020}.
In this regard, Jonas and Hanrahan \cite{jonas_designing_2022} highlighted the significance of researchers and designers acknowledging potential conflicts between their values and those of marginalized participants. They advocated for a value-inclusive approach within PD to facilitate researchers' reflection on their work. The authors asserted that addressing and integrating marginalized perspectives is needed for design efforts to be grounded in a deep understanding of research contexts.
Regarding PD work in healthcare, ~\citet{rothmann_participatory_2016} highlighted how tensions might arise due to incompatibilities between the differing socio-demographic backgrounds and needs of patients, organizational realities of medical institutions, and the increasing complexity of healthcare technology. 
The design of digital healthcare technology presents a complex context concerned with sensitive data. Therefore, researchers and designers need to acknowledge their responsibility to carefully anticipate the values of vulnerable or stigmatized groups to ensure their agency and inclusion in the processes that guide health technology design. 

Hence, VSD and PD provide approaches to ground such design efforts in the empirical investigation into vulnerable stakeholder values, i.e., patients as end-users of consent interfaces~\cite{jonas_designing_2022, friedman_value_2019,mcdonald_privacy_2020}, that are frequently disregarded due to asymmetries of influence between stakeholder groups (e.g. patients and healthcare professionals)~\cite{dahl_facilitating_2020}. Following these methodological approaches aims to inform and enhance technology design in ways that align with the interests of its (direct and indirect) stakeholders.

\subsection{The Influence of Consent Interfaces on Data-Sharing Decisions}
\label{sec:decisionmaking}
\citet{friedman_value_2019} highlights privacy as one of the human values of ethical import, often implicated in system design, defining it as \textit{``a claim, an entitlement, or a right of an individual to determine what information about himself or herself can be communicated to others''}~\cite{friedman_value_2019}[p.28].
Beyond, privacy has previously been identified as a key concern for people sharing medical data for secondary research purposes~\cite{gomez_ortega_towards_2021}.
Hence, in approaching a consent interface design from a value-sensitive design perspective, we first review how individuals make data-sharing decisions along with existing interface approaches for mediating data-sharing decisions.
Data-sharing decisions involve balancing the costs and benefits of sharing personal information. Yet, these trade-offs tend to be ambiguous, complex, and nuanced~\cite{acquisti_nudging_2009}. 
Ideally, people would reflect upon the risks and benefits of a data-sharing decision in light of the information provided and their values and then make their choices accordingly. In reality, people typically can only allocate limited mental resources (e.g., through time constraints), leading them to commonly fall back to heuristics or automatic decision-making processes~\cite{acquisti_nudges_2018}.
These cognitive processes are denoted within dual-process theories as \textit{Type~1} processing, describing heuristic, effective, and automatic processing, which stands in contrast to \textit{Type~2} thinking, referring to reflective, analytical thinking that requires the involvement of working memory~\cite{kitkowska_psychological_2020}.
Recent research has demonstrated people's data-sharing decisions in online contexts to be primarily based on such heuristics~\cite{waldman_cognitive_2020, schneider_digital_2018}.
Users of privacy notices have been shown, for example, to agree to privacy notices without reviewing their decisions to avoid delaying their primary goal of accessing online services~\cite{nouwens_dark_2020}. 
While this automatic processing simplifies routine decisions and reduces the effort expenditure for secondary tasks, it risks regrettable decision outcomes in critical contexts like healthcare (e.g., concerning sensitive data)~\cite{jackson_addressing_2018}. 
In the context of healthcare, a systematic literature review focusing on human factors concerning the privacy of healthcare data conducted by \citet{taziSoKAnalyzingPrivacy2024} found that, although patients and the public express privacy concerns in healthcare, there is a lack of in-depth investigations into patients' preferences and information needs in data-sharing decisions. Furthermore, the authors emphasized a scarcity of user studies addressing the sociotechnical aspects of healthcare privacy and security.
Moreover, it has been argued that existing data-sharing consent approaches focus on facilitating the flow of data between healthcare organizations rather than supporting patient privacy~\cite{mooreConfidentialityPrivacyHealth2007}.
When engaging with consent interfaces, patients have previously been demonstrated to miss important information ~\cite{pearmanUserfriendlyRarelyRead2022}.
As the activation of reflection, i.e., \textit{Type~2} thinking, can create awareness regarding the rationality of a decision at hand~\cite{kitkowska_online_2022}, its activation presents a promising approach for supporting users in actively engaging in value-congruent decision-making~\cite{leimstadtner_investigating_2023}.
Recent work in HCI highlighted the role reflection can play in supporting individuals' data-sharing decisions:
\citet{cho_reflection_2022} outlined a gap concerning technology that engages user-driven reflection for supporting people with meaning-making in data-sharing situations. The authors described how supporting user reflection in these situations helps people understand the potential consequences of sharing their data.
\citet{gomez_ortega_towards_2021} posited that, within the context of making data available for research purposes, active engagement with the purposes of data-sharing can help people reach an informed decision through exploring their attitudes. They highlighted the use of PD approaches, integrating stakeholders, i.e., data subjects, to co-design interfaces that support end-users in decision-making~\cite{gomez_ortega_towards_2021}.
~\citet{baumer_reflective_2015} posited reflection as a tool to empower those disadvantaged by a given socio-technical circumstance. In the healthcare context, reflection could hence empower patients to align their data-sharing decisions with their values.

In summary, the complexity, frequency, and nuance of data-sharing decisions experienced by users of consent interfaces lead them toward reverting to effective yet error-prone \textit{Type~1} thinking, which works through heuristics and automatic processing. Recent work~\cite{kitkowska_online_2022,leimstadtner_investigating_2023} posited the activation of reflection to support deliberate and value-congruent decision-making.

\subsection{Fostering Value-Congruent Decision-Making in Consent Interface Design}
\label{sec:reflectionInterfaces}
One approach to foster user reflection through user interface design is using \emph{cognitive-forcing functions}~\cite{croskerry_cognitive_2003}. This umbrella term describes a group of interface interventions that use mechanisms to disrupt heuristic processing. These de-biasing techniques introduce self-monitoring into decision-making to minimize or avoid errors by preventing people from relying on heuristic processes~\cite{croskerry_cognitive_2003, bucinca_trust_2021}. 
In this regard, \emph{design friction} describes the use of momentary disturbances in a user's interaction with a system~\cite{cox_design_2016,terpstra_improving_2019} as a strategy to activate \emph{Type 2} thinking and disrupt heuristic decision-making~\cite{croskerry_cognitive_2003}. 
Comparably, \citet{cox_design_2016} discussed \emph{micro boundaries}, describing interventions that \emph{design friction} into user's interaction with technology by purposely introducing small obstacles into the user flow to disrupt automatic interactions and create brief moments of reflection.
~\citet{terpstra_improving_2019} posited the use of friction to improve privacy choices. They argued that friction serves as a disorienting dilemma to foster reflective thinking. Friction can thereby be realized by including an interface element that disrupts user flow and (temporarily) prevents them from achieving their primary goal, leading to reflective thinking as it escapes habitual processes.
Comparably, ~\citet{hallnasSlowTechnologyDesigning2001} described the use of \textit{slow technology} on a theoretical level. Therein, an intentional integration of slowness serves to contrast technology design's common target of increasing efficiency to enable space for reflection.
~\citet{jackson_addressing_2018} used a pop-up interface intervention to encourage users to reflect on the discrepancy between their values and privacy behavior. The intervention prompts users to reconsider privacy behaviors that appear inconsistent with their privacy preferences. The authors found that this intervention improves the quality of their decision outcomes by aligning privacy attitudes and decision outcomes.
\citet{gould_task_2015} described the use of a design friction function, namely the task-lockout, which halts users from progressing in a given task for a set timeframe to counteract errors made through inattentiveness, i.e., \emph{Type~1} processing, and thereby improve task accuracy. The authors further argued that \emph{design friction} can support value-led behavior in individuals, as they recognized the value of having an obstacle that allows them to align their behavior with their values.
In this regard, ~\citet{leimstadtner_investigating_2023} demonstrated the use of a task lock-out to help users make value-congruent decisions by counteracting the dissuasive influence of a default nudge within a consent interface.
Evaluating privacy consent interfaces for a US health insurance website, \citet{pearmanUserfriendlyRarelyRead2022} found that users missed important information despite improving usability until asked to review it again. Furthermore, the results indicated that patients are overconfident about the privacy and security of healthcare data and its legal protection. The authors concluded that to arrive at meaningful improvement in informed consent, alternate approaches to health data-sharing consent forms are needed, advocating for the adoption of approaches from a clinical research context, using multimedia formats, quizzes, and conversational approaches.

In summary, we found that people faced with data-sharing decisions tend to adopt ways of thinking that reduce cognitive load through heuristic decision-making. Common consent interfaces promote the values of data collectors, for example, by exploiting these cognitive patterns to dissuade users from acting in their best interests. Human-centered approaches present an avenue for designing consent interfaces that support people's values in data-sharing decisions.
Adopting friction in interface design to disrupt automatic decision patterns can help patients make value-congruent data-sharing decisions. Building upon this research, we seek to examine further the use of such design interventions to support people's reflective deliberation to derive value-congruent decision outcomes.
%

\section{Research Context: Data Donation through the Broad Consent}
\label{sec:context}

\textit{Data donation}~\cite{gomez_ortega_towards_2021} is a form of medical data-sharing suitable for making routinely collected health data available for secondary research purposes~\cite{duckert_protecting_2022}. 
Its peculiarity lies in the openness of future data uses a data subject agrees to~\cite{maus_enhancing_2020}. 
From an ethical perspective, this breadth of future applications, along with the sensitivity of health data, render obtaining informed and voluntary consent from patients critical~\cite{gomezortegaDataTransactionsFramework2023, tsengDataStewardshipClinical2024}.
These factors position data donation as a complex and sensitive use-case, where current consent forms are currently \textit{``complex and hard to understand in its full impact, especially when it is a digital consent, that just requires the press of a button''}~\cite{appenzeller_cpiq_2021}. Digital consent interfaces for health data-sharing are a relatively new concept~\cite{appenzeller_cpiq_2021} that has been shown to improve comprehension for patients sharing digital health data compared to paper formats~\cite{kassamPatientPerspectivesPreferences2023}.
Hence, we approach the complexities of digital health data-sharing decisions from a patient perspective by focusing our research on the situated application context of medical data donation at a German university hospital. 

Across all German university hospitals, data donation is facilitated through the so-called \textit{``broad consent''} (BC). It presents the new national standard for retrieving consent for medical data donations since a unified, legally conforming BC form was developed to be used across all national university hospitals~\cite{zenker_data_2021}. 
This consent form is currently used in paper format and spans six to nine pages (depending on optional modules for bio-samples and health insurance data) to be presented to a patient upon admission to a university hospital.
Yet, the ethical justification and acceptance of the BC remain controversial (see~\cite{grady_broad_2015}). For example, doubts are repeatedly raised whether such consent can be considered \textit{`informed'} and fulfill the function of effective permission for the processing of personal data. 
During the development of the standardized German BC form, extensive consultations were held to inform the development of the form, which involved \emph{all} national data protection authorities and \emph{all} national ethics committees for medical research. However, only \emph{one} consultation group of patient representatives was established~\cite{zenker_data_2021}. 
Under such circumstances, asymmetries of influence may likely cause the patients' perspective and values not to be sufficiently considered when designing digital health technologies, as discussed by~\citet{dahl_facilitating_2020}.
In this regard, prior work in CSCW highlighted the need for human-centered approaches to examine such socio-technical issues related to the emergence of new practices introduced in digital healthcare (e.g., the BC for data donation)~\cite{ciolfiComputerSupportedCooperativeWork2023}, primarily focusing on those stakeholders, who's needs may be underrepresented in the system design~\cite{jacobsComparingHealthInformation2015,zenker_data_2021}.

For this research project, we partnered with the psychosomatic unit at a German university hospital, an early adopter of the BC, to develop a digital consent interface that facilitates data donation in line with patient values. This collaboration enabled us to conduct our research with inpatients for our participatory workshops and user interface evaluations. The hospital also provided local facilities for these activities. 
Further, we collaborated closely with patients throughout the various study phases to develop a consent interface that aligns with their values. This included a final evaluation of the design outcomes with patients in a real-life application context.
Such close collaboration with a hospital ensures that the research activities directly benefit the stakeholders, i.e., patients involved~\cite{tsengDataStewardshipClinical2024}. 

\section{Method}
\subsection{Exploratory Sequential Mixed-Methods Approach}
\label{sec:method}
To base our investigation into consent interface design that supports patient values in digital health data donation, we adopt an \textit{exploratory sequential mixed-methods approach}~\cite{creswell_designing_2017}. 
Through this methodological choice, we address the expressed need for mixed-methods approaches to investigate and address patient consent decisions within a real-world application context, highlighted by~\citet{kassamPatientPerspectivesPreferences2023} in their recent state-of-the-art review of research on patient perspectives on digital health consent.
This inductive approach combines an initial qualitative phase to inform the design of a consent interface with a subsequent quantitative evaluation of the derived design. 
Therein, the exploratory nature of the initial phase serves to ground the development of a novel consent interface in the setting and perspective of participants~\cite{creswell_designing_2017}. 
Epistemologically, this approach ascribes to a pragmatic paradigm, integrating constructivist principles in the qualitative phase and postpositivist assumptions in the quantitative phase.\footnote{For an in-depth discussion on paradigms in mixed-methods research, we refer to Creswell and Plano~\cite{creswell_designing_2017}.}
Hence, following the exploratory sequential approach allows us to derive an empirically evaluated consent interface design grounded within the values of patients from our situated research context~\cite{creswell_designing_2017, bergAccumulatingCoordinatingOccasions1999b}.
In line with VSDs~\cite{friedman_survey_2017} tripartite methodology, such an iterative integration of qualitative and quantitative methods combines \emph{conceptual}, \emph{technical}, and \emph{empirical} investigations (see \autoref{sec:valuesindesign}) to derive a value-based interface design grounded in context-specific investigation.
Such context-specific development of technology increases the likelihood of design outcomes relevant to the group being studied~\cite{creswell_designing_2017}, i.e., patients consenting via the BC in German hospitals.
Through such systematic integration of diverse data sources, methods, and perspectives, we seek to gain a holistic understanding of consent interfaces for data donation, extrapolating insights to inform a larger discourse on the treatment of individuals in algorithmic data practices~\cite{wilcoxAIConsentFutures2023,cajander_electronic_2019,huh-yooItWildWild2020,scheuermanDatasetsHavePolitics2021}. 

\begin{figure*}[h]
 \includegraphics[width=\textwidth]{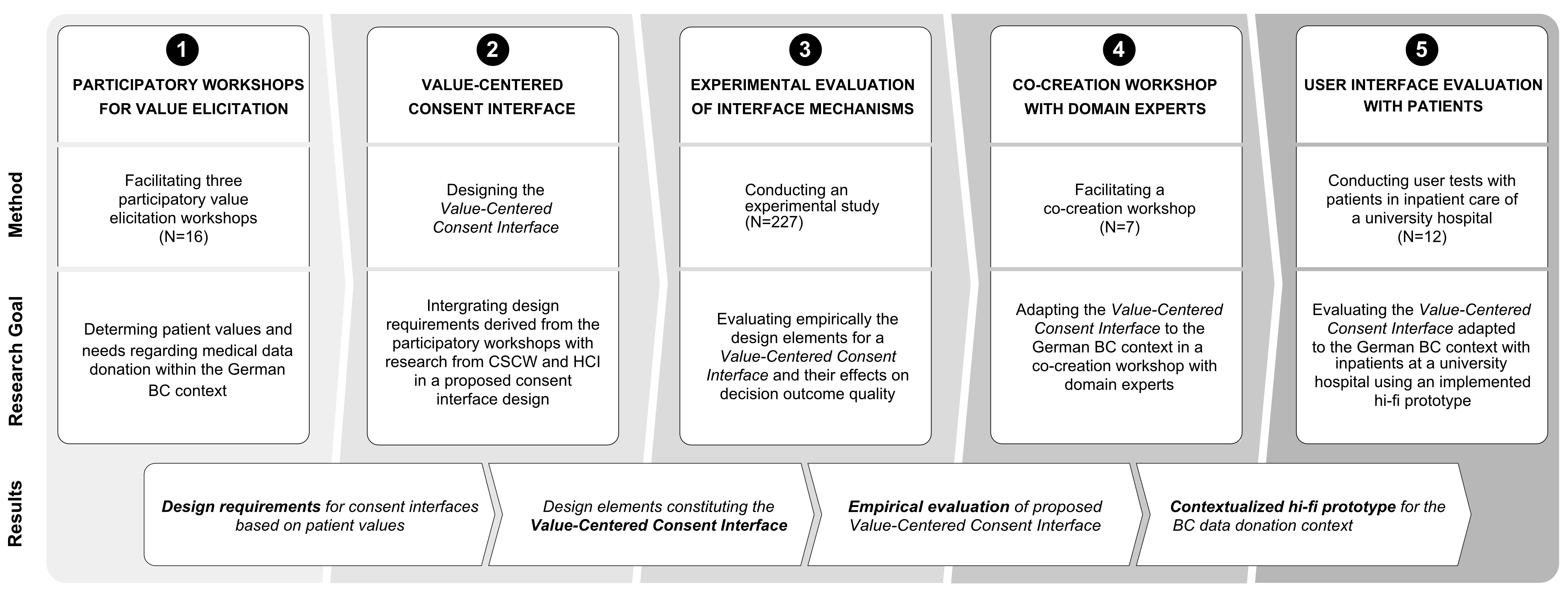}
  \caption{Overview of the five subsequent phases of the exploratory sequential mixed-methods approach for design and evaluation of the \textit{Value-Centered Consent Interface}.}
  \label{fig:mixedmethods}
  \Description[]{This image shows the exploratory sequential mixed-methods approach consisting of five phases, namely the qualitative phase, design phase, quantitative phase, an expert evaluation phase, and user interface evaluations with patients.}
\end{figure*}

For our initial qualitative investigation, we conducted \textcircled{\raisebox{0.0pt}{\footnotesize{1}}} participatory workshops to elicit patient values on health data-sharing within the research context, i.e., German BC~\cite{fitzpatrickReview25Years2013a, bergAccumulatingCoordinatingOccasions1999b}. 
These findings, along with insights from literature (see \autoref{sec:relatedwork}), informed our proposed \textcircled{\raisebox{0.0pt}{\footnotesize{2}}} \emph{Value-Centered Consent Interface}.
To examine the effects of our proposed interface design on decision-making, we conducted an \textcircled{\raisebox{0.0pt}{\footnotesize{3}}} online experiment in the subsequent quantitative phase, evaluating the effect of our devised design in supporting individual decisions and providing insights informing the consent interface design beyond our research context~\cite{fitzpatrickReview25Years2013a}. 
Finally, we returned to the research context to contextualize and evaluate our devised consent interface for supporting data donation decisions at German university hospitals through a co-creation workshop with domain experts \textcircled{\raisebox{0.0pt}{\footnotesize{4}}} and user interface evaluations with inpatients \textcircled{\raisebox{0.0pt}{\footnotesize{5}}}. 

\subsection{Positionality Statement}
\label{sec:positionality}
The author team comprises three HCI researchers active in a German university's computer science institute. 
The first author has a background in communication science, the second author has a background in product and interaction design, specializing in the approaches of VSD and PD, and the third author has a long-standing background in computer science, focusing on human-computer interaction in healthcare.
The research group, of which all three authors are members, pursues a human-centered approach to investigate the underlying assumptions and implications of technology design in digital health, focusing on supporting vulnerable stakeholder groups.
We, the authors, are all white and Western European; as such, we acknowledge that our cultural and socioeconomic background may influence our understanding and framing of the issues studied in this article.
Additionally, none of the authors has personal experience as a patient with psychosomatic conditions. 
This positionality is essential to acknowledge, as it may shape the interpretation of findings, especially when working with patients from diverse and potentially marginalized backgrounds. We have sought to mitigate these limitations by working closely with patients in the inpatient care of a psychosomatic unit, patient representatives, and domain experts, carefully utilizing participatory approaches.
We conducted our research within the German social and cultural context, where policymakers regulate digital healthcare, emphasizing patient autonomy and privacy. At the same time, psychosomatic conditions often carry a stigma, which might shape participants' willingness to engage and share openly. These dynamics influenced participant recruitment, data collection, and the interpretation of findings. However, we strive to encounter these complexities by maintaining reflexivity in our research, incorporating and acknowledging participant voices, and contextualizing their findings carefully within a broader body of related work within HCI and CSCW. By explicitly reflecting on these assumptions and influences, we aim to promote transparency and critically engage with the implications of our research process and outcomes.
Medical professionals from a collaborating university hospital's psychosomatic unit were responsible for recruiting patients, gaining ethics approval for all phases including patient subjects, and being available for patients throughout all processes related to this study.

\section{Phase 1: Participatory Workshops for Value Elicitation}
\label{sec:valueelecitationworkshops}
We conducted participatory workshops to ground our approach to designing a consent interface in a contextualized understanding of patients' perspectives and values regarding medical data-sharing.
For this, we developed a workshop concept for value elicitation aimed at encouraging patients to systematically explore, articulate, and reflect on their values regarding medical data-sharing through the BC and consequently arrive at design requirements informing our proposed consent interface.
Ethical approval was obtained through the ethics committee of the collaborating German university hospital.

The following section presents a condensed version of the workshop approach and findings, focused on the overarching research effort. All details are provided in \citet{sorriesAdvocatingValuesMeaningful2024a}.

\subsection{Method}
\begin{figure*}[h]
  \includegraphics[width=\textwidth]{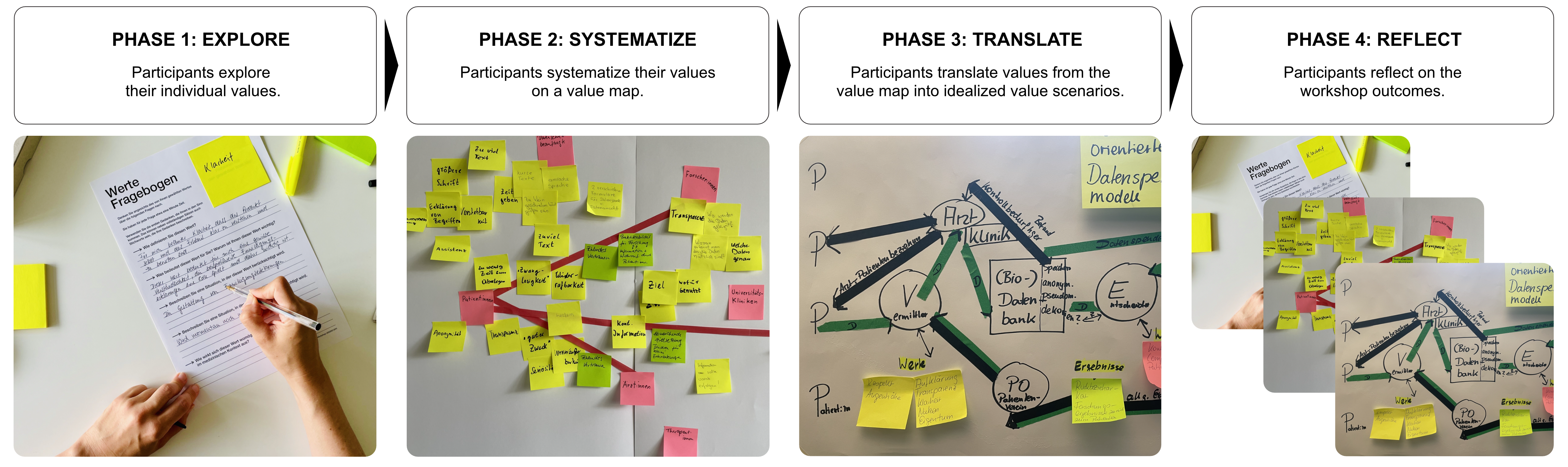}
  \caption{Overview of the participatory value elicitation workshop procedure consisting of four sequential phases. Representations adapted from \citet{sorriesAdvocatingValuesMeaningful2024a}.}
  \label{fig:workshopprocedure}
  \Description[Overview of the participatory value elicitation workshop procedure consisting of four sequential phases.]{This image shows an overview of the participatory value elicitation workshop procedure consisting of four sequential phases: explore, systematize, translate, and reflect.}
\end{figure*}

The workshop consisted of four phases following the goal of exploring patients' values regarding data donation (see~\autoref{fig:workshopprocedure}). In the first phase, each participant completed a questionnaire to externalize their values regarding data donation, followed by a group discussion~\cite{sorriesAdvocatingValuesMeaningful2024a}. In the second phase, these values were contextualized using a \textit{value map} presenting relationships between stakeholders.\footnote{We define a value map as a medium illustrating a contextualized understanding of participants' values which helps to identify direct or indirect stakeholders~\cite{nathan_envisioning_2008} and unfolds stakeholders' relationships and potential conflicts~\cite{miller_value_2007}.} In the third phase, participants used these insights to construct an idealized data donation scenario aligned with their values. Finally, the workshops concluded with a group discussion reflecting on the workshop outcomes.

Three consecutive workshops were conducted: The first workshop (August 2022) included five patient representatives who qualified as patients due to their own or a family member's rare disease and were recruited through the network of local patient advocacy groups. The following two workshops (September and October 2022) involved eleven inpatients from a university hospital's psychosomatic unit recruited through collaborating medical researchers. The three-hour workshops were scheduled on Sundays and did not interfere with treatment activities. The workshops were audio-recorded and transcribed verbatim.
The collected data were analyzed through a two-step qualitative content analysis procedure~\cite{mayring_qualitative_2014}: First, a coding scheme was developed using inductive coding from the values represented on the value maps to gain deeper insights into participants' contextualization of their values. Second, this coding scheme served to analyze how participants materialized values in their data donation scenarios. From this deductive coding, we derived design requirements to be adopted in our proposed design (see~\autoref{sec:design}). 

\subsection{Results}
%
%
While participants generally viewed medical data-sharing as essential for healthcare, they nonetheless expressed concerns about medical institutions (e.g., hospitals) repeatedly asking for their consent, leading to consent fatigue and a perceived loss of ownership of their data. Patients indicated the need to facilitate a better understanding of how their data are stored and processed (e.g., the research purposes or the possibility of third-party access). In addition, participants experienced significant barriers when confronted with consent forms, such as important information buried in long, complex paragraphs or insufficient time to consider a decision within the hospital setting.
Beyond this, our results highlight a need for \textit{`supportive'} measures within a consent interface. First, participants called for restructuring the large text corpus in a fashion that allows access to specific topics of particular importance for a given patient, i.e., for medical research purposes and data processing. Second, participants expressed the need for measures to render BC processes more accessible, for example, by including visual elements.

\subsection{Design Requirements}
Our analysis of the participatory workshops resulted in the following requirements for designing consent interfaces for medical data-sharing:
\begin{itemize}
    \item \textbf{Transparency:} Consent interfaces should ensure transparency regarding medical research purposes to help patients perceive, execute, and maintain control over their shared data. In this regard, consent interfaces need to support patients in navigating what is seen as complex data practices and opaque medical research purposes. We assert that mechanisms are required to help patients make data-sharing decisions that align with their values by providing a well-structured overview of contextualized information, for example, in categories with appropriate descriptions. 
    \item \textbf{Autonomy:} Consent interfaces should facilitate autonomous decisions based on patient's capabilities and needs. Our results highlight patients' information needs when sensitive data is concerned, aggravated through pre-existing skepticism toward medical institutions or due to stigmatization of medical conditions. To address this, integrating a decision facilitator role into healthcare data practices can provide a \textit{`neutral'} resource that helps users navigate complex consent processes while accommodating their diverse information needs. This role could be embodied in a consent interface designed to assist patients in data-sharing by answering their questions and educating them about existing data practices, for example, through tailored descriptions according to varying information needs.
    \item \textbf{Reflection:} Consent interfaces should help patients actively assess the potential consequences of their data-sharing and, thus, provide opportunities to reflect on their consent before deciding. For example, decisions concerning health data-sharing could be separated from the moment of medical treatment, or patients should be provided an appropriate time frame to consider potential consequences. 
\end{itemize}

Overall, our participatory workshops for value elicitation suggested that a consent interface representative of patient values should provide support to transparently address individual information needs concerning the complex consent process and help them engage in active deliberation on the consequences of sharing medical data, enabling value-aligned decision-making.

\begin{figure*}[h]
  \includegraphics[width=\textwidth]{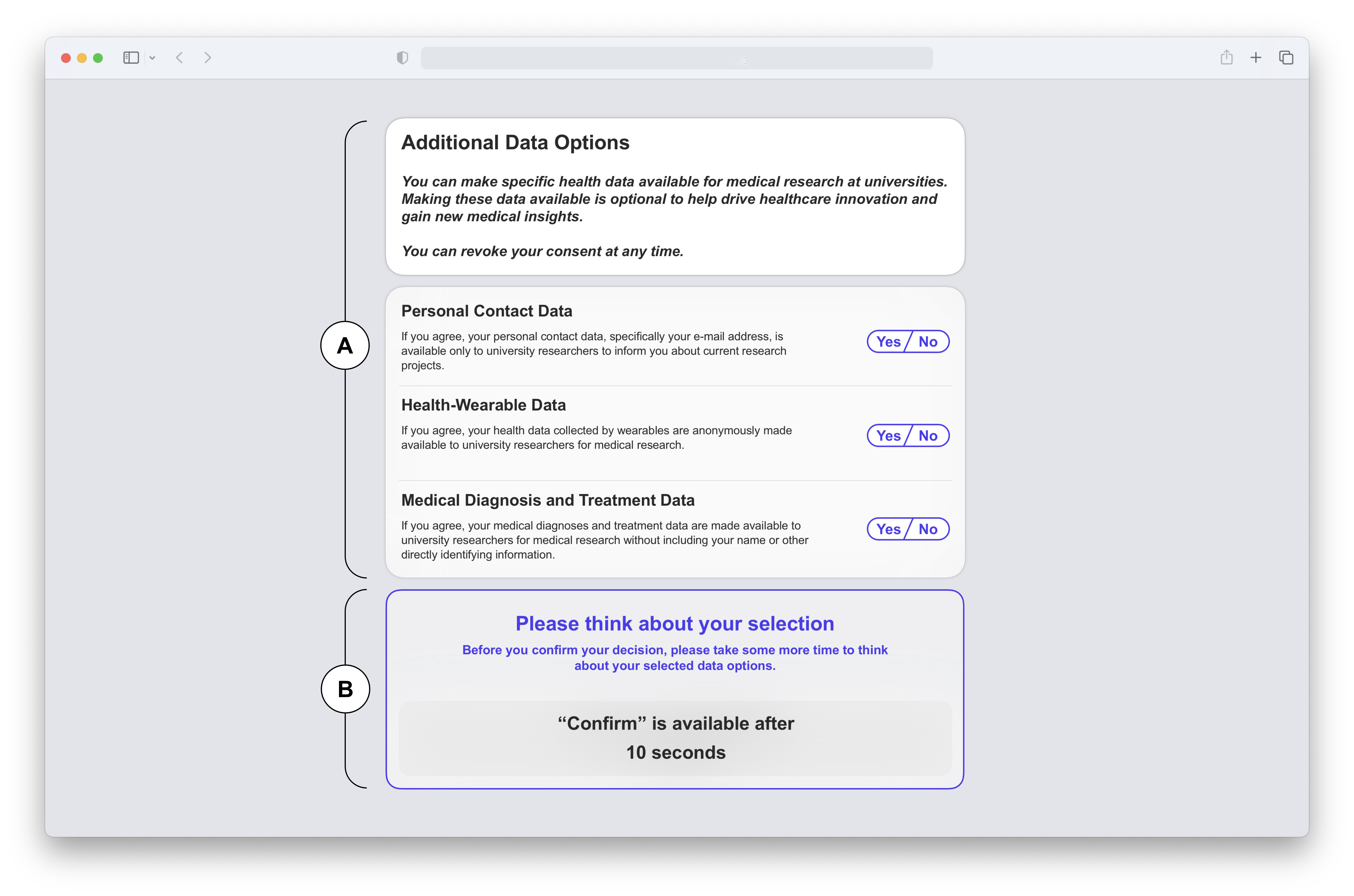}
  \caption{{Consent interface including the \emph{Disclosure Options \textcircled{\raisebox{-0.2pt}{\footnotesize{A}}}} selector and the \emph{Reflection Prompt \textcircled{\raisebox{-0.4pt}{\footnotesize{B}}}}.}}
  \label{fig:interface}
  \Description[]{The data-sharing consent interface consists of three data-sharing options titled Disclosure Options and the reflection prompt, asking participants to reflect on their decision while a count-down of ten seconds runs down before the button allowing to continue is enabled.}
\end{figure*}

\section{Phase 2: Designing a Value-Centered Consent Interface}
\label{sec:design}
Following the \emph{exploratory sequential mixed-methods approach}~\cite{creswell_designing_2017}, we use the qualitative insights gained in the workshops along with theoretical insights (see section \autoref{sec:relatedwork}) from previous research to propose a novel concept for a consent interface design, namely the \emph{Value-Centered Consent Interface}. 
In the following, we detail the implemented design elements, highlighting how these realize the identified design requirements.

\begin{table}
    \centering
    \caption{Mapping design requirement and the respective realizations in the proposed interface (\autoref{fig:interface}).}
    \begin{tabular}[t]{p{10em} p{20em}}
         \toprule
         \textbf{Design Requirement} & \textbf{Interface Element} \\
         \midrule
         \textit{\textbf{Transparency}} & Selector with structured information provision (A)  \\
         \midrule
         \textit{\textbf{Autonomy}} & Disclosure options requiring active selection (A)  \\
         \midrule
         \textit{\textbf{Reflection}} & \textit{Value-centered reflection prompt} (B) \\
         \bottomrule
    \end{tabular}
    \label{tab:requirementstable}
\end{table}

\subsection{Fostering \textit{Transparency} and \textit{Autonomy} in Consent Interface Design}
To achieve a transparent consent interface that enables autonomous decision-making for end-users, we realize a design corresponding with the design requirements identified in our participatory workshops, based on recommendations for the design of digital consent interfaces~\cite{utz_informed_2019,habib_toggles_2021,kassamPatientPerspectivesPreferences2023}.
For this, we've created a selector concept that uses simple language and clear structure and requires active decisions (contrasting the ubiquitous use of default nudges through pre-selection in consent interface ~\cite{fassl_stop_2021,leimstadtner_investigating_2023}).
This approach implemented patients' need for transparent data-sharing forms~\cite{kassamPatientPerspectivesPreferences2023} that present data-sharing decisions in a simple and contextual form expressed in our participatory workshops as well as legal requirements of the European General Data Protection Regulation (GDPR)~\cite{european_commission_regulation_2016}.
The selector elements are as follows:
The \emph{Disclosure Options} selector is the central element presenting the data-sharing options to the user (see~\autoref{fig:interface}, \textcircled{\raisebox{0.0pt}{\footnotesize{A}}}). The options are structured as context-sensitive, categorized information sections~\cite{utz_informed_2019}. 
Keeping sentence construction consistent across options allows for quicker understanding of \textit{who}, would get access to \textit{what data}, and for \textit{which purpose}. This information structure corresponds to the basic information requirements for data-sharing procedures defined by the GDPR~\cite{nunes_vilaza_futures_2020,european_commission_regulation_2016}. 
A choice element per option allows for granular and specific consent without prioritizing access to either outcome~\cite{european_commission_regulation_2016}.
For this, we created a \textit{Yes/No} button without a pre-selected default option, based on Habib et al.'s \textit{stylized toggle button}~\cite{habib_toggles_2021}. Users are required to make a selection to proceed.

\subsection{Prompting \textit{Reflection} for Data-Sharing Decisions}
With the goal of prompting users to engage in active reflection on their data-sharing decision, we adapt the concept of \textit{design friction} (see \autoref{sec:reflectionInterfaces}) and realize it in the consent interface design as follows:
The \emph{reflection prompt} (see \autoref{fig:interface},\textcircled{\raisebox{0.0pt}{\footnotesize{B}}}) introduces \textit{design friction} to encourage users to engage analytically with the decision at hand, hence disengaging heuristic processes~\cite{bucinca_trust_2021}.
This is achieved via design friction aimed to enhance users' reflection on their privacy choices by deploying a timed task lockout interrupting the expected user flow (see~\cite{bucinca_trust_2021, croskerry_cognitive_2003, terpstra_improving_2019}) with the appearance of a \emph{reflection prompt} upon the user's initial pressing of the \emph{Accept} button. 
The task lockout is conceptualized through the appearance of a grayed-out, \ie disabled, confirmation button after the user initially attempts to confirm their selection, initiating a 10-second countdown period.\footnote{We decided to use a short task lockout as Gould et al.~\cite{gould_task_2015} found that longer-lasting task lockouts caused participants to switch activities.} Users can adjust their selection throughout this timeframe. After the timer expires, the button is re-enabled, allowing users to continue. 
This breakdown serves to foster reflective thinking ~\cite{baumer_reflective_2015}.
The textual inquiry,  \emph{``Please think about your selection! Before you accept your decision, please take some more time to think about your selected data options,''} further instigates reflection by inviting the user to consider other choices (see~\cite{baumer_reflective_2015}).
In contrast to Leimst{\"a}dtner et al.~\cite{leimstadtner_investigating_2023}'s previous application of reflection prompts, we require users to make an active choice, as required in the GDPR~\cite{european_commission_regulation_2016}.

\begin{figure*}[h]
  \includegraphics[width=\textwidth]{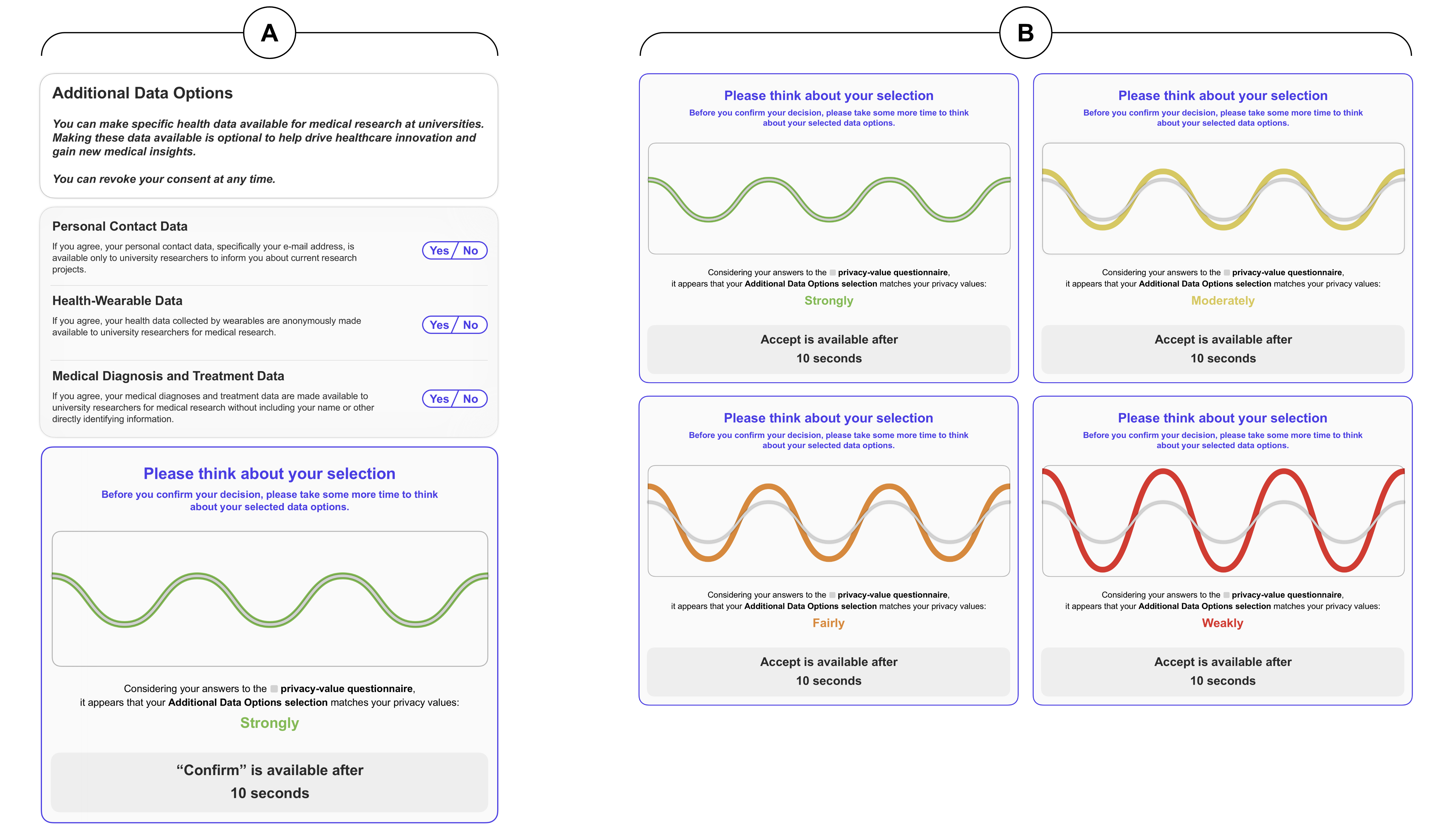}
  \caption{Overview of the data donation consent interface incorporating the \emph{Value-Centered Reflection Prompt} \textcircled{\raisebox{-0.3pt}{\footnotesize{A}}} accompanied by the four possible congruence messages with their accompanying visualization \textcircled{\raisebox{-0.3pt}{\footnotesize{B}}}.}
  \label{fig:matrix}
  \Description[]{This image shows the four states of the \emph{Value-Centered Reflection Prompt}, including a visualization depicting two waves that drift out of synchronicity with increasing levels of value discrepancy.}
\end{figure*}

Further, we extend the reflection prompt into a \emph{Value-Centered Reflection Prompt} by providing a representation of the congruence (or discrepancy) between the users' decision behavior and their values (see Jackson and Wang~\cite{jackson_addressing_2018}, as discussed in \autoref{sec:reflectionInterfaces}), which are collected through a prefacing questionnaire. This realization is context-specific and detailed in \autoref{sec:experimentaleval}.
This \emph{Value-Centered Reflection Prompt} incorporates value clarification regarding a data-sharing decision at hand ~\cite{witteman_design_2016} through two elements: First, a text indicating the level of value congruence, i.e., the \emph{congruence message} (\eg \textit{``Considering your answers to the privacy-value questionnaire, your current selection matches your privacy values: strongly.''}).\footnote{The congruence is thereby indicated by four categories: \textit{Strong}, \textit{Moderate}, \textit{Fair}, and \textit{Weak}. The wording is thereby adapted from common descriptions of correlation coefficients~\cite{akoglu_users_2018}.}
Second, a visualization that recalls the current level of congruence between relevant individual values and the selected \emph{Disclosure Options} provides immediate feedback upon adjusting the decision, acting as a supporting measure by representing users' choices (see a depiction of the interface in \autoref{fig:matrix}).
Using visual cues can serve to attract attention and guide users toward actively exploring the information provided~\cite{rossi_visualization_2017}. 
To conceptualize a visual metaphor, we conducted an interdisciplinary work meeting within our research group (including one researcher from product design, one from communication science, one from digital media studies, one from bioinformatics, and three with a computer science background). The resulting four visualization options were tested as part of our pilot study (see \autoref{sec:experimentaleval}), finding a preference for the visual metaphor of two waves drifting into asynchrony with increasing discrepancy values.

The qualitative results of our workshops revealed the need to support autonomy, transparency, and reflection in consent interfaces. As prior work on reflection posits it as the key toward enabling deliberate decision-making processes~\cite{jackson_addressing_2018}, capable of fostering value congruence in people's decision~\cite{leimstadtner_investigating_2023, munro_choosing_2016}, we focused the following quantitative phase~\cite{croskerry_cognitive_2003} on an examination of value-centered reflection.
Working toward our overall RQ, \textit{``How can a human-centered perspective inform the design of a situated consent interface that supports people in making data-sharing decisions consistent with their values in the medical context?''}, we conducted an experimental study to establish to what degree our proposed \textit{Vale-Centered Consent Interface} succeeds in supporting people in making value-congruent data-sharing decisions through enabling reflection via design frictions.
This experimental study serves to evaluate the effect of the design friction mechanisms within our proposed consent interface with a generalizable sample before approaching the vulnerable target group, i.e., hospital inpatients, within our application context.

\section{Phase 3: Experimental Evaluation}
\label{sec:experimentaleval}
\begin{figure}[h]
  \includegraphics[width=\textwidth]{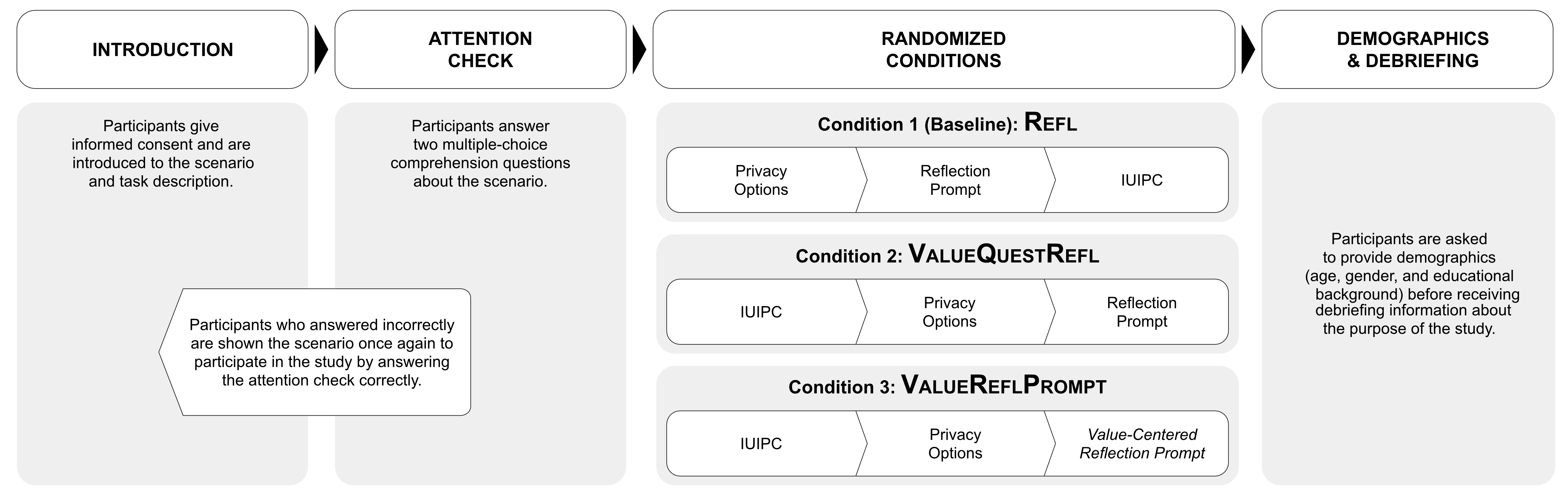}
  \caption{Overview of the experimental study procedure and the three experimental groups, representing differing realizations of the \textit{(Value-Centered) Reflection-Prompt}.}
  \label{fig:studyprocedure}
  \Description[Overview of the study procedure]{This image shows the study procedure, including, first, an introduction to the scenario and task description; second, an attention check with two comprehension questions; third, the condition overview; and finally, questions regarding demographics and a debriefing.}
\end{figure}

To evaluate if our design increases value congruence in data-sharing decisions, we conducted an online study, realizing the \emph{Value-Centered Consent Interface} in a hypothetical health data-sharing scenario, focusing on privacy concerns for the content of the \emph{Value-Centered Reflection Prompt}.\footnote{We focus our experimental setup on privacy, as it has previously been identified as a central concern for potential donors of medical data~\cite{gomez_ortega_towards_2021}.}
In the experimental scenario (see~\autoref{sec:scenariotext}) the participants were asked to test parts of a fictional new health data management platform, with contents modeled after the BC forms~\cite{appenzeller_towards_2022}, the \textit{European Health Data Space}~\cite{marcus_european_2022} and the German electronic health record~\cite{appenzeller_cpiq_2021}.
We recruited participants via convenience sampling on the crowdsourcing platform MTurk. The experimental online study was realized with LimeSurvey and took place over one week in January 2023.
First, a pilot study (N=24) served to evaluate the experimental setup. Based on the average task duration therein, we compensated our participants with \$3.00 (\$12.00/h).\footnote{Our compensation is close to the above-average wage of \$11.00/h~\cite{hara_data-driven_2018}.}  We allowed participants up to 40 minutes to complete the survey to avoid potential rejections in case of slower completions.
Based on an a priori power analysis, we recruited 227 participants (see \autoref{sec:inclusioncriteria}). Descriptive statistics for variables are reported in \autoref{fig:desctable}.
Our study combines within-subject measures of privacy decisions before and after exposure to the reflection prompt with three experimental conditions. \autoref{fig:studyprocedure} shows our study procedure in detail. 
Across all conditions, participants are presented with our consent interface, which contains identical data-sharing options.
The baseline condition \Refl contains a reflection prompt with general instructions.\footnote{We wanted to better understand possible priming effects within the other conditions caused by answering a questionnaire about disclosure intentions and privacy behaviors before choosing the privacy options (see~\cite{alashoorPrimingEffectProminent2017}).} 
After confirming their selection, the participants complete a privacy attitudes questionnaire, \ie the \emph{Internet Users' Information Privacy Concerns} (IUIPC) (see \autoref{sec:measurements}).
In the \ValueQuestRefl condition, participants complete the privacy questionnaire before selecting the data-sharing options, followed by a reflection prompt with general instructions. This condition serves to examine whether engagement with a privacy attitude questionnaire is predictive of value-congruent privacy practice.
In the \ValueReflPrompt condition, participants complete the privacy questionnaire, then select data-sharing options before being presented the \emph{Value-Centered Reflection Prompt}.

We are aware of our responsibility as researchers to collect empirical data from human subjects carefully. As our university did not have an institutional review board for research projects at the time of the study, we followed recommendations for obtaining informed consent in MTurk research~\cite{aguinis_mturk_2021}.

\subsection{Measurements}
\label{sec:measurements}
The following section describes the measurements taken for this online experiment:
\begin{itemize}
    \item \textbf{Privacy attitudes:} We determined participants' privacy attitudes using the IUIPC privacy concern questionnaire~\cite{malhotra_internet_2004}. It consists of ten questions in randomized order, answered on a seven-point Likert scale. 
    \item \textbf{Selection score:} We captured participants' decision behavior through a \emph{selection score}, \ie the sum of the selected privacy options (range 0-4). The selection score is recorded twice: first, after the initial selection, \ie before the reflection prompt (\selA), and second, after the reflection prompt is displayed (\selB).
    \item \textbf{Value discrepancy score:} The \emph{value discrepancy score} describes the level of incongruence between a participant's data-sharing behavior, \ie \emph{selection score}, and \emph{privacy attitudes}, \ie \IUIPC. It is calculated as the difference between the normalized scores for \IUIPC and the selection score.\footnote{We adapted the conceptualization of value congruence as a key component of decision quality along with its calculation from Munro et al.~\cite{munro_choosing_2016}.} A \vds of zero describes the highest level of congruence.
\end{itemize}
    Beyond, we measured \textbf{knowledge} retention of relevant information about the data-sharing options provided to assess whether certain conditions are more transparent regarding information retention on the data-sharing options available to them.\footnote{We adapted this measure from Betzing et al.~\cite{betzing_impact_2020}.} For this, the \knowledge measure is calculated from eight questions presented in a randomized order. Two questions are asked about each data option, and two about the overall interface. Calculation and wording are detailed in \autoref{sec:knowledgequest} and \textbf{demographic variables}, namely age, gender~\cite{spiel_how_2019}, and education as control variables, as prior research has found these to play a role in privacy behavior~\cite{boerman_exploring_2021}. Finally, to ensure the quality of study submission, we included three \textbf{attention checks}~\cite{pei_attention_2020}.

\subsection{Hypotheses}
To examine the effect of the reflection prompt on the value congruence of the decision outcome, we formulated the following three hypotheses: 
Based on prior results on the use of reflection prompts~\cite{leimstadtner_investigating_2023}, we expect (\textbf{H1}) the reflection prompt to increase the value congruence of a decision (\vds) between the initial selection (\selA) and the final selection (\selB) across conditions.
Next, we seek to investigate the potential priming effect of answering a privacy questionnaire (IUIPC) before making a data-sharing decision, as suggested in ~\citet{alashoorPrimingEffectProminent2017}. Based on this, we expect (\textbf{H2}) that participants who answer the IUIPC before the data-sharing decision show a lower \vds for their initial selection (\selA) than those without a preceding questionnaire. 
Finally, building on previous research~\cite{witteman_design_2016,cho_reflection_2022,friedman_survey_2017}, we hypothesized (\textbf{H3}) that introducing value-centered reflection into privacy decisions, i.e., the \ValueReflPrompt condition, would lead to an increased value congruence (\vds) of the decision outcome (\selB). 
Further, we examine the correlations between \knowledge, \vds, and \IUIPC in an exploratory analysis.

\subsection{Results}
\label{sec:results}
Non-parametric tests were performed for all hypotheses, as the prerequisites for parametric testing were not met. All tests were conducted with an alpha level of 0.05. No p-value correction was utilized.\footnote{No p-value correction was utilized, as these are generally criticized for depending on the number of other tests performed and hence increase the likelihood of type two errors \cite{feiseMultipleOutcomeMeasures2002} Instead, all tests that were conducted regarding the main hypotheses, as well as the calculated p-value, are directly reported. Therefore, the results can be interpreted uniquely depending on the reader's preferences. For this, and in alignment with open science principles, the full dataset is provided along with the analysis scripts.
 The dataset and our analysis scripts are available under an open license on \href{https://osf.io/9b4r8/?view_only=0b7f26c7c7194c518b6d0dc0510e4617}{osf.io}. All statistical tests were performed using \textsc{R Studio} (version 2022.12.0), \textsc{R} (version 4.2.2).}. 
We failed to achieve normal distribution for a model including covariance through logarithmic and box-cox transformations\footnote{We recognize two possible reasons for the negative skewness of the IUIPC~\cite{malhotra_internet_2004}: First, MTurk participants might be `experienced' in participating in privacy-related studies and thus familiar with the IUIPC~\cite{hauser_common_2019}. Because of this experience, they might have responded differently, i.e., communicating deeper concerns for privacy. Second, the IUIPC uses loaded words such as \enquote{autonomy} and \enquote{privacy}, which, according to Groß~\cite{gros_validity_2021}, leads to a social desirability bias toward higher IUIPC scores.}. 
Hence, descriptive statistics of all measurements are presented in \autoref{fig:desctable}.

\paragraph{\textbf{H1:} Reflection Prompt}
To test whether the reflection prompt reduced \vds between \selA and \selB in each condition, three separate one-way, paired Wilcoxon signed-rank tests were conducted, assuming lower \vds for \selB.
None of them reached statistical significance. Hence, regarding \textbf{H1}, we did not find reflection prompts to lead to increased value congruence of decision outcomes in situations requiring an initial active decision.
All statistical tests for \textbf{H1} are reported in the \autoref{tab:h1table}.

\paragraph{\textbf{H2:} Priming}
A Shapiro-Wilk test showed that the assumption of a normal distribution must be rejected for \vds of \selA within the combined groups showing the \IUIPC before \selA (W = 0.90, p-value < 0.01). The independent samples Mann-Whitney U test did not show a significant difference between the groups (U = 6752.5, p = 0.059).
As no differences in the value congruence of initial decisions between the groups could be found (\textbf{H2}), we cannot confirm a priming effect due to a preceding privacy survey.

\paragraph{\textbf{H3:} Value-Centered Reflection Prompt}
To test whether the \vds of the resulting decision \selB is lower in the \ValueReflPrompt condition than in the \Refl and \ValueQuestRefl conditions, two unpaired samples Mann-Whitney-U tests are performed.
The comparison between \ValueReflPrompt (M=0.54, SD=0.29) and \Refl (M=0.62, SD=0.27) showed that the \vds of \selB was significantly lower in \ValueReflPrompt (U = 3996.5, $n_1$ < $n_2$ = 164, p = 0.018).
In contrast, the comparison between \ValueReflPrompt (M=0.54, SD=0.29) and \ValueQuestRefl(M=0.56, SD=0.30) revealed that the distributions of \selB's \vds were not significantly different between the two groups (U = 2663, $n_1$ < $n_2$ = 143, p = 0.281).
Supporting \textbf{H3}, we found the value-centered reflection prompt, \ValueReflPrompt, to increase congruence between the data-sharing decision and participants' general privacy values, when compared with the \Refl condition. However, if participants in \ValueQuestRefl engage with their privacy values by filling in a respective questionnaire, this difference compared with \ValueReflPrompt is of no statistical significance.

\begin{table}[ht]
  \centering
  \caption{Descriptive statistics per experimental condition. Reported as mean and standard deviation unless stated otherwise.}
  \label{tab:descriptive}
  \begin{tabular}{l c c c}
    \toprule
      & \textbf{(1) Refl}
      & \textbf{(2) ValueQuestRefl}
      & \textbf{(3) ValueReflPrompt} \\
    \midrule
  \textit{ValueDiscrepancyScore} of \textit{Selection 1} & 0.61 (0.28) &  0.55 (0.30) &  0.56 (0.29)\\  
  \textit{ValueDiscrepancyScore} of \textit{Selection 2} &  0.63 (0.27) &  0.57 (0.30) &  0.54,  (0.29)\\
  IUIPC & 5.76 (0.71) & 5.88 (0.69) &  5.85 (0.65) \\
  Knowledge Retainment Score &  2.83 (3.15) &  1.87 (2.92) &  2.81 (3.10) \\ 
  Age &  40.0 (11.9) &  38.34 (13.29) &  39.65 (12.28) \\
  \midrule
  \textit{\textbf{Gender (N)}} &&&\\
   \hline
  Male & 35 & 31 & 41 \\ 
  Female & 49 & 32 & 39 \\ 
  Nonbinary \& Prefer to self-describe & 0 & 0 & 0 \\ 
  \midrule
  \textit{\textbf{Education (N)}} &&&\\
   \hline
  Bachelor degree's or higher & 63 & 55 & 65\\ 
  Highschool or Associate Degree& 20 & 6 & 14\\ 
  Other & 1 & 2 & 1\\
  \bottomrule

\end{tabular}
\label{fig:desctable}
\end{table}

\paragraph{\textbf{Exploratory Analysis}}
To derive a deeper understanding of the role of individual privacy preferences and knowledge in achieving value-congruent data-sharing decisions, we conducted the following exploratory analyses:

First, in regards to \textbf{individual privacy preference}s, we investigated whether the proposed interface was more effective for people with high privacy concerns, we divided the sample into two groups using the overall mean of the \IUIPC scores as threshold: \enquote{high privacy concern,} \ie participants with an above average \IUIPC, and \enquote{low privacy concern,} \ie participants with a below average \IUIPC. 
For each group, unpaired samples Mann-Whitney U tests were used to examine whether the \vds of the decision outcome \selB is lower in \ValueReflPrompt than in \Refl.
While no significant difference was found in the low privacy concerns group (U = 799, $n_1$ < $n_2$ = 87, p = 0.244), the high privacy concern group exhibits a significant difference between \Refl (M = 0.66, SD = 0.328) and \ValueReflPrompt (M = 0.525, SD = 0.335), (U = 1270.5, $n_1$ < $n_2$ = 77, p = 0.003), suggesting that the value congruence interface is effective in assisting people who give high priority to privacy toward a value-congruent decision outcome.

Concerning \textbf{knowledge}, a Shapiro-Wilk test reveals a non-normal distribution of the knowledge retention score (W = $0.96$, p < $0.000$). 
Spearman's rank correlation was computed to assess the relationship between \vds of \selB and the \knowledge, revealing a negative correlation between the two variables (r(226) = -0.193, p = 0.003),  suggesting that higher levels of \knowledge, \ie being informed, are associated with higher value congruence of decision outcomes.
\knowledge was further found to exhibit a positive correlation with the \IUIPC (r(226) = 0.344, p < 0.000), suggesting that higher privacy concerns are associated with greater retention of knowledge about the disclosure options.
To compare the \knowledge scores between the experimental conditions, unpaired samples Mann-Whitney U tests are performed, revealing a significant difference in \knowledge only between \Refl and \ValueQuestRefl (U = 3109, $n_1$ < $n_2$ = 147, p = 0.033).

\subsection{Design Implications}
\label{sec:findingsexperiment}

\paragraph{Reflection Prompts in Active Decisions} 
In our experimental evaluation, we found no evidence that including a reflection prompt increases the value congruence of the decision outcome if an active selection by the user is required. This finding is notable, as it stands in contrast to previous empirical research regarding the use of reflection prompts, such as ~\citet{leimstadtner_investigating_2023}, which showed reflection prompts' capability to foster active, reflected decision-making and improve their value congruence by counteracting cognitive biases (\eg as exerted through an opt-out nudge).\footnote{An opt-out nudge leverages the status quo bias, that is, people’s affinity for defaults. It is among the most researched and effective types of nudge interventions~\cite{acquisti_nudges_2017,nouwens_dark_2020}}
These findings may indicate that an equivalent presentation of options, following the principles outlined in the GDPR that \textit{``it shall be as easy to withdraw as to give consent''}~\cite{european_commission_regulation_2016} could suffice to foster deliberate decision-making.

\paragraph{Value-Centered Reflection Increases Decision Quality}
Our results suggest that integrating value-centered reflection within consent interface design can increase the value congruence of active privacy choices, particularly for those users who value privacy highly.
This illustrates the role of values in decision processes and how reflecting on them could help decision-makers derive higher-quality decision outcomes for their given situation. 
In contrast, we did not find evidence of the \textit{Value-Centered Consent Interface} improving value congruence for participants with below-average privacy concerns. As an active decision was required, this may suggest that factors beyond privacy concerns had a greater influence on the decision for these participants (\eg trust~\cite{appenzeller_towards_2022,maus_enhancing_2020} or perceived control~\cite{kroll_digital_2021}).
Hence, it is vital to identify what values are central to individuals when applying design friction interventions to support them in considering the values most important to them. While our exemplary use of privacy provided insights into the relationship between reflection prompts and values in decision-making, future applications should seek to elicit those values important to an individual before introducing them as the subject of reflection. 

\paragraph{The Role of Knowledge}
Our exploratory analysis highlighted the central role of knowledge in achieving value-congruent, data-sharing decisions. The negative correlation between knowledge and the value-discrepancy score of the final decision outcome suggests that increased awareness of the decision is associated with value-congruent data-sharing behavior (see also~\cite{witteman_design_2016, munro_choosing_2016}). 
These findings underscore the importance of considering end-users' differing information needs when designing consent interfaces~\cite{munro_choosing_2016}, for example, by allowing access to information at varying levels of detail, enabling individuals to access the desired level of information.\\ 

In summary, we found no evidence that fostering reflection through \emph{design friction} during a decision process significantly increases the decision quality in terms of value congruence if an active decision is required. However, if values are introduced as the subject of reflection, the decision outcomes are more in line with people's values--especially if the value in question, namely, privacy, is vital to a given user. 
These online experiments served to delineate the design space. The design implications were consequently translated into a paper prototype for a consent interface for data donation in the application context, i.e., German BC. 
This prototype provided the basis for a co-creation workshop with domain experts, in which the findings of all preceding phases were integrated into a situated \textit{Value-Centered Consent Interface}, adapted for application toward BC data donation at German university hospitals.
The integration of domain experts serves to provide broader insight into the socio-technical context~\cite{huh-yooItWildWild2020,wong_bringing_2019} in which the proposed consent interface would be applied and to ensure ecological validity through the inclusion of multiple relevant research domains, i.e., medicine, ethics, and interaction design, before a high-fidelity prototype of the situated consent interface is implemented and brought into the field for user interface evaluations with patients.

\section{Phase 4: Co-Creation Workshop with Domain Experts}
\label{sec:experteval}
The three-hour co-creation workshop with domain experts took place in April 2023. We recruited a purposive sample of seven experts who brought different perspectives and knowledge related to the German BC context within the overarching research consortium's network. This included two medical practitioners (N=2) as well as researchers specializing in interaction design (N=1), medical informatics (N=1), critical educational technology (N=2), and medical ethics (N=1).
Informed consent was obtained in writing one week before the workshop. The university's responsible data protection officer approved the study information and consent form.

\subsection{Method}
\begin{figure*}[h]
    \includegraphics[width=\textwidth]{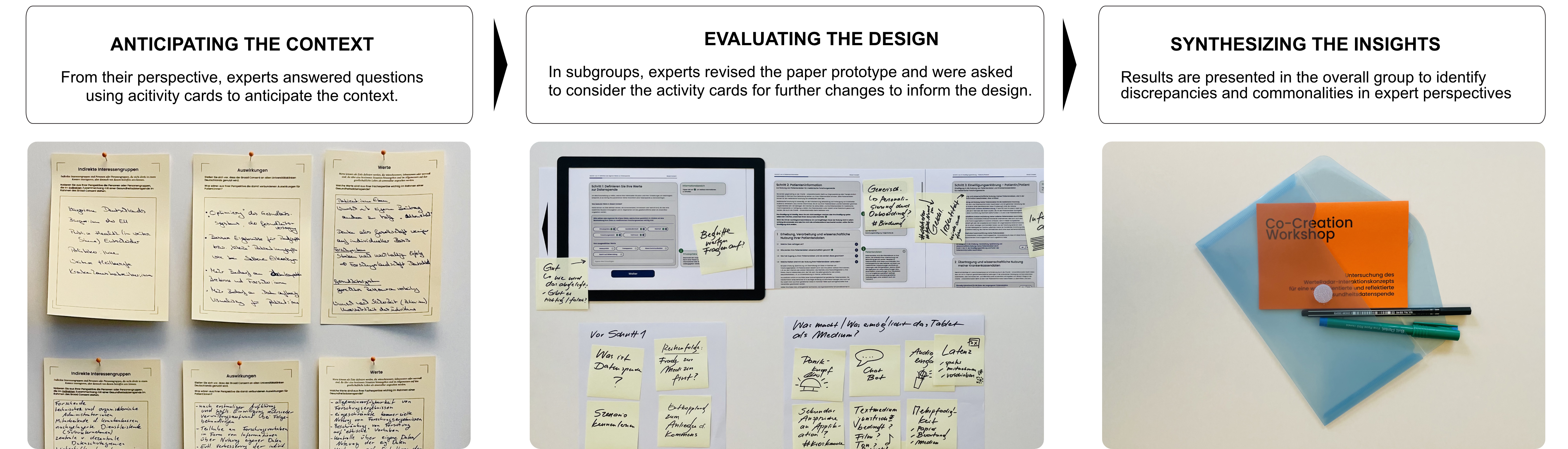}
    \caption{Overview of co-creation workshop procedure. First, activity cards (left) help experts anticipate and assess the context and the proposed \textit{Value-Centered Consent Interface} as a paper prototype (middle) that allows hands-on interaction and direct commentary. Finally, the participating experts presented and discussed the results in the overall group (right).}
    \label{fig:expertevaluation}
    \Description[]{This image shows the co-creation workshop procedure. First, experts used activity cards that helped to anticipate and assess the context, and second, experts revised the data donation user interface using a paper prototype that allows direct commentary. Finally, the participating experts presented and discussed the results in the overall group.}
\end{figure*}

The workshop was divided into three main phases (see \autoref{fig:expertevaluation}): In the first phase, a reflection activity was conducted to capture the experts' general perspectives on the application context. To facilitate this, we utilized a set of five activity cards (see~\autoref{sec:activitycards}), which participants individually responded to considering their specific areas of expertise, adapted from~\citet{nathan_envisioning_2008}. The phase concluded with a group discussion to share and discuss the gathered insights.
In the second phase, we introduced the proposed design through a paper prototype. The experts were divided into groups of two or three and tasked with reviewing and annotating the paper prototype. These groups were encouraged to utilize the activity cards created during the first phase as prompts to suggest adaptations and improvements~\cite{yoo_value_2013}.
In the final phase, the proposed changes from all the groups were synthesized. For this, each subgroup presented its annotated paper prototypes, followed by a discussion within the whole group to identify discrepancies and commonalities in the experts' perspectives.

To analyze the workshop results, we transcribed the presentation and discussion of the annotated paper prototypes verbatim, supplementing it with our notes. Through qualitative content analysis, i.e., inductive category formation following Mayring~\cite{mayring_qualitative_2014}, the first author of this paper examined the proposed changes and questions raised by the experts concerning the user interface design. The resulting codes were then consolidated into categories of improvement described below.

\subsection{Summary of Findings}
\label{sec:workshopfindings}
Based on our analysis, we identified three primary areas of improvement: (1) consideration of contextual factors in the information provision, (2) accessibility of the consent content, and (3) shifting the focus toward contextual and personal value-centered reflection rather than congruence calculations.

\paragraph{Contextual Factors}
The experts unanimously agreed on the need to strongly consider contextual factors when adapting the \textit{Value-Centered Consent Interface} to BC data donation. They highlighted the need to provide patients with information on the benefits and risks of donating medical data, mainly since data donation frequently constitutes a secondary task for those admitted to a hospital. To address this, the experts suggested adapting multi-modal information presentation, for example, by incorporating an explanatory video that can effectively convey the necessary information with reduced cognitive effort compared to the text format.

\paragraph{Accessibility}
Further, criticism was expressed regarding the length and verbosity of the BC data donation form, noting that it may be inaccessible to most patients. This aligns with the challenges identified in our participatory workshops (see \autoref{sec:valueelecitationworkshops}). As a solution, an easy-to-understand summary alongside the original document was recommended. In this regard, some experts proposed utilizing a multi-layer information provision format that accommodates different learning styles, allowing for individual user flows and improving user control.
Concerns were raised, particularly among medical experts, regarding the use of the term \textit{`value'} in the context of medical data donation. This term may be confusing, potentially leading to associations with medical values such as blood values. To address this, the experts suggested using alternative terminology, i.e., speaking of \textit{`important aspects'} to avoid confusion.

\paragraph{Contextual Value-Centered Reflection}
The experts emphasized that the specification of personal values should not be used to calculate a value congruence score, as it would contradict the principle of data minimization\footnote{~According to \S 3 a of the German Federal Data Protection Act (BDSG), public and non-public bodies that handle personal data should continuously operate under the condition that they only store, use, or process as much data as is necessary for the purpose in question.}. Instead, the explication of personal values should serve exclusively for the individuals' reflection rather than for scoring or measurement purposes.

\subsection{Design Implications}
\begin{figure*}[h]
  \includegraphics[width=350pt]{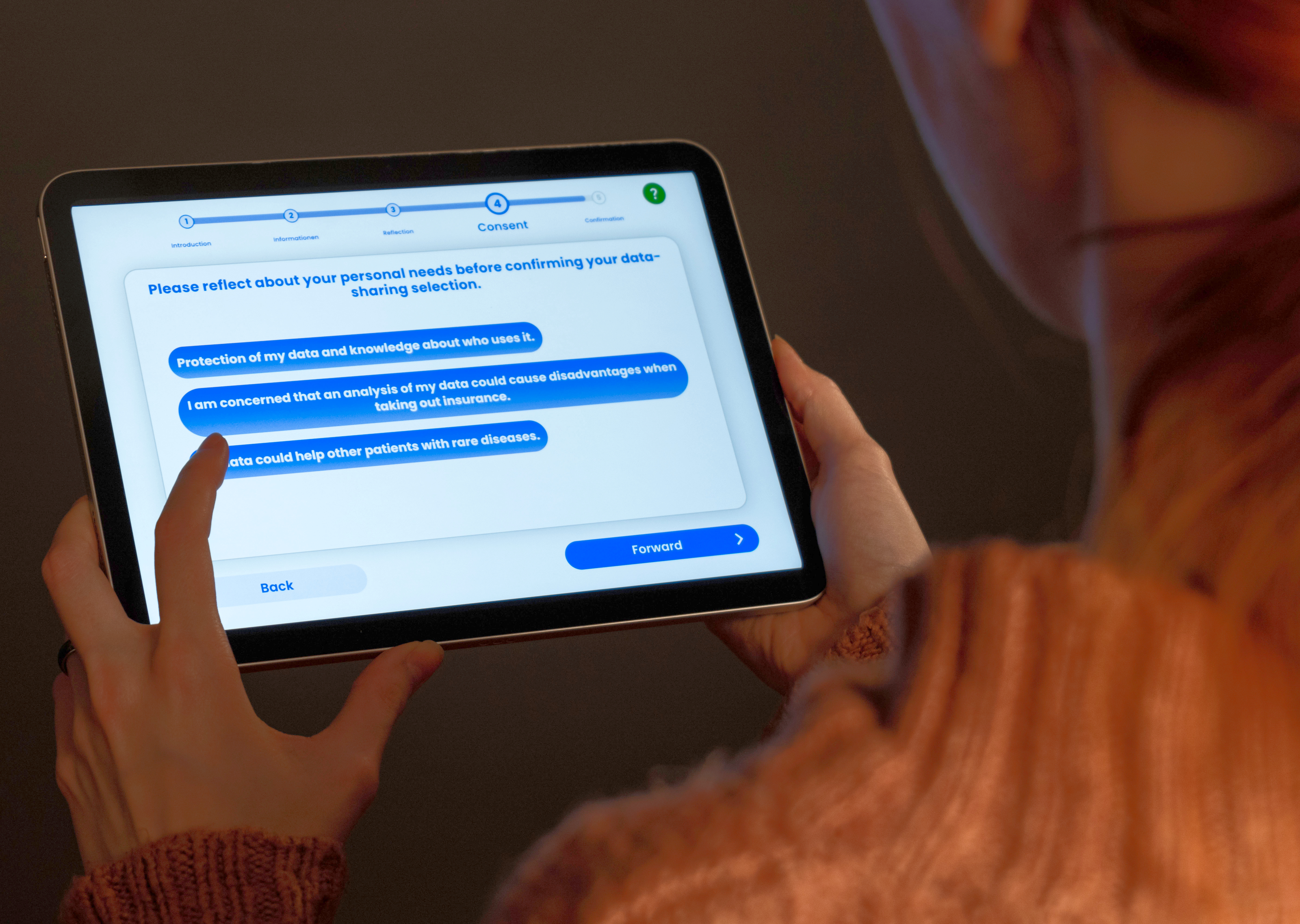}
  \caption{Interaction with the prototype: The \emph{Value-Centered Consent Interface} shows the time-out design friction as adapted to the BC application context after the domain expert co-creation workshop. For this paper, the text has been translated from German into English.}
  \label{fig:final_interface}
  \Description[]{This shows the time-out design friction in the \emph{Value-Centered Consent Interface} adapted to the BC application context after the domain expert co-creation workshop. The values expressed by the patients in response to initial value-elicitation questions are shown following a reflection prompt while a countdown timer runs out. For this figure, the text has been translated into English.}
\end{figure*}

The design iterations resulting from the co-creation expert workshop are summarized in~\autoref{tab:experttable}.
\begin{table}
    \centering
    \caption{Overview of the design recommendations and respective interface iterations.}
    \begin{tabular}[t]{ p{18.5em} p{18.5em}}
         \toprule
         \textbf{Design Recommendation} & \textbf{Interface Realization} \\
         \midrule
         \textit{\textbf{Segmentation}} \\
         \midrule
         \small The provision of information in the form of discrete, succinct units facilitates user engagement with the content at the requisite depth. &  \small The BC text is divided into smaller, self-contained sections on separate sub-pages, with a consistent and simple syntactic and lexical structure. \\
         \midrule
         \textit{\textbf{Multi-Linear Navigation}} \\
         \midrule
         \small Implementing a multi-linear navigation system allows users rapid access to the required level of information provision. & \small The user interface was expanded through navigation tools, i.e., a multi-step menu bar, enabling multi-linear navigation between sub-sections of the BC.  \\
         \midrule
         \textit{\textbf{Multi-Modal Information Provision}} \\
         \midrule
         \small  A multi-modal information provision accommodates individual differences in information needs and learning styles. & \small An explanation video, an FAQ section, and popups explaining uncommon concepts (e.g., pseudonymization) were incorporated into the interface. \\
         \midrule
         \textit{\textbf{Contextual Reflection Prompt}} \\
         \midrule
         \small  The value-reflection should adopt a qualitative approach to allow for contextual reflection inclusive of all possible personal values that may impact data-sharing decisions in a real-life setting.  & \small Users are asked a set of value-elicitation questions~\cite{lim_facilitating_2019} based on the four bioethics principles to support patients' self-determination and informed decision, i.e., beneficence, non-maleficence, autonomy, and justice~\cite{wilcoxAIConsentFutures2023}. The reflection prompt presents patients with their answers via a time-out design friction once a data-sharing decision is made. The questions can be found in \autoref{sec:valuequestions}.  \\
         \bottomrule
    \end{tabular}
    \label{tab:experttable}
\end{table}
By implementing these design recommendations, we aim to enhance the accessibility, clarity, and user experience of the BC data donation interfaces, promoting informed decision-making and value alignment.
Hence, after an exploration of patient values through participatory workshops (see ~\autoref{sec:valueelecitationworkshops}), a quantitative examination of \emph{design friction} as intervention for increasing value congruence through fostering reflection (see ~\autoref{sec:experimentaleval}), and a co-creation workshop with domain experts (see ~\autoref{sec:experteval}), the findings were realized through the implementation of a functional, high-fidelity prototype (illustrated in \autoref{fig:final_interface}).
This prototype was brought back to the application context to be tested with inpatients of a psychosomatic unit at a German university hospital for examining whether patients perceived the \textit{Value-Centered Consent Interface} as supporting their health data donation decisions.

\section{Phase 5: User Interface Evaluations with Patients}
\label{sec:usertests}
For the user interface evaluation of the situated \textit{Value-Centered Consent Interface}, participants, i.e., inpatients, followed a data donation decision process using the high-fidelity prototype, followed by a semi-structured interview and a usability survey. This study aimed to examine the extent to which the \textit{Value-Centered Consent Interface} is perceived as supporting value-congruent decisions by end-users within a socio-technical context, namely health data donation for patients of a university hospital.
Ethical approval was obtained through the ethics committee of the collaborating German university hospital. Medical researchers at the hospital's psychosomatic unit were responsible for recruiting participants. No monetary compensation was offered, but drinks and snacks were served. The evaluation was scheduled on Sundays to avoid interfering with therapy schedules. The treating physicians, i.e., the collaborating researchers, remained available for patients throughout the session. 

\subsection{Method}
The user interface evaluation was conducted in-lab with twelve inpatients from the university hospital's psychosomatic unit in December 2023 (N=6) and January 2024 (N=6). The participants' average age was 46,75 years (SD = 16,29).
The study procedure was as follows:
After providing informed consent, the participants received instructions on the thinking-aloud procedure by the conducting researcher (see \autoref{sec:usertestintructions}) and read the scenario text (see \autoref{sec:usertestscenario}). Each participant interacted with the prototype on a device to make a hypothetical decision regarding a health data donation. This interaction took, on average, $18.1$ minutes (SD = $4.92$).
Next, a semi-structured interview was conducted in an adjacent room by a separate researcher reviewing the experience of interacting with the prototype, followed by a short survey including UEQ-S\footnote{The short version of the \emph{User Experience Questionnaire} (UEQ). This questionnaire is a standard measure of the usability of a user interface. In contrast to other common usability questionnaires, the UEQ measures both hedonic and pragmatic usability~\cite{schreppComparisonSUSUMUXLITE2023}.} and demographic variables. The interview questions are reported in \autoref{sec:usertestinterview}. The interview sessions were audio recorded and transcribed verbatim. 
The transcripts were analyzed following Mayring's~\cite{mayring_qualitative_2014} procedure for inductive category formation in qualitative content analysis, which is especially suitable for material retrieved from open interview processes.
All resulting categories can be found in~\autoref{sec:userstudycategories}.
Based on the theoretical (see~\autoref{sec:relatedwork}, ~\autoref{sec:design}) and empirical (see~\autoref{sec:experimentaleval}) considerations, it was the goal of this analysis to, first, examine how participants' decision processes take place while interacting with the \textit{Value-Centered Consent Interface} and, second, how they perceive the overall consent interface, and finally, the value reflection prompt (task lock-out design friction) in particular, to be supportive to users' decision-making processes.  

\subsection{Summary of Findings}
\label{sec:userevalfindings}

Overall, participants emphasized the comprehensibility of the data donation decision process, perceiving the provided information as well-structured and sufficiently detailed. The segmentation of the overall information was positively received, highlighting the utility of making additional information available through pop-ups.
This is underscored by the results of the UEQ-S, showing an overall positive perception of the proposed interface's usability, rating it to be \textit{supportive} (M=6, SD=1.76), \textit{easy} (M=6.67, SD=0.89), \textit{clear} (M=6, SD=1.41), and \textit{interesting} (M=$6.6$, SD=$0.78$). For the perceived level of innovation, the patients' ratings were marginally lower and more varied: \textit{exciting} (M=$5.5$, SD=$2.28$), \textit{leading edge} (M=$5.92$, SD=$1.84$), and \textit{inventing} (M=$6$, SD=$1.21$). The lowest score concerned \textit{efficiency} (M=$5.75$, SD=$2.14$), reflecting some participants' concerns about text length and high page count.
The multi-modal presentation of information, suggested in our co-creation expert workshop and realized via explanatory video, was described as particularly helpful for understanding the decision context at hand. This finding echoes patients' calls for the inclusion of multimedia elements in our value-elicitation workshops (see \autoref{sec:valueelecitationworkshops}), highlighting the increased efficiency of such elements compared to textual descriptions.
Further, some patients suggested more specialized information provision measures, such as a support hotline. 

\paragraph{Value-Centered Reflection Prompt}
The participants indicated that engaging with the value-centered reflection prompt assisted them in comprehending and organizing their intentions in light of their specified values. In particular, reviewing decision outcomes alongside their initial values gave them a greater sense of control and deliberation.
Participant (P2) described encountering the \textit{Value-Centered Reflection Prompt} as follows: \textit{``I thought it was good because I actually read through the options again and then briefly considered changing it.''}
In this respect, the reflection questions helped participants to explicate their intentions and gain a broader perspective. The wording of the questions was generally perceived as easy to understand and helpful. However, one participant voiced concerns that specific vocabulary may not be accessible to those speaking German as a secondary language, pledging to use the simplest wording wherever possible. 

\paragraph{Values Guide Individual Decision Behaviour}
Overall, the findings of our qualitative content analysis show that values play a central role in guiding people's decision-making behavior: 
Patient values like altruism, medical progress, or the possibility of personal health gain motivate them to make their data available for secondary research purposes. At the same time, values related to privacy (e.g., anonymity and data treatment) are the key considerations dissuading patients from data donations. This central role of privacy within data donation decisions aligns with previous findings~\cite{gomez_ortega_towards_2021}. 
The values that are important to an individual consequently shape their information needs. If information needs regarding a particular value (or a salient topic related to it) cannot be satisfied, uncertainty results, and people may be dissuaded from making value-congruent decisions, for example, they don't share data even though they would want to support medical advancements, as they choose the decision perceived as reducing the uncertain risk.
Specific examples of this dissuasive role of uncertainty related to privacy include implications for health insurance, i.e., the need to disclose life-threatening chance findings, doubt about the extent of data disclosure, or the long timescales involved, i.e., 30 years of data storage. 
If such uncertainty cannot be mitigated by the information offered, patients will refrain from donating their data despite holding contrasting values. One participant (P4) illustrates this as follows: \textit{``So the most important thing that stuck with me is that my data donation helps others. That was actually the most important thing for me. The other thing [referring to uncertainty regarding health insurance involvement] ruined it all.''}
Furthermore, we found values to shape patients' perception of the presented consent interface:
For a subgroup of participants intrinsically motivated to donate data based on long-term values (e.g., altruism), the effort of engaging with the interface was perceived as redundant since their decision was determined from the first interaction (e.g., on the welcome page). The following provision of in-depth information was deemed unnecessary and time-consuming, preventing the timely confirmation of a preconceived decision.
One participant (P7) describes this as follows: \textit{``I'm asked: `Would you like to share your data for research?', along with the reasoning: ``It will help you and others. I respond: `Yes, no problem! Where do I have to sign?' I won't look at the privacy policy because, personally, it's way too much for me to take 15 minutes to understand it. You want my data? Let me sign it.''}
Overall, these findings highlight the need for a tailored approach to providing information at a level of detail that aligns with individuals' specific information requirements.

\section{Discussion}
\label{sec:discussion}
As recent shifts in digital health data infrastructures~\cite{scheuermanDatasetsHavePolitics2021} exacerbated power differentials disadvantaging the values of vulnerable communities like patients~\cite{cajander_electronic_2019}, contextualized approaches for supporting their values in consent mechanisms are required~\cite{tsengDataStewardshipClinical2024}.
Hence, we adopted a human-centered design perspective to conceptualize and evaluate a \textit{Value-Centered Consent Interface} grounded in patient values within our research context, i.e., data donation through the BC at German university hospitals.
For this, we began our human-centered design processes by exploring patients' values through participatory workshops. 
The analysis of the workshop findings informed design requirements, namely the need for a consent interface that supports patients in terms of providing \textit{transparency} on the consent process, enabling \textit{reflection} on the consequences of their actions, and consequently 
achieving \textit{autonomy} in their decision-making (see \autoref{sec:valueelecitationworkshops}).
We integrated these requirements with insights from literature on value clarification techniques~\cite{lim_facilitating_2019,berry_supporting_2021} and the use of reflection to foster deliberate decision-making~\cite{jackson_addressing_2018,terpstra_improving_2019,leimstadtner_investigating_2023,baumer_reflective_2015,cox_design_2016} to conceptualize the \textit{Value-Centered Consent Interfaces}. 
This consent interface takes the role of a decision facilitator that supports data subject values in the consent process by both fostering value-congruent decision-making through reflection (see \autoref{sec:design}) and creating transparency regarding the consent process through a composite information approach (see \autoref{sec:experteval}).
To evaluate the proposed interface, we conducted online experiments examining the effectiveness of reflection in fostering congruent decisions. Next, co-creation expert workshops developed a contextualized high-fidelity prototype for the German BC data donation context. Finally, we returned to said context to evaluate the resulting consent interface with inpatients at a university hospital. 

In the following discussion, we integrate insights gained throughout this three-year research effort on (1) the roles of patients' values and information needs in decision-making processes, (2) how reflection can support patients therein, and (3) review how our mixed-methods approach served to address power differentials in health technology design.

\subsection{Supporting Patient Values and Information Needs in Data-Sharing Decisions}
Previous work (e.g., ~\cite{okane_non-static_2013, jacobsComparingHealthInformation2015})  has highlighted how patients' willingness to share medical information depends on context factors, including temporal, i.e., willingness to share may change throughout the progression of a health condition, or interpersonal factors, i.e., depending on the professional role of the person providing the information.
In our initial value elicitation workshops, patients expressed values that required a consent interface to provide them with transparency, autonomy, and the possibility to reflect on the potential consequences of making their medical data available to researchers.
After interacting with the situated \textit{Value-Centered Consent Interface}, patients brought up values such as altruism or belief in scientific progress as the critical factors motivating their data donations. 
These findings align with previous research by Ortega et al.~\cite{gomezortegaDataTransactionsFramework2023}, which found that the positive feeling of contributing to science or the prospect of future benefits motivates data donors. The authors argue that these factors contribute to the perception of data donation as a reciprocal transaction in which both parties receive value~\cite{gomezortegaDataTransactionsFramework2023}.
In light of the context-specific nature of digital health technologies~\cite{kassamPatientPerspectivesPreferences2023}, we posit that patient-centered consent interfaces need to (1) help data subjects deliberate the relevant values for a given situation and (2) consider the alignment between their values and their decision outcome. 

Beyond decision outcome, the values an individual holds important are a central factor shaping their information needs concerning medical data-sharing: Motivating values (e.g.,\textit{possible health benefits, societal gain}) reduce information needs, while others increase them (e.g., privacy). 
Crucially, uncertainty arises when information needs relating to a value cannot be met by the provided information. In such instances, individuals may revert to a decision associated with the least perceived risk, regardless of its alignment with their values~\cite{munro_choosing_2016}. Our experimental findings substantiate this relationship between values and information needs, demonstrating a positive correlation between decision quality, i.e., value alignment, and information acquisition. This highlights the importance of addressing individuals' specific information needs to achieve value-congruent decisions. 
Within the research context of BC data donations, we found critical challenges related to unmet information needs: Our participatory workshops indicated that the long text form and dense legal language of the original BC form presented significant barriers to patients, rendering them unable to locate relevant information.
Informed by the insights from our expert workshop, we addressed this issue by adopting a multi-modal approach to information provision, i.e., through a visual instructional video, along with a multi-layered layout segmenting the patient information into smaller chunks.
Adopting a multimodal approach has been indicated to effectively promote reflexive processes, as integrating various media can enhance a deeper understanding of the complexity of an experience~\cite{espositoCounselingReflexiveProcesses2018}.
Furthermore, recent findings in personal informatics support using layered approaches to provide users control over the received information toward supporting decision-making processes through the promotion of a deeper understanding of a decision context (see~\cite{bentvelzen_designing_2022}) and increasing procedural transparency~\cite{huh-yooItWildWild2020}.

Adaptive interface strategies can serve to address differing information needs on a secondary level~\cite{maus_enhancing_2020}: Those whose decisions are impeded by uncertainty require access to further details to accommodate their information needs. Individuals who are inclined to make immediate decisions based on overarching values (e.g., altruism) require only the most essential information to avoid dissatisfaction, fatigue, or even process abandonment.
In line with our propositions, \citet{kassamPatientPerspectivesPreferences2023} argued that the design consideration concerning consent design needs to look beyond document length and focus on innovations regarding the presentation format of consent information. The authors highlighted how customizable electronic formats can facilitate informed decisions by giving users control over the required level of information.
A promising approach for future work that takes an adaptive approach to give users control over their consent based on their information needs is the so-called \textit{``meta-consent''}~\cite{cumynMetaconsentSecondaryUse2021}. This type of consent interface takes data-sharing decisions out of the respective decision context. It presents these decisions on a centralized decision platform, which provides users control over the granularity regarding data uses and frequency of requests for a particular topic at which consent is required. Such an approach may counteract consent fatigue~\cite{utz_informed_2019} and heuristic biases arising from data-sharing decisions regularly constituting secondary tasks~\cite{acquisti_nudges_2018}. Within HCI, existing efforts have investigated similar concepts, for example, ~\citet{faith_cranor_platform_2011}'s approach to automatically applying stored privacy preferences or ~\citet{colnago_informing_2020}'s personalized privacy assistants. 

In summary, we find that values shape people's willingness to provide data for research. These values are closely linked to information needs that ought to be accommodated to enable value-congruent decision-making while avoiding uncertainty or consent fatigue. In this regard, a multifaceted and multimodal approach can support patients by providing them with control over the (level of) information they receive.

\subsection{Designing for Reflection toward Value-Congruent Decision-Making}
As common consent interfaces disadvantage data subjects through leveraging heuristic decision patterns toward increased data-sharing rates (see~\cite{utz_informed_2019, fassl_stop_2021,cajander_electronic_2019, wilcoxAIConsentFutures2023,scheuermanDatasetsHavePolitics2021}), we approached designing for reflection by aiming to empower data subjects to act in accordance with their values~\cite{baumer_reflective_2015}.  
Through introducing values as the subject of reflection before a decision, we expand on prior research, fostering reflection within interface design to aid data-sharing decisions~\cite{wang_field_2014,leimstadtner_investigating_2023,zhang_nudge_2021}. 
Our experimental findings indicated that value-centered reflection could increase value congruence of decision outcomes, especially for those for whom it matters most: people who value privacy highly (see \autoref{sec:findingsexperiment}). 
This finding underscores the potential of reflection toward helping people gain self-knowledge about their values~\cite{berry_supporting_2021}, further positing reflection-based interventions as an approach toward supporting value-congruent decision-making in sensitive contexts. 
In this regard, we provide further insight into the use of cognitive-forcing functions~\cite{bucinca_trust_2021,gould_task_2015} as a means of counteracting people's demonstrated tendency to make inappropriate choices, disregarding good knowledge and conflicting personal values (see, e.g., ~\cite{munro_choosing_2016}), especially in privacy-related contexts~\cite{waldman_cognitive_2020, jackson_addressing_2018}.
The findings of our user interface evaluations further corroborate the beneficial impact of value-centered reflection. Patients reported that engaging in value-centered reflection facilitated a sense of clarity and control concerning their data donation. This perception starkly contrasts with those observed during our initial participatory workshops for value elicitation (see \autoref{sec:valueelecitationworkshops}), where patients expressed feeling a loss of data ownership and confusion when faced with the original paper-based BC form.

Taken together, our findings demonstrate how integrating value clarification~\cite{munro_choosing_2016} with design frictions prompting reflection~\cite{croskerry_cognitive_2003} can support patients' decision quality in terms of value congruence~\cite{leimstadtner_investigating_2023,witteman_design_2016}.
Our proposal of a \textit{Value-Centered Consent Interface} seeks to stimulate future research that posits consent interfaces as a means to empower end-users toward deliberate and informed decision-making rather than increase uninformed data-collection rates. This approach responds to the call for context-specific approaches for improving the mechanisms for supporting patient values in obtaining informed consent for secondary research purposes within CSCW~\cite{tsengDataStewardshipClinical2024,scheuermanDatasetsHavePolitics2021}.
Future patient-centered design endeavors in the digital health realm can build on our insights into the use of design friction for supporting reflection in patient decisions for use in critical situations beyond BC, for example, in shared decision-making~\cite{charlesSharedDecisionmakingMedical1997}. 
In addition, future research should consider the influence of the context in which patient medical data donors interact with such a consent interface. As reported in our patient workshops, data donation consent forms are commonly encountered as secondary tasks while undergoing healthcare procedures. These circumstances may critically influence the amount of attention and time devoted to deriving a deliberate decision. In this respect, a digital solution such as the \textit{Value-Centered Consent Interface} may help to take patients' decisions out of the clinical context and into more reflective, temporally and spatially distanced environments, for example, accessing the consent form online from home. 

\subsection{Methodological Lessons for Addressing Patient Values in Health Data Practices}
Work in HCI and CSCW has followed the expansion of knowledge infrastructures from data as a static resource toward recent paradigms of fluid knowledge within \textit{big data} and \textit{open science}~\cite{tsengDataStewardshipClinical2024}. 
Ethical considerations must accompany these changes, particularly regarding the way in which individuals grant consent to the data collections fueling these advancements~\cite{tsengDataStewardshipClinical2024, wilcoxAIConsentFutures2023}. As future uses of disclosed data become increasingly opaque, modes of consent applicable to this new breadth of data uses are needed (e.g., the broad consent~\cite{zenker_data_2021}).
Recent work on disciplinary values within the practices of data-driven domains by Scheuerman et al.~\cite{scheuermanDatasetsHavePolitics2021} showcased how values can become embedded in datasets. Yet, these domains lack a reflexive culture to consider whose values find representation~\cite{scheuermanDatasetsHavePolitics2021}. Beyond, the authors identified dehumanizing practices in machine learning that deal with highly sensitive data, such as disregarding ethical considerations when humans constitute data instances~\cite{scheuermanDatasetsHavePolitics2021}. 
This extends to the domain of digital health, where an absence of consideration of the ethical implications of data collection practices has been described, particularly concerning the treatment of human data subjects, rendering their role \textit{``undervalued and understudied''}~\cite{wilcoxAIConsentFutures2023}. The neglect of data subjects' values in the technology mediating consent processes can entail far-reaching discrimination consequences: If the modes of data collection lead to the exclusion of vulnerable data subjects (e.g., through unmet information needs and uncertainty), this may result in an underrepresentation of marginalized groups in the resulting dataset, risking their systemic discrimination in clinical decision support systems~\cite{tsengDataStewardshipClinical2024}. 
This absence of consideration for data subjects' values perpetuates a power imbalance inherent to current forms of consent interfaces~\cite{utz_informed_2019, fassl_stop_2021,cajander_electronic_2019, wilcoxAIConsentFutures2023,scheuermanDatasetsHavePolitics2021}.
It is, therefore, imperative to investigate how power differentials shape the design of technology and the implementation of measures to guarantee the representation of the values of vulnerable stakeholder groups in domains such as digital health, where sensitive data and vulnerable populations are handled~\cite{jackson_addressing_2018}. 

Our engagements with patients and experts in our research context, the BC at German university hospitals, demonstrated how the existing standardized and legally compliant paper version of the BC data donation form employed across German university hospitals presented patients with significant barriers toward reaching an informed data-sharing decision. We argue that this stems partly from a lack of patient engagement in the development process of the current BC form (discussed in \autoref{sec:context}).
Recent work has argued for the need to bring human-centered design methodologies and PD approaches to topics related to the ethical construction of datasets~\cite{birhanePowerPeopleOpportunities2022,cooperSystematicReviewThematic2022,wilcoxAIConsentFutures2023}, to empower data subject groups toward deliberate and value-congruent data-sharing and build toward ethical and inclusive data practices in digital health~\cite{scheuermanDatasetsHavePolitics2021,wilcoxAIConsentFutures2023}, rather than merely satisfying regulatory requirements when vulnerable end-user groups, such as patients, are involved ~\cite{fassl_stop_2021}.
In light of the situated nature of digital health applications, qualitative and contextualized PD approaches present a means to ground patient-centered design of technology in the sociotechnical reality in which end-users will engage with it~\cite{kassamPatientPerspectivesPreferences2023}.

The present study presents an example of how adopting an \emph{sequential mixed-methods approach}~\cite{creswell_designing_2017} can be employed toward technology design representative of stakeholder values~\cite{birhanePowerPeopleOpportunities2022,wong_bringing_2019}, building toward a larger goal of ethical data practices within digital health technology~\cite{scheuermanDatasetsHavePolitics2021}. 
The distinctive feature of this approach is that the initial qualitative phase focuses on the values of the targeted vulnerable stakeholder group, i.e., patients, situated within their socio-technical context, i.e., BC data donation at German university hospitals~\cite{sodergrenEnsuringInclusivityWellbeing2024a}. 
Putting such an open exploration of stakeholder values as the starting point for technology design enables researchers and designers to identify where existing technological infrastructure might embed value conflicts between different stakeholders~\cite{scheuermanDatasetsHavePolitics2021,friedman_survey_2017}.
It thereby avoids the pitfalls in PD associated with basing participatory activities on existing technologies (and the values embedded therein) or the simultaneous engagement of stakeholder groups with different values~\cite{parkPowerDynamicsValue2022}. Both approaches potentially disadvantage vulnerable stakeholder groups in negotiations over design requirements due to power imbalances and differences in influence~\cite{tschiderConsentMythImproving2018,tsengDataStewardshipClinical2024,cajander_electronic_2019}.
By conceptualizing the \textit{Value-Centered Consent Interface} based on the patient values elicited in the initial workshops and evaluating the interface mechanisms in an online experiment, we find in our concluding user interface evaluations with patients that the resulting interface concepts succeed in supporting them in providing transparency on the data donation process and fosters informed decision-making through supporting reflection on relevant values for a given decision-context (see \autoref{sec:userevalfindings}).

In summary, research in the field of CSCW has long focused on healthcare as a workplace, where \textit{``patients were invisible apart from being the object of the information and coordination efforts''}~\cite{fitzpatrickReview25Years2013a}. However, the advent of new data collection methods has necessitated the development of ethical consent frameworks in digital healthcare, empowering data subjects to act according to their values.  
Our present study contributes to a reflective approach to the design of dataset infrastructures in digital healthcare~\cite{scheuermanDatasetsHavePolitics2021}. To this end, the participatory involvement of end-users and situated case studies is essential for designing technology that supports new data practices and challenges the power structures embedded into the current data infrastructure.

\section{Limitations}

\label{sec:limitations}
In the following, we review the limitations across the various phases of our mixed-methods approach.

\paragraph{Participatory Workshops for Value Elicitation} Our initial qualitative examination of the application context using participatory workshops and value clarification techniques (\eg \cite{berry_supporting_2021,lim_facilitating_2019}) constituted an in-depth exploration of participants' experiences, perceptions, and values. Due to the sensitive nature of the participant group, it was organizationally only possible to conduct a limited number of workshops, which may have resulted in a failure to achieve theoretical saturation (see~\cite{saunders_saturation_2018}). Yet, through the inclusion of patient representatives in one workshop, presenting a broader and integrative patient perspective, we posit that the collected data provided an appropriate and valuable perspective for a user interface design in the given research context, enhancing our understanding of its complexities and diverse patient needs~\cite{huh-yooItWildWild2020,wong_bringing_2019}. In this phase, we also recognized the importance of ongoing engagement with our participants as stakeholders in the design process beyond the workshops to align with the values we advocate for in our work. In this regard, we continued close engagement with patient advocates by publishing the workshop results in their community magazines and providing updates on our research progress to patient advocates.

\paragraph{Online Experiment} The quantitative phase evaluated the design derived from the qualitative phase's findings~\cite{creswell_designing_2017}. To extend the examination of the cognitive effects involved in prompting reflection to a larger (and potentially generalizable sample~\cite{redmilesHowWellMy2019}), we conducted a crowdsourcing experiment on the platform MTurk. While this platform is commonly used for online experiments (see, \eg ~\cite{kittur_crowdsourcing_2008}), we encountered a large number of data quality issues (see \autoref{sec:inclusioncriteria}). These challenges may be attributed to the availability of automated completion and texting tools on MTurk, which might impact data quality~\cite{webb_too_2022}. For future research, we recommend the consideration of alternative platforms with stricter account validation procedures, such as \textit{Prolific} (see~\cite{peer_beyond_2017}), to address these challenges.
Next, the online experiment was conducted on a fictional health data platform. Hence, the recorded decisions describe intentions rather than real-life behavior. ~\citet{kassamPatientPerspectivesPreferences2023}'s literature review of patient perspectives on digital health found such use of fictional scenarios to be prevalent in studies examining behavior involving sensitive data, arguing for more research that combines qualitative and quantitative approaches to understand individual preferences and behaviors better.
Given the open-ended definition of values in Friedman's conceptualization~\cite{friedman_survey_2017} (see~\autoref{sec:informeddecision}), we referenced existing research identifying privacy as a key value in data-sharing decisions to operationalize value-congruence for the quantitative evaluation of the proposed consent interface. Other values may have influenced participant behavior, potentially confounding our results. Through our expansion of the exploratory sequential mixed-method~\cite{creswell_designing_2017} approach by a concluding return to the application context, we addressed these limitations of the quantitative phase by evaluating the proposed interaction concept in a realistic healthcare setting under an openness of values that might be of interest to a patient. 

\paragraph{User Interface Evaluations} 
Finally, for our concluding user interface evaluations with patients, we implemented measures to counteract possible limitations due to demand characteristic bias~\cite{dellYoursBetterParticipant2012}, namely a neutral formulation of the user tasks (see \autoref{sec:usertestintructions}) and separating the roles of the user tests conductor and interviewer. 
Nevertheless, a confounding influence on participants' responses on the preferability of the interface design, i.e., UEQ-S, cannot be excluded. 
However, we assume that the core research objective, i.e., the role of values in data donation decision processes, lies outside the perceptions affected by such social desirability biases.

\section{Conclusions and Future Work}
\label{sec:conclusions}
Our research highlights the broader need to investigate the implications of new data practices within their socio-technical context. A reflexive approach to technology design requires critical consideration of the values embedded within the technologies mediating data, seeking to support vulnerable stakeholders whose values may be overlooked~\cite{scheuermanDatasetsHavePolitics2021}, as digital healthcare requires a balance between different stakeholder values to safeguard patient privacy while realizing the benefits of data-driven medical innovation~\cite{kassamPatientPerspectivesPreferences2023}. 
Specifically, our research contributes to the design of consent interfaces in digital health by establishing the barriers present in current solutions and providing directions for the future design of consent interfaces that support patients in making value-congruent decisions through the introduction and evaluation of the \emph{Value-Centered Reflection Prompt}. By introducing and evaluating this novel interface intervention, we demonstrate the central role that individual values play in shaping people's data-sharing decisions and information needs. 
Focusing on the research context of health data donation, we demonstrate how a human-centered design approach can help realize a design grounded in the perspective of a vulnerable group, i.e., patients, following an \emph{exploratory sequential mixed-methods approach}. 
We encourage further exploration and adoption of mixed-methods approaches in CSCW, particularly for digital health. Insights gained through the combination of qualitative and quantitative methods present a valuable resource for informing the discourse between different scientific paradigms at the center of interdisciplinary collaboration central to advancing digital health efforts (e.g., medicine, design, ethics, bioinformatics). Further, this can enable researchers to gain richer insights into the complex interplay between the multiple stakeholders' perspectives and values within clinical realities.
In this regard, we posit that a human-centered design approach presents a way to steer against the tendency toward \textit{solutionism} within digital health, enabling digital health practices that empower end-users by creating technology that supports them in acting in alignment with their values.

\begin{acks}
We thank the patients and domain experts for their time and valuable contributions to our workshops and user interface evaluations, which have helped inform our research. We also appreciate the reviewers for their thoughtful and constructive feedback. This work was supported by the Federal Ministry of Education and Research (grant 16SV8463: WerteRadar).
\end{acks}

\bibliographystyle{ACM-Reference-Format}
\bibliography{references}


\begin{thebibliography}{116}


\ifx \showCODEN    \undefined \def \showCODEN     #1{\unskip}     \fi
\ifx \showDOI      \undefined \def \showDOI       #1{#1}\fi
\ifx \showISBNx    \undefined \def \showISBNx     #1{\unskip}     \fi
\ifx \showISBNxiii \undefined \def \showISBNxiii  #1{\unskip}     \fi
\ifx \showISSN     \undefined \def \showISSN      #1{\unskip}     \fi
\ifx \showLCCN     \undefined \def \showLCCN      #1{\unskip}     \fi
\ifx \shownote     \undefined \def \shownote      #1{#1}          \fi
\ifx \showarticletitle \undefined \def \showarticletitle #1{#1}   \fi
\ifx \showURL      \undefined \def \showURL       {\relax}        \fi
\providecommand\bibfield[2]{#2}
\providecommand\bibinfo[2]{#2}
\providecommand\natexlab[1]{#1}
\providecommand\showeprint[2][]{arXiv:#2}

\bibitem[Acquisti(2009)]%
        {acquisti_nudging_2009}
\bibfield{author}{\bibinfo{person}{Alessandro Acquisti}.} \bibinfo{year}{2009}\natexlab{}.
\newblock \showarticletitle{Nudging {Privacy}: {The} {Behavioral} {Economics} of {Personal} {Information}}.
\newblock \bibinfo{journal}{\emph{IEEE Security \& Privacy Magazine}} \bibinfo{volume}{7}, \bibinfo{number}{6} (\bibinfo{date}{Nov.} \bibinfo{year}{2009}), \bibinfo{pages}{82--85}.
\newblock
\showISSN{1540-7993}
\urldef\tempurl%
\url{https://doi.org/10.1109/MSP.2009.163}
\showDOI{\tempurl}


\bibitem[Acquisti et~al\mbox{.}(2017)]%
        {acquisti_nudges_2017}
\bibfield{author}{\bibinfo{person}{Alessandro Acquisti}, \bibinfo{person}{Idris Adjerid}, \bibinfo{person}{Rebecca Balebako}, \bibinfo{person}{Laura Brandimarte}, \bibinfo{person}{Lorrie~Faith Cranor}, \bibinfo{person}{Saranga Komanduri}, \bibinfo{person}{Pedro~Giovanni Leon}, \bibinfo{person}{Norman Sadeh}, \bibinfo{person}{Florian Schaub}, \bibinfo{person}{Manya Sleeper}, \bibinfo{person}{Yang Wang}, {and} \bibinfo{person}{Shomir Wilson}.} \bibinfo{year}{2017}\natexlab{}.
\newblock \showarticletitle{Nudges for {Privacy} and {Security}: {Understanding} and {Assisting} {Users}’ {Choices} {Online}}.
\newblock \bibinfo{journal}{\emph{Comput. Surveys}} \bibinfo{volume}{50}, \bibinfo{number}{3} (\bibinfo{year}{2017}).
\newblock
\showISSN{0360-0300}
\urldef\tempurl%
\url{https://doi.org/10.1145/3054926}
\showDOI{\tempurl}
\newblock
\shownote{Place: New York, NY, USA Publisher: Association for Computing Machinery}.


\bibitem[Acquisti et~al\mbox{.}(2018)]%
        {acquisti_nudges_2018}
\bibfield{author}{\bibinfo{person}{Alessandro Acquisti}, \bibinfo{person}{Idris Adjerid}, \bibinfo{person}{Rebecca Balebako}, \bibinfo{person}{Laura Brandimarte}, \bibinfo{person}{Lorrie~Faith Cranor}, \bibinfo{person}{Saranga Komanduri}, \bibinfo{person}{Pedro~Giovanni Leon}, \bibinfo{person}{Norman Sadeh}, \bibinfo{person}{Florian Schaub}, \bibinfo{person}{Manya Sleeper}, \bibinfo{person}{Yang Wang}, {and} \bibinfo{person}{Shomir Wilson}.} \bibinfo{year}{2018}\natexlab{}.
\newblock \showarticletitle{Nudges for {Privacy} and {Security}: {Understanding} and {Assisting} {Users}’ {Choices} {Online}}.
\newblock \bibinfo{journal}{\emph{Comput. Surveys}} \bibinfo{volume}{50}, \bibinfo{number}{3} (\bibinfo{date}{May} \bibinfo{year}{2018}), \bibinfo{pages}{1--41}.
\newblock
\showISSN{0360-0300, 1557-7341}
\urldef\tempurl%
\url{https://doi.org/10.1145/3054926}
\showDOI{\tempurl}


\bibitem[Aguinis et~al\mbox{.}(2021)]%
        {aguinis_mturk_2021}
\bibfield{author}{\bibinfo{person}{Herman Aguinis}, \bibinfo{person}{Isabel Villamor}, {and} \bibinfo{person}{Ravi~S. Ramani}.} \bibinfo{year}{2021}\natexlab{}.
\newblock \showarticletitle{{MTurk} {Research}: {Review} and {Recommendations}}.
\newblock \bibinfo{journal}{\emph{Journal of Management}} \bibinfo{volume}{47}, \bibinfo{number}{4} (\bibinfo{date}{April} \bibinfo{year}{2021}), \bibinfo{pages}{823--837}.
\newblock
\showISSN{0149-2063, 1557-1211}
\urldef\tempurl%
\url{https://doi.org/10.1177/0149206320969787}
\showDOI{\tempurl}


\bibitem[Akoglu(2018)]%
        {akoglu_users_2018}
\bibfield{author}{\bibinfo{person}{Haldun Akoglu}.} \bibinfo{year}{2018}\natexlab{}.
\newblock \showarticletitle{User's guide to correlation coefficients}.
\newblock \bibinfo{journal}{\emph{Turkish Journal of Emergency Medicine}} \bibinfo{volume}{18}, \bibinfo{number}{3} (\bibinfo{date}{Aug.} \bibinfo{year}{2018}), \bibinfo{pages}{91--93}.
\newblock
\showISSN{2452-2473}
\urldef\tempurl%
\url{https://doi.org/10.1016/j.tjem.2018.08.001}
\showDOI{\tempurl}


\bibitem[Alashoor et~al\mbox{.}(2017)]%
        {alashoorPrimingEffectProminent2017}
\bibfield{author}{\bibinfo{person}{Tawfiq Alashoor}, \bibinfo{person}{Grace Fox}, {and} \bibinfo{person}{H Smith}.} \bibinfo{year}{2017}\natexlab{}.
\newblock \showarticletitle{The {Priming} {Effect} of {Prominent} {IS} {Privacy} {Concerns} {Scales} on {Disclosure} {Outcomes}: {An} {Empirical} {Examination}}. In \bibinfo{booktitle}{\emph{Pre-{ICIS} {Workshop} on {Information} {Security} and {Privacy}}}. \bibinfo{address}{Seoul}.
\newblock


\bibitem[Appenzeller et~al\mbox{.}(2021)]%
        {appenzeller_cpiq_2021}
\bibfield{author}{\bibinfo{person}{Arno Appenzeller}, \bibinfo{person}{Thomas Kadow}, \bibinfo{person}{Erik Krempel}, {and} \bibinfo{person}{Jürgen Beyerer}.} \bibinfo{year}{2021}\natexlab{}.
\newblock \showarticletitle{{CPIQ} - {A} {Privacy} {Impact} {Quantification} for {Digital} {Medical} {Consent}}. In \bibinfo{booktitle}{\emph{The 14th {PErvasive} {Technologies} {Related} to {Assistive} {Environments} {Conference}}}. \bibinfo{publisher}{ACM}, \bibinfo{address}{Corfu Greece}, \bibinfo{pages}{534--543}.
\newblock
\showISBNx{978-1-4503-8792-7}
\urldef\tempurl%
\url{https://doi.org/10.1145/3453892.3461653}
\showDOI{\tempurl}


\bibitem[Appenzeller et~al\mbox{.}(2022)]%
        {appenzeller_towards_2022}
\bibfield{author}{\bibinfo{person}{Arno Appenzeller}, \bibinfo{person}{Nick Terzer}, \bibinfo{person}{Erik Krempel}, {and} \bibinfo{person}{Jürgen Beyerer}.} \bibinfo{year}{2022}\natexlab{}.
\newblock \showarticletitle{Towards {Private} {Medical} {Data} {Donations} by {Using} {Privacy} {Preserving} {Technologies}}. In \bibinfo{booktitle}{\emph{The15th {International} {Conference} on {PErvasive} {Technologies} {Related} to {Assistive} {Environments}}}. \bibinfo{publisher}{ACM}, \bibinfo{address}{Corfu Greece}, \bibinfo{pages}{446--454}.
\newblock
\showISBNx{978-1-4503-9631-8}
\urldef\tempurl%
\url{https://doi.org/10.1145/3529190.3534768}
\showDOI{\tempurl}


\bibitem[Bannon(1995)]%
        {bannon_human_1995}
\bibfield{author}{\bibinfo{person}{Liam~J. Bannon}.} \bibinfo{year}{1995}\natexlab{}.
\newblock \showarticletitle{From human factors to human actors: {The} role of psychology and human-computer interaction studies in system design}.
\newblock In \bibinfo{booktitle}{\emph{Readings in human–computer interaction}}, \bibfield{editor}{\bibinfo{person}{RONALD~M. BAECKER}, \bibinfo{person}{JONATHAN GRUDIN}, \bibinfo{person}{WILLIAM~A.S. BUXTON}, {and} \bibinfo{person}{SAUL GREENBERG}} (Eds.). \bibinfo{publisher}{Morgan Kaufmann}, \bibinfo{pages}{205--214}.
\newblock
\showISBNx{978-0-08-051574-8}
\urldef\tempurl%
\url{https://doi.org/10.1016/B978-0-08-051574-8.50024-8}
\showDOI{\tempurl}


\bibitem[Baumer(2015)]%
        {baumer_reflective_2015}
\bibfield{author}{\bibinfo{person}{Eric~P.S. Baumer}.} \bibinfo{year}{2015}\natexlab{}.
\newblock \showarticletitle{Reflective {Informatics}: {Conceptual} {Dimensions} for {Designing} {Technologies} of {Reflection}}. In \bibinfo{booktitle}{\emph{Proceedings of the 33rd {Annual} {ACM} {Conference} on {Human} {Factors} in {Computing} {Systems}}} \emph{(\bibinfo{series}{{CHI} '15})}. \bibinfo{publisher}{Association for Computing Machinery}, \bibinfo{address}{New York, NY, USA}, \bibinfo{pages}{585--594}.
\newblock
\showISBNx{978-1-4503-3145-6}
\urldef\tempurl%
\url{https://doi.org/10.1145/2702123.2702234}
\showDOI{\tempurl}


\bibitem[Bentvelzen et~al\mbox{.}(2022)]%
        {bentvelzen_designing_2022}
\bibfield{author}{\bibinfo{person}{Marit Bentvelzen}, \bibinfo{person}{Jasmin Niess}, {and} \bibinfo{person}{Paweł~W. Woźniak}.} \bibinfo{year}{2022}\natexlab{}.
\newblock \showarticletitle{Designing {Reflective} {Derived} {Metrics} for {Fitness} {Trackers}}.
\newblock \bibinfo{journal}{\emph{Proceedings of the ACM on Interactive, Mobile, Wearable and Ubiquitous Technologies}} \bibinfo{volume}{6}, \bibinfo{number}{4} (\bibinfo{date}{Dec.} \bibinfo{year}{2022}), \bibinfo{pages}{1--19}.
\newblock
\showISSN{2474-9567}
\urldef\tempurl%
\url{https://doi.org/10.1145/3569475}
\showDOI{\tempurl}


\bibitem[Berg(1999)]%
        {bergAccumulatingCoordinatingOccasions1999b}
\bibfield{author}{\bibinfo{person}{Marc Berg}.} \bibinfo{year}{1999}\natexlab{}.
\newblock \showarticletitle{Accumulating and {Coordinating}: {Occasions} for {Information} {Technologies} in {Medical} {Work}}.
\newblock \bibinfo{journal}{\emph{Computer Supported Cooperative Work (CSCW)}} \bibinfo{volume}{8}, \bibinfo{number}{4} (\bibinfo{date}{Dec.} \bibinfo{year}{1999}), \bibinfo{pages}{373--401}.
\newblock
\showISSN{0925-9724, 1573-7551}
\urldef\tempurl%
\url{https://doi.org/10.1023/A:1008757115404}
\showDOI{\tempurl}


\bibitem[Berry et~al\mbox{.}(2021)]%
        {berry_supporting_2021}
\bibfield{author}{\bibinfo{person}{Andrew~B.L. Berry}, \bibinfo{person}{Catherine~Y. Lim}, \bibinfo{person}{Calvin~A. Liang}, \bibinfo{person}{Andrea~L. Hartzler}, \bibinfo{person}{Tad Hirsch}, \bibinfo{person}{Dawn~M. Ferguson}, \bibinfo{person}{Zoë~A. Bermet}, {and} \bibinfo{person}{James~D. Ralston}.} \bibinfo{year}{2021}\natexlab{}.
\newblock \showarticletitle{Supporting {Collaborative} {Reflection} on {Personal} {Values} and {Health}}.
\newblock \bibinfo{journal}{\emph{Proceedings of the ACM on Human-Computer Interaction}} \bibinfo{volume}{5}, \bibinfo{number}{CSCW2} (\bibinfo{date}{Oct.} \bibinfo{year}{2021}), \bibinfo{pages}{1--39}.
\newblock
\showISSN{2573-0142}
\urldef\tempurl%
\url{https://doi.org/10.1145/3476040}
\showDOI{\tempurl}


\bibitem[Betzing et~al\mbox{.}(2020)]%
        {betzing_impact_2020}
\bibfield{author}{\bibinfo{person}{Jan~Hendrik Betzing}, \bibinfo{person}{Matthias Tietz}, \bibinfo{person}{Jan vom Brocke}, {and} \bibinfo{person}{Jörg Becker}.} \bibinfo{year}{2020}\natexlab{}.
\newblock \showarticletitle{The impact of transparency on mobile privacy decision making}.
\newblock \bibinfo{journal}{\emph{Electronic Markets}} \bibinfo{volume}{30}, \bibinfo{number}{3} (\bibinfo{date}{Sept.} \bibinfo{year}{2020}), \bibinfo{pages}{607--625}.
\newblock
\showISSN{1019-6781, 1422-8890}
\urldef\tempurl%
\url{https://doi.org/10.1007/s12525-019-00332-3}
\showDOI{\tempurl}


\bibitem[Birhane et~al\mbox{.}(2022)]%
        {birhanePowerPeopleOpportunities2022}
\bibfield{author}{\bibinfo{person}{Abeba Birhane}, \bibinfo{person}{William Isaac}, \bibinfo{person}{Vinodkumar Prabhakaran}, \bibinfo{person}{Mark Diaz}, \bibinfo{person}{Madeleine~Clare Elish}, \bibinfo{person}{Iason Gabriel}, {and} \bibinfo{person}{Shakir Mohamed}.} \bibinfo{year}{2022}\natexlab{}.
\newblock \showarticletitle{Power to the {People}? {Opportunities} and {Challenges} for {Participatory} {AI}}. In \bibinfo{booktitle}{\emph{Equity and {Access} in {Algorithms}, {Mechanisms}, and {Optimization}}}. \bibinfo{publisher}{ACM}, \bibinfo{address}{Arlington VA USA}, \bibinfo{pages}{1--8}.
\newblock
\showISBNx{978-1-4503-9477-2}
\urldef\tempurl%
\url{https://doi.org/10.1145/3551624.3555290}
\showDOI{\tempurl}


\bibitem[Boerman et~al\mbox{.}(2021)]%
        {boerman_exploring_2021}
\bibfield{author}{\bibinfo{person}{Sophie~C. Boerman}, \bibinfo{person}{Sanne Kruikemeier}, {and} \bibinfo{person}{Frederik~J. Zuiderveen~Borgesius}.} \bibinfo{year}{2021}\natexlab{}.
\newblock \showarticletitle{Exploring {Motivations} for {Online} {Privacy} {Protection} {Behavior}: {Insights} {From} {Panel} {Data}}.
\newblock \bibinfo{journal}{\emph{Communication Research}} \bibinfo{volume}{48}, \bibinfo{number}{7} (\bibinfo{date}{Oct.} \bibinfo{year}{2021}), \bibinfo{pages}{953--977}.
\newblock
\showISSN{0093-6502}
\urldef\tempurl%
\url{https://doi.org/10.1177/0093650218800915}
\showDOI{\tempurl}
\newblock
\shownote{Publisher: SAGE Publications Inc}.


\bibitem[Borning and Muller(2012)]%
        {borning_next_2012}
\bibfield{author}{\bibinfo{person}{Alan Borning} {and} \bibinfo{person}{Michael Muller}.} \bibinfo{year}{2012}\natexlab{}.
\newblock \showarticletitle{Next {Steps} for {Value} {Sensitive} {Design}}. In \bibinfo{booktitle}{\emph{Proceedings of the {SIGCHI} {Conference} on {Human} {Factors} in {Computing} {Systems}}}. \bibinfo{publisher}{Association for Computing Machinery}, \bibinfo{address}{New York, NY, USA}, \bibinfo{pages}{1125--1134}.
\newblock
\showISBNx{978-1-4503-1015-4}
\urldef\tempurl%
\url{https://doi.org/10.1145/2207676.2208560}
\showDOI{\tempurl}


\bibitem[Bratteteig and Wagner(2016)]%
        {bratteteigUnpackingNotionParticipation2016b}
\bibfield{author}{\bibinfo{person}{Tone Bratteteig} {and} \bibinfo{person}{Ina Wagner}.} \bibinfo{year}{2016}\natexlab{}.
\newblock \showarticletitle{Unpacking the {Notion} of {Participation} in {Participatory} {Design}}.
\newblock \bibinfo{journal}{\emph{Computer Supported Cooperative Work (CSCW)}} \bibinfo{volume}{25}, \bibinfo{number}{6} (\bibinfo{date}{Dec.} \bibinfo{year}{2016}), \bibinfo{pages}{425--475}.
\newblock
\showISSN{0925-9724, 1573-7551}
\urldef\tempurl%
\url{https://doi.org/10.1007/s10606-016-9259-4}
\showDOI{\tempurl}


\bibitem[Buçinca et~al\mbox{.}(2021)]%
        {bucinca_trust_2021}
\bibfield{author}{\bibinfo{person}{Zana Buçinca}, \bibinfo{person}{Maja~Barbara Malaya}, {and} \bibinfo{person}{Krzysztof~Z. Gajos}.} \bibinfo{year}{2021}\natexlab{}.
\newblock \showarticletitle{To {Trust} or to {Think}: {Cognitive} {Forcing} {Functions} {Can} {Reduce} {Overreliance} on {AI} in {AI}-assisted {Decision}-making}.
\newblock \bibinfo{journal}{\emph{Proceedings of the ACM on Human-Computer Interaction}} \bibinfo{volume}{5}, \bibinfo{number}{CSCW1} (\bibinfo{date}{April} \bibinfo{year}{2021}), \bibinfo{pages}{1--21}.
\newblock
\showISSN{2573-0142}
\urldef\tempurl%
\url{https://doi.org/10.1145/3449287}
\showDOI{\tempurl}


\bibitem[Cajander and Grünloh(2019)]%
        {cajander_electronic_2019}
\bibfield{author}{\bibinfo{person}{Åsa Cajander} {and} \bibinfo{person}{Christiane Grünloh}.} \bibinfo{year}{2019}\natexlab{}.
\newblock \showarticletitle{Electronic {Health} {Records} {Are} {More} {Than} a {Work} {Tool}: {Conflicting} {Needs} of {Direct} and {Indirect} {Stakeholders}}. In \bibinfo{booktitle}{\emph{Proceedings of the 2019 {CHI} {Conference} on {Human} {Factors} in {Computing} {Systems}}}. \bibinfo{publisher}{ACM}, \bibinfo{address}{Glasgow, SC, UK}, \bibinfo{pages}{1--13}.
\newblock
\urldef\tempurl%
\url{https://doi.org/10.1145/3290605.3300865}
\showDOI{\tempurl}


\bibitem[Charles et~al\mbox{.}(1997)]%
        {charlesSharedDecisionmakingMedical1997}
\bibfield{author}{\bibinfo{person}{Cathy Charles}, \bibinfo{person}{Amiram Gafni}, {and} \bibinfo{person}{Tim Whelan}.} \bibinfo{year}{1997}\natexlab{}.
\newblock \showarticletitle{Shared decision-making in the medical encounter: {What} does it mean? (or it takes at least two to tango)}.
\newblock \bibinfo{journal}{\emph{Social Science \& Medicine}} \bibinfo{volume}{44}, \bibinfo{number}{5} (\bibinfo{date}{March} \bibinfo{year}{1997}), \bibinfo{pages}{681--692}.
\newblock
\showISSN{0277-9536}
\urldef\tempurl%
\url{https://doi.org/10.1016/S0277-9536(96)00221-3}
\showDOI{\tempurl}


\bibitem[Cho et~al\mbox{.}(2022)]%
        {cho_reflection_2022}
\bibfield{author}{\bibinfo{person}{Janghee Cho}, \bibinfo{person}{Tian Xu}, \bibinfo{person}{Abigail Zimmermann-Niefield}, {and} \bibinfo{person}{Stephen Voida}.} \bibinfo{year}{2022}\natexlab{}.
\newblock \showarticletitle{Reflection in {Theory} and {Reflection} in {Practice}: {An} {Exploration} of the {Gaps} in {Reflection} {Support} among {Personal} {Informatics} {Apps}}. In \bibinfo{booktitle}{\emph{{CHI} {Conference} on {Human} {Factors} in {Computing} {Systems}}}. \bibinfo{publisher}{ACM}, \bibinfo{address}{New Orleans LA USA}, \bibinfo{pages}{1--23}.
\newblock
\showISBNx{978-1-4503-9157-3}
\urldef\tempurl%
\url{https://doi.org/10.1145/3491102.3501991}
\showDOI{\tempurl}


\bibitem[Ciolfi et~al\mbox{.}(2023)]%
        {ciolfiComputerSupportedCooperativeWork2023}
\bibfield{author}{\bibinfo{person}{Luigina Ciolfi}, \bibinfo{person}{Myriam Lewkowicz}, {and} \bibinfo{person}{Kjeld Schmidt}.} \bibinfo{year}{2023}\natexlab{}.
\newblock \showarticletitle{Computer-{Supported} {Cooperative} {Work}}.
\newblock In \bibinfo{booktitle}{\emph{Handbook of {Human} {Computer} {Interaction}}}, \bibfield{editor}{\bibinfo{person}{Jean Vanderdonckt}, \bibinfo{person}{Philippe Palanque}, {and} \bibinfo{person}{Marco Winckler}} (Eds.). \bibinfo{publisher}{Springer International Publishing}, \bibinfo{address}{Cham}, \bibinfo{pages}{1--26}.
\newblock
\showISBNx{978-3-319-27648-9}
\urldef\tempurl%
\url{https://doi.org/10.1007/978-3-319-27648-9_30-1}
\showDOI{\tempurl}


\bibitem[Colnago et~al\mbox{.}(2020)]%
        {colnago_informing_2020}
\bibfield{author}{\bibinfo{person}{Jessica Colnago}, \bibinfo{person}{Yuanyuan Feng}, \bibinfo{person}{Tharangini Palanivel}, \bibinfo{person}{Sarah Pearman}, \bibinfo{person}{Megan Ung}, \bibinfo{person}{Alessandro Acquisti}, \bibinfo{person}{Lorrie~Faith Cranor}, {and} \bibinfo{person}{Norman Sadeh}.} \bibinfo{year}{2020}\natexlab{}.
\newblock \showarticletitle{Informing the {Design} of a {Personalized} {Privacy} {Assistant} for the {Internet} of {Things}}. In \bibinfo{booktitle}{\emph{Proceedings of the 2020 {CHI} {Conference} on {Human} {Factors} in {Computing} {Systems}}} \emph{(\bibinfo{series}{{CHI} '20})}. \bibinfo{publisher}{Association for Computing Machinery}, \bibinfo{address}{New York, NY, USA}, \bibinfo{pages}{1--13}.
\newblock
\showISBNx{978-1-4503-6708-0}
\urldef\tempurl%
\url{https://doi.org/10.1145/3313831.3376389}
\showDOI{\tempurl}
\newblock
\shownote{event-place: Honolulu, HI, USA}.


\bibitem[Cooper et~al\mbox{.}(2022)]%
        {cooperSystematicReviewThematic2022}
\bibfield{author}{\bibinfo{person}{Ned Cooper}, \bibinfo{person}{Tiffanie Horne}, \bibinfo{person}{Gillian~R Hayes}, \bibinfo{person}{Courtney Heldreth}, \bibinfo{person}{Michal Lahav}, \bibinfo{person}{Jess Holbrook}, {and} \bibinfo{person}{Lauren Wilcox}.} \bibinfo{year}{2022}\natexlab{}.
\newblock \showarticletitle{A {Systematic} {Review} and {Thematic} {Analysis} of {Community}-{Collaborative} {Approaches} to {Computing} {Research}}. In \bibinfo{booktitle}{\emph{Proceedings of the 2022 {CHI} {Conference} on {Human} {Factors} in {Computing} {Systems}}} \emph{(\bibinfo{series}{{CHI} '22})}. \bibinfo{publisher}{Association for Computing Machinery}, \bibinfo{address}{New York, NY, USA}, \bibinfo{pages}{1--18}.
\newblock
\showISBNx{978-1-4503-9157-3}
\urldef\tempurl%
\url{https://doi.org/10.1145/3491102.3517716}
\showDOI{\tempurl}


\bibitem[Cox et~al\mbox{.}(2016)]%
        {cox_design_2016}
\bibfield{author}{\bibinfo{person}{Anna~L. Cox}, \bibinfo{person}{Sandy~J.J. Gould}, \bibinfo{person}{Marta~E. Cecchinato}, \bibinfo{person}{Ioanna Iacovides}, {and} \bibinfo{person}{Ian Renfree}.} \bibinfo{year}{2016}\natexlab{}.
\newblock \showarticletitle{Design {Frictions} for {Mindful} {Interactions}: {The} {Case} for {Microboundaries}}. In \bibinfo{booktitle}{\emph{Proceedings of the 2016 {CHI} {Conference} {Extended} {Abstracts} on {Human} {Factors} in {Computing} {Systems}}}. \bibinfo{publisher}{ACM}, \bibinfo{address}{San Jose California USA}, \bibinfo{pages}{1389--1397}.
\newblock
\showISBNx{978-1-4503-4082-3}
\urldef\tempurl%
\url{https://doi.org/10.1145/2851581.2892410}
\showDOI{\tempurl}


\bibitem[Cranor(2011)]%
        {faith_cranor_platform_2011}
\bibfield{author}{\bibinfo{person}{Lorrie~Faith Cranor}.} \bibinfo{year}{2011}\natexlab{}.
\newblock \showarticletitle{Platform for {Privacy} {Preferences} ({P3P})}.
\newblock In \bibinfo{booktitle}{\emph{Encyclopedia of {Cryptography} and {Security}}}, \bibfield{editor}{\bibinfo{person}{Henk C.~A. van Tilborg} {and} \bibinfo{person}{Sushil Jajodia}} (Eds.). \bibinfo{publisher}{Springer US}, \bibinfo{address}{Boston, MA}, \bibinfo{pages}{940--941}.
\newblock
\showISBNx{978-1-4419-5906-5}
\urldef\tempurl%
\url{https://doi.org/10.1007/978-1-4419-5906-5_759}
\showDOI{\tempurl}


\bibitem[Creswell and Plano~Clark(2018)]%
        {creswell_designing_2017}
\bibfield{author}{\bibinfo{person}{John~W. Creswell} {and} \bibinfo{person}{Vicki~L. Plano~Clark}.} \bibinfo{year}{2018}\natexlab{}.
\newblock \bibinfo{booktitle}{\emph{Designing and conducting mixed methods research} (\bibinfo{edition}{third edition, international student edition} ed.)}.
\newblock \bibinfo{publisher}{Sage}, \bibinfo{address}{Los Angeles London New Delhi Singapore Washington DC Melbourne}.
\newblock
\showISBNx{978-1-5063-8662-1}


\bibitem[Croskerry(2003)]%
        {croskerry_cognitive_2003}
\bibfield{author}{\bibinfo{person}{Pat Croskerry}.} \bibinfo{year}{2003}\natexlab{}.
\newblock \showarticletitle{Cognitive forcing strategies in clinical decisionmaking}.
\newblock \bibinfo{journal}{\emph{Annals of Emergency Medicine}} \bibinfo{volume}{41}, \bibinfo{number}{1} (\bibinfo{date}{Jan.} \bibinfo{year}{2003}), \bibinfo{pages}{110--120}.
\newblock
\showISSN{01960644}
\urldef\tempurl%
\url{https://doi.org/10.1067/mem.2003.22}
\showDOI{\tempurl}


\bibitem[Cumyn et~al\mbox{.}(2021)]%
        {cumynMetaconsentSecondaryUse2021}
\bibfield{author}{\bibinfo{person}{Annabelle Cumyn}, \bibinfo{person}{Adrien Barton}, \bibinfo{person}{Roxanne Dault}, \bibinfo{person}{Nissrine Safa}, \bibinfo{person}{Anne-Marie Cloutier}, {and} \bibinfo{person}{Jean-François Ethier}.} \bibinfo{year}{2021}\natexlab{}.
\newblock \showarticletitle{Meta-consent for the secondary use of health data within a learning health system: a qualitative study of the public’s perspective}.
\newblock \bibinfo{journal}{\emph{BMC Medical Ethics}} \bibinfo{volume}{22}, \bibinfo{number}{1} (\bibinfo{date}{June} \bibinfo{year}{2021}), \bibinfo{pages}{81}.
\newblock
\showISSN{1472-6939}
\urldef\tempurl%
\url{https://doi.org/10.1186/s12910-021-00647-x}
\showDOI{\tempurl}


\bibitem[Dahl and Svanæs(2020)]%
        {dahl_facilitating_2020}
\bibfield{author}{\bibinfo{person}{Yngve Dahl} {and} \bibinfo{person}{Dag Svanæs}.} \bibinfo{year}{2020}\natexlab{}.
\newblock \showarticletitle{Facilitating {Democracy}: {Concerns} from {Participatory} {Design} with {Asymmetric} {Stakeholder} {Relations} in {Health} {Care}}. In \bibinfo{booktitle}{\emph{Proceedings of the 2020 {CHI} {Conference} on {Human} {Factors} in {Computing} {Systems}}}. \bibinfo{publisher}{ACM}, \bibinfo{address}{Honolulu HI USA}, \bibinfo{pages}{1--13}.
\newblock
\showISBNx{978-1-4503-6708-0}
\urldef\tempurl%
\url{https://doi.org/10.1145/3313831.3376805}
\showDOI{\tempurl}


\bibitem[De~Sutter et~al\mbox{.}(2020)]%
        {desutterImplementationElectronicInformed2020}
\bibfield{author}{\bibinfo{person}{Evelien De~Sutter}, \bibinfo{person}{Drieda Zaçe}, \bibinfo{person}{Stefania Boccia}, \bibinfo{person}{Maria~Luisa Di~Pietro}, \bibinfo{person}{David Geerts}, \bibinfo{person}{Pascal Borry}, {and} \bibinfo{person}{Isabelle Huys}.} \bibinfo{year}{2020}\natexlab{}.
\newblock \showarticletitle{Implementation of {Electronic} {Informed} {Consent} in {Biomedical} {Research} and {Stakeholders}’ {Perspectives}: {Systematic} {Review}}.
\newblock \bibinfo{journal}{\emph{Journal of Medical Internet Research}} \bibinfo{volume}{22}, \bibinfo{number}{10} (\bibinfo{date}{Oct.} \bibinfo{year}{2020}), \bibinfo{pages}{e19129}.
\newblock
\showISSN{1438-8871}
\urldef\tempurl%
\url{https://doi.org/10.2196/19129}
\showDOI{\tempurl}


\bibitem[Dell et~al\mbox{.}(2012)]%
        {dellYoursBetterParticipant2012}
\bibfield{author}{\bibinfo{person}{Nicola Dell}, \bibinfo{person}{Vidya Vaidyanathan}, \bibinfo{person}{Indrani Medhi}, \bibinfo{person}{Edward Cutrell}, {and} \bibinfo{person}{William Thies}.} \bibinfo{year}{2012}\natexlab{}.
\newblock \showarticletitle{"{Yours} is better!": participant response bias in {HCI}}. In \bibinfo{booktitle}{\emph{Proceedings of the {SIGCHI} {Conference} on {Human} {Factors} in {Computing} {Systems}}}. \bibinfo{publisher}{ACM}, \bibinfo{address}{Austin Texas USA}, \bibinfo{pages}{1321--1330}.
\newblock
\showISBNx{978-1-4503-1015-4}
\urldef\tempurl%
\url{https://doi.org/10.1145/2207676.2208589}
\showDOI{\tempurl}


\bibitem[Duckert and Barkhuus(2022)]%
        {duckert_protecting_2022}
\bibfield{author}{\bibinfo{person}{Melanie Duckert} {and} \bibinfo{person}{Louise Barkhuus}.} \bibinfo{year}{2022}\natexlab{}.
\newblock \showarticletitle{Protecting {Personal} {Health} {Data} through {Privacy} {Awareness}: {A} study of perceived data privacy among people with chronic or long-term illness}.
\newblock \bibinfo{journal}{\emph{Proceedings of the ACM on Human-Computer Interaction}} \bibinfo{volume}{6}, \bibinfo{number}{GROUP} (\bibinfo{date}{Jan.} \bibinfo{year}{2022}), \bibinfo{pages}{1--22}.
\newblock
\showISSN{2573-0142}
\urldef\tempurl%
\url{https://doi.org/10.1145/3492830}
\showDOI{\tempurl}


\bibitem[Dupuis et~al\mbox{.}(2022)]%
        {dupuis_crowdsourcing_2022}
\bibfield{author}{\bibinfo{person}{Marc Dupuis}, \bibinfo{person}{Karen Renaud}, {and} \bibinfo{person}{Rosalind Searle}.} \bibinfo{year}{2022}\natexlab{}.
\newblock \showarticletitle{Crowdsourcing {Quality} {Concerns}: {An} {Examination} of {Amazon}’s {Mechanical} {Turk}}. In \bibinfo{booktitle}{\emph{The 23rd {Annual} {Conference} on {Information} {Technology} {Education}}}. \bibinfo{publisher}{ACM}, \bibinfo{address}{Chicago IL USA}, \bibinfo{pages}{127--129}.
\newblock
\showISBNx{978-1-4503-9391-1}
\urldef\tempurl%
\url{https://doi.org/10.1145/3537674.3555783}
\showDOI{\tempurl}


\bibitem[Eardley et~al\mbox{.}(2023)]%
        {eardleyExplanationAdoptionSupporting2023}
\bibfield{author}{\bibinfo{person}{Rachel Eardley}, \bibinfo{person}{Emma~L. Tonkin}, \bibinfo{person}{Ewan Soubutts}, \bibinfo{person}{Amid Ayobi}, \bibinfo{person}{Gregory J.~L. Tourte}, \bibinfo{person}{Rachael Gooberman-Hill}, \bibinfo{person}{Ian Craddock}, {and} \bibinfo{person}{Aisling~Ann O'Kane}.} \bibinfo{year}{2023}\natexlab{}.
\newblock \showarticletitle{Explanation before {Adoption}: {Supporting} {Informed} {Consent} for {Complex} {Machine} {Learning} and {IoT} {Health} {Platforms}}.
\newblock \bibinfo{journal}{\emph{Proceedings of the ACM on Human-Computer Interaction}} \bibinfo{volume}{7}, \bibinfo{number}{CSCW1} (\bibinfo{date}{April} \bibinfo{year}{2023}), \bibinfo{pages}{49:1--49:25}.
\newblock
\urldef\tempurl%
\url{https://doi.org/10.1145/3579482}
\showDOI{\tempurl}


\bibitem[Ertner et~al\mbox{.}(2010)]%
        {ertner_five_2010}
\bibfield{author}{\bibinfo{person}{Marie Ertner}, \bibinfo{person}{Anne~Mie Kragelund}, {and} \bibinfo{person}{Lone Malmborg}.} \bibinfo{year}{2010}\natexlab{}.
\newblock \showarticletitle{Five {Enunciations} of {Empowerment} in {Participatory} {Design}}. In \bibinfo{booktitle}{\emph{Proceedings of the 11th {Biennial} {Participatory} {Design} {Conference}}}. \bibinfo{publisher}{Association for Computing Machinery}, \bibinfo{address}{New York, NY, USA}, \bibinfo{pages}{191--194}.
\newblock
\showISBNx{978-1-4503-0131-2}
\urldef\tempurl%
\url{https://doi.org/10.1145/1900441.1900475}
\showDOI{\tempurl}


\bibitem[Esposito and Freda(2018)]%
        {espositoCounselingReflexiveProcesses2018}
\bibfield{author}{\bibinfo{person}{Giovanna Esposito} {and} \bibinfo{person}{Maria~Francesca Freda}.} \bibinfo{year}{2018}\natexlab{}.
\newblock \showarticletitle{Counseling and reflexive processes: {Role} of multimodal narrative devices in promoting reflection and reflexivity in a university context.}
\newblock In \bibinfo{booktitle}{\emph{Counseling and coaching in times of crisis and transition: {From} research to practice.}} \bibinfo{publisher}{Routledge/Taylor \& Francis Group}, \bibinfo{address}{New York, NY, US}, \bibinfo{pages}{15--27}.
\newblock
\showISBNx{978-1-138-29008-2 (Hardcover); 978-1-315-26659-6 (Digital (undefined format))}
\urldef\tempurl%
\url{https://doi.org/10.4324/9781315266596-3}
\showDOI{\tempurl}


\bibitem[{European Commission}(2016)]%
        {european_commission_regulation_2016}
\bibfield{author}{\bibinfo{person}{{European Commission}}.} \bibinfo{year}{2016}\natexlab{}.
\newblock \bibinfo{title}{Regulation ({EU}) 2016/679 of the {European} {Parliament} and of the {Council} of 27 {April} 2016 on the protection of natural persons with regard to the processing of personal data and on the free movement of such data, and repealing {Directive} 95/46/{EC} ({General} {Data} {Protection} {Regulation})}.
\newblock
\newblock
\urldef\tempurl%
\url{https://eur-lex.europa.eu/eli/reg/2016/679/oj}
\showURL{%
\tempurl}


\bibitem[Fassl et~al\mbox{.}(2021)]%
        {fassl_stop_2021}
\bibfield{author}{\bibinfo{person}{Matthias Fassl}, \bibinfo{person}{Lea~Theresa Gröber}, {and} \bibinfo{person}{Katharina Krombholz}.} \bibinfo{year}{2021}\natexlab{}.
\newblock \showarticletitle{Stop the {Consent} {Theater}}. In \bibinfo{booktitle}{\emph{Extended {Abstracts} of the 2021 {CHI} {Conference} on {Human} {Factors} in {Computing} {Systems}}}. \bibinfo{publisher}{ACM}, \bibinfo{address}{Yokohama Japan}, \bibinfo{pages}{1--7}.
\newblock
\showISBNx{978-1-4503-8095-9}
\urldef\tempurl%
\url{https://doi.org/10.1145/3411763.3451230}
\showDOI{\tempurl}


\bibitem[Feise(2002)]%
        {feiseMultipleOutcomeMeasures2002}
\bibfield{author}{\bibinfo{person}{Ronald~J. Feise}.} \bibinfo{year}{2002}\natexlab{}.
\newblock \showarticletitle{Do multiple outcome measures require p-value adjustment?}
\newblock \bibinfo{journal}{\emph{BMC Medical Research Methodology}} \bibinfo{volume}{2}, \bibinfo{number}{1} (\bibinfo{date}{June} \bibinfo{year}{2002}), \bibinfo{pages}{8}.
\newblock
\showISSN{1471-2288}
\urldef\tempurl%
\url{https://doi.org/10.1186/1471-2288-2-8}
\showDOI{\tempurl}


\bibitem[Fitzpatrick and Ellingsen(2013)]%
        {fitzpatrickReview25Years2013a}
\bibfield{author}{\bibinfo{person}{Geraldine Fitzpatrick} {and} \bibinfo{person}{Gunnar Ellingsen}.} \bibinfo{year}{2013}\natexlab{}.
\newblock \showarticletitle{A {Review} of 25 {Years} of {CSCW} {Research} in {Healthcare}: {Contributions}, {Challenges} and {Future} {Agendas}}.
\newblock \bibinfo{journal}{\emph{Computer Supported Cooperative Work (CSCW)}} \bibinfo{volume}{22}, \bibinfo{number}{4-6} (\bibinfo{date}{Aug.} \bibinfo{year}{2013}), \bibinfo{pages}{609--665}.
\newblock
\showISSN{0925-9724, 1573-7551}
\urldef\tempurl%
\url{https://doi.org/10.1007/s10606-012-9168-0}
\showDOI{\tempurl}


\bibitem[Friedman and Hendry(2019)]%
        {friedman_value_2019}
\bibfield{author}{\bibinfo{person}{Batya Friedman} {and} \bibinfo{person}{David~G. Hendry}.} \bibinfo{year}{2019}\natexlab{}.
\newblock \bibinfo{booktitle}{\emph{Value {Sensitive} {Design}: {Shaping} {Technology} with {Moral} {Imagination}}}.
\newblock \bibinfo{publisher}{MIT Press}, \bibinfo{address}{Cambridge, MA, USA}.
\newblock


\bibitem[Friedman et~al\mbox{.}(2017)]%
        {friedman_survey_2017}
\bibfield{author}{\bibinfo{person}{Batya Friedman}, \bibinfo{person}{David~G. Hendry}, {and} \bibinfo{person}{Alan Borning}.} \bibinfo{year}{2017}\natexlab{}.
\newblock \showarticletitle{A {Survey} of {Value} {Sensitive} {Design} {Methods}}.
\newblock \bibinfo{journal}{\emph{Foundations and Trends® in Human–Computer Interaction}} \bibinfo{volume}{11}, \bibinfo{number}{2} (\bibinfo{year}{2017}), \bibinfo{pages}{63--125}.
\newblock
\showISSN{1551-3955, 1551-3963}
\urldef\tempurl%
\url{https://doi.org/10.1561/1100000015}
\showDOI{\tempurl}


\bibitem[Ghanouni et~al\mbox{.}(2016)]%
        {ghanouni_common_2016}
\bibfield{author}{\bibinfo{person}{Alex Ghanouni}, \bibinfo{person}{Cristina Renzi}, \bibinfo{person}{Susanne~F Meisel}, {and} \bibinfo{person}{Jo Waller}.} \bibinfo{year}{2016}\natexlab{}.
\newblock \showarticletitle{Common methods of measuring ‘informed choice’ in screening participation: {Challenges} and future directions}.
\newblock \bibinfo{journal}{\emph{Preventive Medicine Reports}}  \bibinfo{volume}{4} (\bibinfo{date}{Dec.} \bibinfo{year}{2016}), \bibinfo{pages}{601--607}.
\newblock
\showISSN{22113355}
\urldef\tempurl%
\url{https://doi.org/10.1016/j.pmedr.2016.10.017}
\showDOI{\tempurl}


\bibitem[Gomez~Ortega et~al\mbox{.}(2023)]%
        {gomezortegaDataTransactionsFramework2023}
\bibfield{author}{\bibinfo{person}{Alejandra Gomez~Ortega}, \bibinfo{person}{Jacky Bourgeois}, \bibinfo{person}{Wiebke~Toussaint Hutiri}, {and} \bibinfo{person}{Gerd Kortuem}.} \bibinfo{year}{2023}\natexlab{}.
\newblock \showarticletitle{Beyond data transactions: a framework for meaningfully informed data donation}.
\newblock \bibinfo{journal}{\emph{AI \& SOCIETY}} \bibinfo{volume}{40}, \bibinfo{number}{2} (\bibinfo{date}{Aug.} \bibinfo{year}{2023}), \bibinfo{pages}{1--18}.
\newblock
\showISSN{1435-5655}
\urldef\tempurl%
\url{https://doi.org/10.1007/s00146-023-01755-5}
\showDOI{\tempurl}


\bibitem[Gomez~Ortega et~al\mbox{.}(2021)]%
        {gomez_ortega_towards_2021}
\bibfield{author}{\bibinfo{person}{Alejandra Gomez~Ortega}, \bibinfo{person}{Jacky Bourgeois}, {and} \bibinfo{person}{Gerd Kortuem}.} \bibinfo{year}{2021}\natexlab{}.
\newblock \showarticletitle{Towards {Designerly} {Data} {Donation}}. In \bibinfo{booktitle}{\emph{Adjunct {Proceedings} of the 2021 {ACM} {International} {Joint} {Conference} on {Pervasive} and {Ubiquitous} {Computing} and {Proceedings} of the 2021 {ACM} {International} {Symposium} on {Wearable} {Computers}}}. \bibinfo{publisher}{Association for Computing Machinery}, \bibinfo{address}{New York, NY, USA}, \bibinfo{pages}{496--501}.
\newblock
\showISBNx{978-1-4503-8461-2}
\urldef\tempurl%
\url{https://doi.org/10.1145/3460418.3479362}
\showDOI{\tempurl}


\bibitem[Gould et~al\mbox{.}(2015)]%
        {gould_task_2015}
\bibfield{author}{\bibinfo{person}{Sandy~J.J. Gould}, \bibinfo{person}{Anna~L. Cox}, {and} \bibinfo{person}{Duncan~P. Brumby}.} \bibinfo{year}{2015}\natexlab{}.
\newblock \showarticletitle{Task {Lockouts} {Induce} {Crowdworkers} to {Switch} to {Other} {Activities}}. In \bibinfo{booktitle}{\emph{Proceedings of the 33rd {Annual} {ACM} {Conference} {Extended} {Abstracts} on {Human} {Factors} in {Computing} {Systems}}} \emph{(\bibinfo{series}{{CHI} {EA} '15})}. \bibinfo{publisher}{Association for Computing Machinery}, \bibinfo{address}{New York, NY, USA}, \bibinfo{pages}{1785--1790}.
\newblock
\showISBNx{978-1-4503-3146-3}
\urldef\tempurl%
\url{https://doi.org/10.1145/2702613.2732709}
\showDOI{\tempurl}


\bibitem[Grady et~al\mbox{.}(2015)]%
        {grady_broad_2015}
\bibfield{author}{\bibinfo{person}{Christine Grady}, \bibinfo{person}{Lisa Eckstein}, \bibinfo{person}{Ben Berkman}, \bibinfo{person}{Dan Brock}, \bibinfo{person}{Robert Cook-Deegan}, \bibinfo{person}{Stephanie~M. Fullerton}, \bibinfo{person}{Hank Greely}, \bibinfo{person}{Mats~G. Hansson}, \bibinfo{person}{Sara Hull}, \bibinfo{person}{Scott Kim}, {and} \bibinfo{person}{{others}}.} \bibinfo{year}{2015}\natexlab{}.
\newblock \showarticletitle{Broad {Consent} for {Research} with {Biological} {Samples}: {Workshop} {Conclusions}}.
\newblock \bibinfo{journal}{\emph{The American Journal of Bioethics}} \bibinfo{volume}{15}, \bibinfo{number}{9} (\bibinfo{year}{2015}), \bibinfo{pages}{34--42}.
\newblock
\newblock
\shownote{Publisher: Taylor \& Francis}.


\bibitem[Groß(2021)]%
        {gros_validity_2021}
\bibfield{author}{\bibinfo{person}{Thomas Groß}.} \bibinfo{year}{2021}\natexlab{}.
\newblock \showarticletitle{Validity and {Reliability} of the {Scale} {Internet} {Users}’ {Information} {Privacy} {Concerns} ({IUIPC})}.
\newblock \bibinfo{journal}{\emph{Proceedings on Privacy Enhancing Technologies}} \bibinfo{volume}{2021}, \bibinfo{number}{2} (\bibinfo{date}{April} \bibinfo{year}{2021}), \bibinfo{pages}{235--258}.
\newblock
\showISSN{2299-0984}
\urldef\tempurl%
\url{https://doi.org/10.2478/popets-2021-0026}
\showDOI{\tempurl}


\bibitem[Habib et~al\mbox{.}(2021)]%
        {habib_toggles_2021}
\bibfield{author}{\bibinfo{person}{Hana Habib}, \bibinfo{person}{Yixin Zou}, \bibinfo{person}{Yaxing Yao}, \bibinfo{person}{Alessandro Acquisti}, \bibinfo{person}{Lorrie Cranor}, \bibinfo{person}{Joel Reidenberg}, \bibinfo{person}{Norman Sadeh}, {and} \bibinfo{person}{Florian Schaub}.} \bibinfo{year}{2021}\natexlab{}.
\newblock \showarticletitle{Toggles, {Dollar} {Signs}, and {Triangles}: {How} to ({In}){Effectively} {Convey} {Privacy} {Choices} with {Icons} and {Link} {Texts}}. In \bibinfo{booktitle}{\emph{Proceedings of the 2021 {CHI} {Conference} on {Human} {Factors} in {Computing} {Systems}}} \emph{(\bibinfo{series}{{CHI} '21})}. \bibinfo{publisher}{Association for Computing Machinery}, \bibinfo{address}{New York, NY, USA}.
\newblock
\showISBNx{978-1-4503-8096-6}
\urldef\tempurl%
\url{https://doi.org/10.1145/3411764.3445387}
\showDOI{\tempurl}


\bibitem[Hallnäs and Redström(2001)]%
        {hallnasSlowTechnologyDesigning2001}
\bibfield{author}{\bibinfo{person}{Lars Hallnäs} {and} \bibinfo{person}{Johan Redström}.} \bibinfo{year}{2001}\natexlab{}.
\newblock \showarticletitle{Slow {Technology} – {Designing} for {Reflection}}.
\newblock \bibinfo{journal}{\emph{Personal and Ubiquitous Computing}} \bibinfo{volume}{5}, \bibinfo{number}{3} (\bibinfo{date}{Aug.} \bibinfo{year}{2001}), \bibinfo{pages}{201--212}.
\newblock
\showISSN{1617-4909}
\urldef\tempurl%
\url{https://doi.org/10.1007/PL00000019}
\showDOI{\tempurl}


\bibitem[Hara et~al\mbox{.}(2018)]%
        {hara_data-driven_2018}
\bibfield{author}{\bibinfo{person}{Kotaro Hara}, \bibinfo{person}{Abigail Adams}, \bibinfo{person}{Kristy Milland}, \bibinfo{person}{Saiph Savage}, \bibinfo{person}{Chris Callison-Burch}, {and} \bibinfo{person}{Jeffrey~P. Bigham}.} \bibinfo{year}{2018}\natexlab{}.
\newblock \showarticletitle{A {Data}-{Driven} {Analysis} of {Workers}' {Earnings} on {Amazon} {Mechanical} {Turk}}. In \bibinfo{booktitle}{\emph{Proceedings of the 2018 {CHI} {Conference} on {Human} {Factors} in {Computing} {Systems}}} \emph{(\bibinfo{series}{{CHI} '18})}. \bibinfo{publisher}{Association for Computing Machinery}, \bibinfo{address}{New York, NY, USA}, \bibinfo{pages}{1--14}.
\newblock
\showISBNx{978-1-4503-5620-6}
\urldef\tempurl%
\url{https://doi.org/10.1145/3173574.3174023}
\showDOI{\tempurl}


\bibitem[Hauser et~al\mbox{.}(2019)]%
        {hauser_common_2019}
\bibfield{author}{\bibinfo{person}{David Hauser}, \bibinfo{person}{Gabriele Paolacci}, {and} \bibinfo{person}{Jesse Chandler}.} \bibinfo{year}{2019}\natexlab{}.
\newblock \showarticletitle{Common {Concerns} with {MTurk} as a {Participant} {Pool}: {Evidence} and {Solutions}}.
\newblock In \bibinfo{booktitle}{\emph{Handbook of {Research} {Methods} in {Consumer} {Psychology}}}. \bibinfo{publisher}{Routledge}.
\newblock
\showISBNx{978-1-351-13771-3}


\bibitem[Huh-Yoo and Rader(2020)]%
        {huh-yooItWildWild2020}
\bibfield{author}{\bibinfo{person}{Jina Huh-Yoo} {and} \bibinfo{person}{Emilee Rader}.} \bibinfo{year}{2020}\natexlab{}.
\newblock \showarticletitle{It's the {Wild}, {Wild} {West}: {Lessons} {Learned} {From} {IRB} {Members}' {Risk} {Perceptions} {Toward} {Digital} {Research} {Data}}.
\newblock \bibinfo{journal}{\emph{Proceedings of the ACM on Human-Computer Interaction}} \bibinfo{volume}{4}, \bibinfo{number}{CSCW1} (\bibinfo{date}{May} \bibinfo{year}{2020}), \bibinfo{pages}{1--22}.
\newblock
\showISSN{2573-0142}
\urldef\tempurl%
\url{https://doi.org/10.1145/3392868}
\showDOI{\tempurl}


\bibitem[Jackson and Wang(2018)]%
        {jackson_addressing_2018}
\bibfield{author}{\bibinfo{person}{Corey~Brian Jackson} {and} \bibinfo{person}{Yang Wang}.} \bibinfo{year}{2018}\natexlab{}.
\newblock \showarticletitle{Addressing {The} {Privacy} {Paradox} through {Personalized} {Privacy} {Notifications}}.
\newblock \bibinfo{journal}{\emph{Proceedings of the ACM on Interactive, Mobile, Wearable and Ubiquitous Technologies}} \bibinfo{volume}{2}, \bibinfo{number}{2} (\bibinfo{date}{July} \bibinfo{year}{2018}), \bibinfo{pages}{1--25}.
\newblock
\showISSN{2474-9567}
\urldef\tempurl%
\url{https://doi.org/10.1145/3214271}
\showDOI{\tempurl}


\bibitem[Jacobs et~al\mbox{.}(2015)]%
        {jacobsComparingHealthInformation2015}
\bibfield{author}{\bibinfo{person}{Maia~L. Jacobs}, \bibinfo{person}{James Clawson}, {and} \bibinfo{person}{Elizabeth~D. Mynatt}.} \bibinfo{year}{2015}\natexlab{}.
\newblock \showarticletitle{Comparing {Health} {Information} {Sharing} {Preferences} of {Cancer} {Patients}, {Doctors}, and {Navigators}}. In \bibinfo{booktitle}{\emph{Proceedings of the 18th {ACM} {Conference} on {Computer} {Supported} {Cooperative} {Work} \& {Social} {Computing}}} \emph{(\bibinfo{series}{{CSCW} '15})}. \bibinfo{publisher}{Association for Computing Machinery}, \bibinfo{address}{New York, NY, USA}, \bibinfo{pages}{808--818}.
\newblock
\showISBNx{978-1-4503-2922-4}
\urldef\tempurl%
\url{https://doi.org/10.1145/2675133.2675252}
\showDOI{\tempurl}


\bibitem[Jonas and Hanrahan(2022)]%
        {jonas_designing_2022}
\bibfield{author}{\bibinfo{person}{Rebecca~M. Jonas} {and} \bibinfo{person}{Benjamin~V. Hanrahan}.} \bibinfo{year}{2022}\natexlab{}.
\newblock \showarticletitle{Designing for {Shared} {Values}: {Exploring} {Ethical} {Dilemmas} of {Conducting} {Values} {Inclusive} {Design} {Research}}.
\newblock \bibinfo{journal}{\emph{Proceeding of the ACM on Human-Computer Interaction}} \bibinfo{volume}{6}, \bibinfo{number}{CSCW2} (\bibinfo{year}{2022}).
\newblock
\urldef\tempurl%
\url{https://doi.org/10.1145/3555182}
\showDOI{\tempurl}
\newblock
\shownote{Place: New York, NY, USA Publisher: Association for Computing Machinery}.


\bibitem[Kassam et~al\mbox{.}(2023)]%
        {kassamPatientPerspectivesPreferences2023}
\bibfield{author}{\bibinfo{person}{Iman Kassam}, \bibinfo{person}{Daria Ilkina}, \bibinfo{person}{Jessica Kemp}, \bibinfo{person}{Heba Roble}, \bibinfo{person}{Abigail Carter-Langford}, {and} \bibinfo{person}{Nelson Shen}.} \bibinfo{year}{2023}\natexlab{}.
\newblock \showarticletitle{Patient {Perspectives} and {Preferences} for {Consent} in the {Digital} {Health} {Context}: {State}-of-the-art {Literature} {Review}}.
\newblock \bibinfo{journal}{\emph{Journal of Medical Internet Research}}  \bibinfo{volume}{25} (\bibinfo{date}{Feb.} \bibinfo{year}{2023}), \bibinfo{pages}{e42507}.
\newblock
\showISSN{1438-8871}
\urldef\tempurl%
\url{https://doi.org/10.2196/42507}
\showDOI{\tempurl}


\bibitem[Kaye et~al\mbox{.}(2015)]%
        {kayeDynamicConsentPatient2015}
\bibfield{author}{\bibinfo{person}{Jane Kaye}, \bibinfo{person}{Edgar~A. Whitley}, \bibinfo{person}{David Lund}, \bibinfo{person}{Michael Morrison}, \bibinfo{person}{Harriet Teare}, {and} \bibinfo{person}{Karen Melham}.} \bibinfo{year}{2015}\natexlab{}.
\newblock \showarticletitle{Dynamic consent: a patient interface for twenty-first century research networks}.
\newblock \bibinfo{journal}{\emph{European Journal of Human Genetics}} \bibinfo{volume}{23}, \bibinfo{number}{2} (\bibinfo{date}{Feb.} \bibinfo{year}{2015}), \bibinfo{pages}{141--146}.
\newblock
\showISSN{1476-5438}
\urldef\tempurl%
\url{https://doi.org/10.1038/ejhg.2014.71}
\showDOI{\tempurl}
\newblock
\shownote{Publisher: Nature Publishing Group}.


\bibitem[Kitkowska et~al\mbox{.}(2022)]%
        {kitkowska_online_2022}
\bibfield{author}{\bibinfo{person}{Agnieszka Kitkowska}, \bibinfo{person}{Johan Högberg}, {and} \bibinfo{person}{Erik Wästlund}.} \bibinfo{year}{2022}\natexlab{}.
\newblock \showarticletitle{Online {Terms} and {Conditions}: {Improving} {User} {Engagement}, {Awareness}, and {Satisfaction} through {UI} {Design}}. In \bibinfo{booktitle}{\emph{{CHI} {Conference} on {Human} {Factors} in {Computing} {Systems}}}. \bibinfo{publisher}{ACM}, \bibinfo{address}{New Orleans LA USA}, \bibinfo{pages}{1--22}.
\newblock
\showISBNx{978-1-4503-9157-3}
\urldef\tempurl%
\url{https://doi.org/10.1145/3491102.3517720}
\showDOI{\tempurl}


\bibitem[Kitkowska et~al\mbox{.}(2020)]%
        {kitkowska_psychological_2020}
\bibfield{author}{\bibinfo{person}{Agnieszka Kitkowska}, \bibinfo{person}{Yefim Shulman}, \bibinfo{person}{Leonardo~A. Martucci}, {and} \bibinfo{person}{Erik Wästlund}.} \bibinfo{year}{2020}\natexlab{}.
\newblock \showarticletitle{Psychological {Effects} and {Their} {Role} in {Online} {Privacy} {Interactions}: {A} {Review}}.
\newblock \bibinfo{journal}{\emph{IEEE Access}}  \bibinfo{volume}{8} (\bibinfo{year}{2020}), \bibinfo{pages}{21236--21260}.
\newblock
\showISSN{2169-3536}
\urldef\tempurl%
\url{https://doi.org/10.1109/ACCESS.2020.2969562}
\showDOI{\tempurl}
\newblock
\shownote{Conference Name: IEEE Access}.


\bibitem[Kittur et~al\mbox{.}(2008)]%
        {kittur_crowdsourcing_2008}
\bibfield{author}{\bibinfo{person}{Aniket Kittur}, \bibinfo{person}{Ed~H. Chi}, {and} \bibinfo{person}{Bongwon Suh}.} \bibinfo{year}{2008}\natexlab{}.
\newblock \showarticletitle{Crowdsourcing user studies with {Mechanical} {Turk}}. In \bibinfo{booktitle}{\emph{Proceedings of the {SIGCHI} {Conference} on {Human} {Factors} in {Computing} {Systems}}}. \bibinfo{publisher}{ACM}, \bibinfo{address}{Florence Italy}, \bibinfo{pages}{453--456}.
\newblock
\showISBNx{978-1-60558-011-1}
\urldef\tempurl%
\url{https://doi.org/10.1145/1357054.1357127}
\showDOI{\tempurl}


\bibitem[Kroll and Stieglitz(2021)]%
        {kroll_digital_2021}
\bibfield{author}{\bibinfo{person}{Tobias Kroll} {and} \bibinfo{person}{Stefan Stieglitz}.} \bibinfo{year}{2021}\natexlab{}.
\newblock \showarticletitle{Digital nudging and privacy: improving decisions about self-disclosure in social networks}.
\newblock \bibinfo{journal}{\emph{Behaviour \& Information Technology}} \bibinfo{volume}{40}, \bibinfo{number}{1} (\bibinfo{date}{Jan.} \bibinfo{year}{2021}), \bibinfo{pages}{1--19}.
\newblock
\showISSN{0144-929X, 1362-3001}
\urldef\tempurl%
\url{https://doi.org/10.1080/0144929X.2019.1584644}
\showDOI{\tempurl}


\bibitem[Le~Dantec et~al\mbox{.}(2009)]%
        {le_dantec_values_2009}
\bibfield{author}{\bibinfo{person}{Christopher~A. Le~Dantec}, \bibinfo{person}{Erika~Shehan Poole}, {and} \bibinfo{person}{Susan~P. Wyche}.} \bibinfo{year}{2009}\natexlab{}.
\newblock \showarticletitle{Values as lived experience: evolving value sensitive design in support of value discovery}. In \bibinfo{booktitle}{\emph{Proceedings of the {SIGCHI} {Conference} on {Human} {Factors} in {Computing} {Systems}}}. \bibinfo{publisher}{ACM}, \bibinfo{address}{Boston MA USA}, \bibinfo{pages}{1141--1150}.
\newblock
\showISBNx{978-1-60558-246-7}
\urldef\tempurl%
\url{https://doi.org/10.1145/1518701.1518875}
\showDOI{\tempurl}


\bibitem[Leimstädtner et~al\mbox{.}(2023)]%
        {leimstadtner_investigating_2023}
\bibfield{author}{\bibinfo{person}{David Leimstädtner}, \bibinfo{person}{Peter Sörries}, {and} \bibinfo{person}{Claudia Müller-Birn}.} \bibinfo{year}{2023}\natexlab{}.
\newblock \showarticletitle{Investigating {Responsible} {Nudge} {Design} for {Informed} {Decision}-{Making} {Enabling} {Transparent} and {Reflective} {Decision}-{Making}}. In \bibinfo{booktitle}{\emph{Proceedings of {Mensch} {Und} {Computer} 2023}} \emph{(\bibinfo{series}{{MuC} '23})}. \bibinfo{publisher}{Association for Computing Machinery}, \bibinfo{address}{New York, NY, USA}, \bibinfo{pages}{220--236}.
\newblock
\showISBNx{9798400707711}
\urldef\tempurl%
\url{https://doi.org/10.1145/3603555.3603567}
\showDOI{\tempurl}
\newblock
\shownote{event-place: Rapperswil, Switzerland}.


\bibitem[Lim et~al\mbox{.}(2019)]%
        {lim_facilitating_2019}
\bibfield{author}{\bibinfo{person}{Catherine~Y. Lim}, \bibinfo{person}{Andrew~B.L. Berry}, \bibinfo{person}{Andrea~L. Hartzler}, \bibinfo{person}{Tad Hirsch}, \bibinfo{person}{David~S. Carrell}, \bibinfo{person}{Zoë~A. Bermet}, {and} \bibinfo{person}{James~D. Ralston}.} \bibinfo{year}{2019}\natexlab{}.
\newblock \showarticletitle{Facilitating {Self}-{Reflection} about {Values} and {Self}-{Care} {Among} {Individuals} with {Chronic} {Conditions}}. In \bibinfo{booktitle}{\emph{Proceedings of the 2019 {CHI} {Conference} on {Human} {Factors} in {Computing} {Systems}}}. \bibinfo{publisher}{ACM}, \bibinfo{address}{Glasgow, SC, UK}, \bibinfo{pages}{1--12}.
\newblock
\urldef\tempurl%
\url{https://doi.org/10.1145/3290605.3300885}
\showDOI{\tempurl}


\bibitem[Malhotra et~al\mbox{.}(2004)]%
        {malhotra_internet_2004}
\bibfield{author}{\bibinfo{person}{Naresh~K. Malhotra}, \bibinfo{person}{Sung~S. Kim}, {and} \bibinfo{person}{James Agarwal}.} \bibinfo{year}{2004}\natexlab{}.
\newblock \showarticletitle{Internet {Users}' {Information} {Privacy} {Concerns} ({IUIPC}): {The} {Construct}, the {Scale}, and a {Causal} {Model}}.
\newblock \bibinfo{journal}{\emph{Information Systems Research}} \bibinfo{volume}{15}, \bibinfo{number}{4} (\bibinfo{date}{Dec.} \bibinfo{year}{2004}), \bibinfo{pages}{336--355}.
\newblock
\showISSN{1047-7047, 1526-5536}
\urldef\tempurl%
\url{https://doi.org/10.1287/isre.1040.0032}
\showDOI{\tempurl}


\bibitem[Marcus et~al\mbox{.}(2022)]%
        {marcus_european_2022}
\bibfield{author}{\bibinfo{person}{J.~Scott Marcus}, \bibinfo{person}{Bertin Martens}, \bibinfo{person}{Christophe Carugati}, \bibinfo{person}{Anne Bucher}, {and} \bibinfo{person}{Ilsa Godlovitch}.} \bibinfo{year}{2022}\natexlab{}.
\newblock \bibinfo{title}{The {European} {Health} {Data} {Space}}.
\newblock
\newblock
\urldef\tempurl%
\url{https://doi.org/10.2139/ssrn.4300393}
\showDOI{\tempurl}


\bibitem[Maus et~al\mbox{.}(2020)]%
        {maus_enhancing_2020}
\bibfield{author}{\bibinfo{person}{Benjamin Maus}, \bibinfo{person}{Dario Salvi}, {and} \bibinfo{person}{Carl~Magnus Olsson}.} \bibinfo{year}{2020}\natexlab{}.
\newblock \showarticletitle{Enhancing citizens trust in technologies for data donation in clinical research: validation of a design prototype}. In \bibinfo{booktitle}{\emph{10th {International} {Conference} on the {Internet} of {Things} {Companion}}}. \bibinfo{publisher}{ACM}, \bibinfo{address}{Malmö Sweden}, \bibinfo{pages}{1--8}.
\newblock
\showISBNx{978-1-4503-8820-7}
\urldef\tempurl%
\url{https://doi.org/10.1145/3423423.3423430}
\showDOI{\tempurl}


\bibitem[Mayring(2014)]%
        {mayring_qualitative_2014}
\bibfield{author}{\bibinfo{person}{Philipp Mayring}.} \bibinfo{year}{2014}\natexlab{}.
\newblock \bibinfo{booktitle}{\emph{Qualitative content analysis: theoretical foundation, basic procedures and software solution}}.
\newblock \bibinfo{address}{Klagenfurt}. 143 pages.
\newblock


\bibitem[McDonald et~al\mbox{.}(2020)]%
        {mcdonald_privacy_2020}
\bibfield{author}{\bibinfo{person}{Nora McDonald}, \bibinfo{person}{Karla Badillo-Urquiola}, \bibinfo{person}{Morgan~G. Ames}, \bibinfo{person}{Nicola Dell}, \bibinfo{person}{Elizabeth Keneski}, \bibinfo{person}{Manya Sleeper}, {and} \bibinfo{person}{Pamela~J. Wisniewski}.} \bibinfo{year}{2020}\natexlab{}.
\newblock \showarticletitle{Privacy and {Power}: {Acknowledging} the {Importance} of {Privacy} {Research} and {Design} for {Vulnerable} {Populations}}. In \bibinfo{booktitle}{\emph{Extended {Abstracts} of the 2020 {CHI} {Conference} on {Human} {Factors} in {Computing} {Systems}}}. \bibinfo{publisher}{Association for Computing Machinery}, \bibinfo{address}{New York, NY, USA}, \bibinfo{pages}{1--8}.
\newblock
\showISBNx{978-1-4503-6819-3}
\urldef\tempurl%
\url{https://doi.org/10.1145/3334480.3375174}
\showDOI{\tempurl}


\bibitem[McGraw and Mandl(2021)]%
        {mcgraw_privacy_2021}
\bibfield{author}{\bibinfo{person}{Deven McGraw} {and} \bibinfo{person}{Kenneth~D. Mandl}.} \bibinfo{year}{2021}\natexlab{}.
\newblock \showarticletitle{Privacy protections to encourage use of health-relevant digital data in a learning health system}.
\newblock \bibinfo{journal}{\emph{npj Digital Medicine}} \bibinfo{volume}{4}, \bibinfo{number}{1} (\bibinfo{date}{Jan.} \bibinfo{year}{2021}), \bibinfo{pages}{2}.
\newblock
\showISSN{2398-6352}
\urldef\tempurl%
\url{https://doi.org/10.1038/s41746-020-00362-8}
\showDOI{\tempurl}


\bibitem[Miller et~al\mbox{.}(2007)]%
        {miller_value_2007}
\bibfield{author}{\bibinfo{person}{Jessica~K. Miller}, \bibinfo{person}{Batya Friedman}, \bibinfo{person}{Gavin Jancke}, {and} \bibinfo{person}{Brian Gill}.} \bibinfo{year}{2007}\natexlab{}.
\newblock \showarticletitle{Value {Tensions} in {Design}: {The} {Value} {Sensitive} {Design}, {Development}, and {Appropriation} of a {Corporation}'s {Groupware} {System}}. In \bibinfo{booktitle}{\emph{Proceedings of the 2007 {International} {ACM} {Conference} on {Supporting} {Group} {Work}}}. \bibinfo{publisher}{Association for Computing Machinery}, \bibinfo{address}{New York, NY, USA}, \bibinfo{pages}{281--290}.
\newblock
\showISBNx{978-1-59593-845-9}
\urldef\tempurl%
\url{https://doi.org/10.1145/1316624.1316668}
\showDOI{\tempurl}


\bibitem[Moore et~al\mbox{.}(2007)]%
        {mooreConfidentialityPrivacyHealth2007}
\bibfield{author}{\bibinfo{person}{Ilene~N. Moore}, \bibinfo{person}{Samuel~Leason Snyder}, \bibinfo{person}{Cynthia Miller}, \bibinfo{person}{Angel~Qi An}, \bibinfo{person}{Jennifer~U. Blackford}, \bibinfo{person}{Chuan Zhou}, {and} \bibinfo{person}{Gerald~B. Hickson}.} \bibinfo{year}{2007}\natexlab{}.
\newblock \showarticletitle{Confidentiality and privacy in health care from the patient's perspective: does {HIPAA} help?}
\newblock \bibinfo{journal}{\emph{Health Matrix (Cleveland, Ohio: 1991)}} \bibinfo{volume}{17}, \bibinfo{number}{2} (\bibinfo{year}{2007}), \bibinfo{pages}{215--272}.
\newblock
\showISSN{0748-383X}


\bibitem[Mullen et~al\mbox{.}(2006)]%
        {mullen_measures_2006}
\bibfield{author}{\bibinfo{person}{Patricia~Dolan Mullen}, \bibinfo{person}{Jennifer~Dacey Allen}, \bibinfo{person}{Karen Glanz}, \bibinfo{person}{Maria~E. Fernandez}, \bibinfo{person}{Deborah~J. Bowen}, \bibinfo{person}{Sandi~L. Pruitt}, \bibinfo{person}{Beth~A. Glenn}, {and} \bibinfo{person}{Michael Pignone}.} \bibinfo{year}{2006}\natexlab{}.
\newblock \showarticletitle{Measures used in studies of informed decision making about cancer screening: {A} systematic review}.
\newblock \bibinfo{journal}{\emph{Annals of Behavioral Medicine}} \bibinfo{volume}{32}, \bibinfo{number}{3} (\bibinfo{date}{Dec.} \bibinfo{year}{2006}), \bibinfo{pages}{188--201}.
\newblock
\showISSN{0883-6612, 1532-4796}
\urldef\tempurl%
\url{https://doi.org/10.1207/s15324796abm3203_4}
\showDOI{\tempurl}


\bibitem[Muller and Druin(2007)]%
        {muller_participatory_2007}
\bibfield{author}{\bibinfo{person}{Michael~J. Muller} {and} \bibinfo{person}{Allison Druin}.} \bibinfo{year}{2007}\natexlab{}.
\newblock \bibinfo{booktitle}{\emph{Participatory {Design}: {The} {Third} {Space} in {HCI}}}.
\newblock \bibinfo{publisher}{CRC Press}, \bibinfo{address}{Boca Raton, FL, USA}.
\newblock


\bibitem[Munro et~al\mbox{.}(2016)]%
        {munro_choosing_2016}
\bibfield{author}{\bibinfo{person}{Sarah Munro}, \bibinfo{person}{Dawn Stacey}, \bibinfo{person}{Krystina~B. Lewis}, {and} \bibinfo{person}{Nick Bansback}.} \bibinfo{year}{2016}\natexlab{}.
\newblock \showarticletitle{Choosing treatment and screening options congruent with values: {Do} decision aids help? {Sub}-analysis of a systematic review}.
\newblock \bibinfo{journal}{\emph{Patient Education and Counseling}} \bibinfo{volume}{99}, \bibinfo{number}{4} (\bibinfo{date}{April} \bibinfo{year}{2016}), \bibinfo{pages}{491--500}.
\newblock
\showISSN{07383991}
\urldef\tempurl%
\url{https://doi.org/10.1016/j.pec.2015.10.026}
\showDOI{\tempurl}


\bibitem[Nathan et~al\mbox{.}(2008)]%
        {nathan_envisioning_2008}
\bibfield{author}{\bibinfo{person}{Lisa~P. Nathan}, \bibinfo{person}{Batya Friedman}, \bibinfo{person}{Predrag Klasnja}, \bibinfo{person}{Shaun~K. Kane}, {and} \bibinfo{person}{Jessica~K. Miller}.} \bibinfo{year}{2008}\natexlab{}.
\newblock \showarticletitle{Envisioning {Systemic} {Effects} on {Persons} and {Society} throughout {Interactive} {System} {Design}}. In \bibinfo{booktitle}{\emph{Proceedings of the 7th {ACM} {Conference} on {Designing} {Interactive} {Systems}}}. \bibinfo{publisher}{Association for Computing Machinery}, \bibinfo{address}{New York, NY, USA}, \bibinfo{pages}{1--10}.
\newblock
\showISBNx{978-1-60558-002-9}
\urldef\tempurl%
\url{https://doi.org/10.1145/1394445.1394446}
\showDOI{\tempurl}


\bibitem[Norman and Draper(1986)]%
        {norman_user_1986}
\bibfield{author}{\bibinfo{person}{Donald~A. Norman} {and} \bibinfo{person}{Stephen~W. Draper}.} \bibinfo{year}{1986}\natexlab{}.
\newblock \bibinfo{title}{User {Centered} {System} {Design}: {New} {Perspectives} on {Human}-{Computer} {Interaction}}.
\newblock
\newblock
\newblock
\shownote{Pages: 31–61 Place: Hillsdale, NJ, USA}.


\bibitem[Nouwens et~al\mbox{.}(2020)]%
        {nouwens_dark_2020}
\bibfield{author}{\bibinfo{person}{Midas Nouwens}, \bibinfo{person}{Ilaria Liccardi}, \bibinfo{person}{Michael Veale}, \bibinfo{person}{David Karger}, {and} \bibinfo{person}{Lalana Kagal}.} \bibinfo{year}{2020}\natexlab{}.
\newblock \showarticletitle{Dark {Patterns} after the {GDPR}: {Scraping} {Consent} {Pop}-ups and {Demonstrating} their {Influence}}. In \bibinfo{booktitle}{\emph{Proceedings of the 2020 {CHI} {Conference} on {Human} {Factors} in {Computing} {Systems}}}. \bibinfo{publisher}{ACM}, \bibinfo{address}{Honolulu HI USA}, \bibinfo{pages}{1--13}.
\newblock
\showISBNx{978-1-4503-6708-0}
\urldef\tempurl%
\url{https://doi.org/10.1145/3313831.3376321}
\showDOI{\tempurl}


\bibitem[Nunes~Vilaza et~al\mbox{.}(2020)]%
        {nunes_vilaza_futures_2020}
\bibfield{author}{\bibinfo{person}{Giovanna Nunes~Vilaza}, \bibinfo{person}{Raju Maharjan}, \bibinfo{person}{David Coyle}, {and} \bibinfo{person}{Jakob Bardram}.} \bibinfo{year}{2020}\natexlab{}.
\newblock \showarticletitle{Futures for {Health} {Research} {Data} {Platforms} {From} the {Participants}’ {Perspectives}}. In \bibinfo{booktitle}{\emph{Proceedings of the 11th {Nordic} {Conference} on {Human}-{Computer} {Interaction}: {Shaping} {Experiences}, {Shaping} {Society}}}. \bibinfo{publisher}{ACM}, \bibinfo{address}{Tallinn Estonia}, \bibinfo{pages}{1--14}.
\newblock
\showISBNx{978-1-4503-7579-5}
\urldef\tempurl%
\url{https://doi.org/10.1145/3419249.3420110}
\showDOI{\tempurl}


\bibitem[O'Kane et~al\mbox{.}(2013)]%
        {okane_non-static_2013}
\bibfield{author}{\bibinfo{person}{Aisling~A. O'Kane}, \bibinfo{person}{Helena~M. Mentis}, {and} \bibinfo{person}{Eno Thereska}.} \bibinfo{year}{2013}\natexlab{}.
\newblock \showarticletitle{Non-{Static} {Nature} of {Patient} {Consent}: {Shifting} {Privacy} {Perspectives} in {Health} {Information} {Sharing}}. In \bibinfo{booktitle}{\emph{Proceedings of the 2013 {Conference} on {Computer} {Supported} {Cooperative} {Work}}}. \bibinfo{publisher}{Association for Computing Machinery}, \bibinfo{address}{New York, NY, USA}, \bibinfo{pages}{553--562}.
\newblock
\showISBNx{978-1-4503-1331-5}
\urldef\tempurl%
\url{https://doi.org/10.1145/2441776.2441838}
\showDOI{\tempurl}


\bibitem[Park et~al\mbox{.}(2022)]%
        {parkPowerDynamicsValue2022}
\bibfield{author}{\bibinfo{person}{Joon~Sung Park}, \bibinfo{person}{Karrie Karahalios}, \bibinfo{person}{Niloufar Salehi}, {and} \bibinfo{person}{Motahhare Eslami}.} \bibinfo{year}{2022}\natexlab{}.
\newblock \showarticletitle{Power {Dynamics} and {Value} {Conflicts} in {Designing} and {Maintaining} {Socio}-{Technical} {Algorithmic} {Processes}}.
\newblock \bibinfo{journal}{\emph{Proceedings of the ACM on Human-Computer Interaction}} \bibinfo{volume}{6}, \bibinfo{number}{CSCW1} (\bibinfo{date}{March} \bibinfo{year}{2022}), \bibinfo{pages}{1--21}.
\newblock
\showISSN{2573-0142}
\urldef\tempurl%
\url{https://doi.org/10.1145/3512957}
\showDOI{\tempurl}


\bibitem[Pearman et~al\mbox{.}(2022)]%
        {pearmanUserfriendlyRarelyRead2022}
\bibfield{author}{\bibinfo{person}{Sarah Pearman}, \bibinfo{person}{Ellie Young}, {and} \bibinfo{person}{Lorrie~Faith Cranor}.} \bibinfo{year}{2022}\natexlab{}.
\newblock \showarticletitle{User-friendly yet rarely read: {A} case study on the redesign of an online {HIPAA} authorization}.
\newblock \bibinfo{journal}{\emph{Proceedings on Privacy Enhancing Technologies}} \bibinfo{volume}{2022}, \bibinfo{number}{3} (\bibinfo{date}{July} \bibinfo{year}{2022}), \bibinfo{pages}{558--581}.
\newblock
\showISSN{2299-0984}
\urldef\tempurl%
\url{https://doi.org/10.56553/popets-2022-0086}
\showDOI{\tempurl}


\bibitem[Peer et~al\mbox{.}(2017)]%
        {peer_beyond_2017}
\bibfield{author}{\bibinfo{person}{Eyal Peer}, \bibinfo{person}{Laura Brandimarte}, \bibinfo{person}{Sonam Samat}, {and} \bibinfo{person}{Alessandro Acquisti}.} \bibinfo{year}{2017}\natexlab{}.
\newblock \showarticletitle{Beyond the {Turk}: {Alternative} platforms for crowdsourcing behavioral research}.
\newblock \bibinfo{journal}{\emph{Journal of Experimental Social Psychology}}  \bibinfo{volume}{70} (\bibinfo{date}{May} \bibinfo{year}{2017}), \bibinfo{pages}{153--163}.
\newblock
\showISSN{00221031}
\urldef\tempurl%
\url{https://doi.org/10.1016/j.jesp.2017.01.006}
\showDOI{\tempurl}


\bibitem[Pei et~al\mbox{.}(2020)]%
        {pei_attention_2020}
\bibfield{author}{\bibinfo{person}{Weiping Pei}, \bibinfo{person}{Arthur Mayer}, \bibinfo{person}{Kaylynn Tu}, {and} \bibinfo{person}{Chuan Yue}.} \bibinfo{year}{2020}\natexlab{}.
\newblock \showarticletitle{Attention {Please}: {Your} {Attention} {Check} {Questions} in {Survey} {Studies} {Can} {Be} {Automatically} {Answered}}.
\newblock In \bibinfo{booktitle}{\emph{Proceedings of {The} {Web} {Conference} 2020}}. \bibinfo{publisher}{Association for Computing Machinery}, \bibinfo{address}{New York, NY, USA}, \bibinfo{pages}{1182--1193}.
\newblock
\showISBNx{978-1-4503-7023-3}
\urldef\tempurl%
\url{https://doi.org/10.1145/3366423.3380195}
\showURL{%
\tempurl}


\bibitem[Ramírez et~al\mbox{.}(2021)]%
        {ramirez_state_2021}
\bibfield{author}{\bibinfo{person}{Jorge Ramírez}, \bibinfo{person}{Burcu Sayin}, \bibinfo{person}{Marcos Baez}, \bibinfo{person}{Fabio Casati}, \bibinfo{person}{Luca Cernuzzi}, \bibinfo{person}{Boualem Benatallah}, {and} \bibinfo{person}{Gianluca Demartini}.} \bibinfo{year}{2021}\natexlab{}.
\newblock \showarticletitle{On the {State} of {Reporting} in {Crowdsourcing} {Experiments} and a {Checklist} to {Aid} {Current} {Practices}}.
\newblock \bibinfo{journal}{\emph{Proceedings of the ACM on Human-Computer Interaction}} \bibinfo{volume}{5}, \bibinfo{number}{CSCW2} (\bibinfo{year}{2021}).
\newblock
\urldef\tempurl%
\url{https://doi.org/10.1145/3479531}
\showDOI{\tempurl}
\newblock
\shownote{Place: New York, NY, USA Publisher: Association for Computing Machinery}.


\bibitem[Redmiles et~al\mbox{.}(2019)]%
        {redmilesHowWellMy2019}
\bibfield{author}{\bibinfo{person}{Elissa~M. Redmiles}, \bibinfo{person}{Sean Kross}, {and} \bibinfo{person}{Michelle~L. Mazurek}.} \bibinfo{year}{2019}\natexlab{}.
\newblock \showarticletitle{How {Well} {Do} {My} {Results} {Generalize}? {Comparing} {Security} and {Privacy} {Survey} {Results} from {MTurk}, {Web}, and {Telephone} {Samples}}. In \bibinfo{booktitle}{\emph{2019 {IEEE} {Symposium} on {Security} and {Privacy} ({SP})}}. \bibinfo{publisher}{IEEE}, \bibinfo{address}{San Francisco, CA, USA}, \bibinfo{pages}{1326--1343}.
\newblock
\showISBNx{978-1-5386-6660-9}
\urldef\tempurl%
\url{https://doi.org/10.1109/SP.2019.00014}
\showDOI{\tempurl}


\bibitem[Rossi and Palmirani(2017)]%
        {rossi_visualization_2017}
\bibfield{author}{\bibinfo{person}{Arianna Rossi} {and} \bibinfo{person}{Monica Palmirani}.} \bibinfo{year}{2017}\natexlab{}.
\newblock \showarticletitle{A {Visualization} {Approach} for {Adaptive} {Consent} in the {European} {Data} {Protection} {Framework}}. In \bibinfo{booktitle}{\emph{2017 {Conference} for {E}-{Democracy} and {Open} {Government} ({CeDEM})}}. \bibinfo{publisher}{IEEE}, \bibinfo{address}{Krems, Austria}, \bibinfo{pages}{159--170}.
\newblock
\showISBNx{978-1-5090-6718-3 978-1-5090-6719-0}
\urldef\tempurl%
\url{https://doi.org/10.1109/CeDEM.2017.23}
\showDOI{\tempurl}


\bibitem[Rothmann et~al\mbox{.}(2016)]%
        {rothmann_participatory_2016}
\bibfield{author}{\bibinfo{person}{M.~J. Rothmann}, \bibinfo{person}{D.~B. Danbjørg}, \bibinfo{person}{C.~M. Jensen}, {and} \bibinfo{person}{J. Clemensen}.} \bibinfo{year}{2016}\natexlab{}.
\newblock \showarticletitle{Participatory {Design} in {Health} {Care}: {Participation}, {Power} and {Knowledge}}. In \bibinfo{booktitle}{\emph{Proceedings of the 14th {Participatory} {Design} {Conference}: {Short} {Papers}, {Interactive} {Exhibitions}, {Workshops} - {Volume} 2}}. \bibinfo{publisher}{Association for Computing Machinery}, \bibinfo{address}{New York, NY, USA}, \bibinfo{pages}{127--128}.
\newblock
\showISBNx{978-1-4503-4136-3}
\urldef\tempurl%
\url{https://doi.org/10.1145/2948076.2948106}
\showDOI{\tempurl}


\bibitem[Saunders et~al\mbox{.}(2018)]%
        {saunders_saturation_2018}
\bibfield{author}{\bibinfo{person}{Benjamin Saunders}, \bibinfo{person}{Julius Sim}, \bibinfo{person}{Tom Kingstone}, \bibinfo{person}{Shula Baker}, \bibinfo{person}{Jackie Waterfield}, \bibinfo{person}{Bernadette Bartlam}, \bibinfo{person}{Heather Burroughs}, {and} \bibinfo{person}{Clare Jinks}.} \bibinfo{year}{2018}\natexlab{}.
\newblock \showarticletitle{Saturation in qualitative research: exploring its conceptualization and operationalization}.
\newblock \bibinfo{journal}{\emph{Quality \& Quantity}} \bibinfo{volume}{52}, \bibinfo{number}{4} (\bibinfo{date}{July} \bibinfo{year}{2018}), \bibinfo{pages}{1893--1907}.
\newblock
\showISSN{0033-5177, 1573-7845}
\urldef\tempurl%
\url{https://doi.org/10.1007/s11135-017-0574-8}
\showDOI{\tempurl}


\bibitem[Scheuerman et~al\mbox{.}(2021)]%
        {scheuermanDatasetsHavePolitics2021}
\bibfield{author}{\bibinfo{person}{Morgan~Klaus Scheuerman}, \bibinfo{person}{Alex Hanna}, {and} \bibinfo{person}{Emily Denton}.} \bibinfo{year}{2021}\natexlab{}.
\newblock \showarticletitle{Do {Datasets} {Have} {Politics}? {Disciplinary} {Values} in {Computer} {Vision} {Dataset} {Development}}.
\newblock \bibinfo{journal}{\emph{Proceedings of the ACM on Human-Computer Interaction}} \bibinfo{volume}{5}, \bibinfo{number}{CSCW2} (\bibinfo{date}{Oct.} \bibinfo{year}{2021}), \bibinfo{pages}{1--37}.
\newblock
\showISSN{2573-0142}
\urldef\tempurl%
\url{https://doi.org/10.1145/3476058}
\showDOI{\tempurl}


\bibitem[Schneider et~al\mbox{.}(2018)]%
        {schneider_digital_2018}
\bibfield{author}{\bibinfo{person}{Christoph Schneider}, \bibinfo{person}{Markus Weinmann}, {and} \bibinfo{person}{Jan vom Brocke}.} \bibinfo{year}{2018}\natexlab{}.
\newblock \showarticletitle{Digital {Nudging}: {Guiding} {Online} {User} {Choices} through {Interface} {Design}}.
\newblock \bibinfo{journal}{\emph{Commun. ACM}} \bibinfo{volume}{61}, \bibinfo{number}{7} (\bibinfo{year}{2018}), \bibinfo{pages}{67--73}.
\newblock
\showISSN{0001-0782}
\urldef\tempurl%
\url{https://doi.org/10.1145/3213765}
\showDOI{\tempurl}
\newblock
\shownote{Place: New York, NY, USA Publisher: Association for Computing Machinery}.


\bibitem[Schrepp et~al\mbox{.}(2023)]%
        {schreppComparisonSUSUMUXLITE2023}
\bibfield{author}{\bibinfo{person}{Martin Schrepp}, \bibinfo{person}{Jessica Kollmorgen}, {and} \bibinfo{person}{Jörg Thomaschewski}.} \bibinfo{year}{2023}\natexlab{}.
\newblock \showarticletitle{A {Comparison} of {SUS}, {UMUX}-{LITE}, and {UEQ}-{S}}.
\newblock \bibinfo{journal}{\emph{J. User Exper.}} \bibinfo{volume}{18}, \bibinfo{number}{2} (\bibinfo{date}{June} \bibinfo{year}{2023}), \bibinfo{pages}{86--104}.
\newblock
\showISSN{1931-3357}
\urldef\tempurl%
\url{https://doi.org/10.5555/3604890.3604893}
\showDOI{\tempurl}


\bibitem[Schuler and Namioka(1993)]%
        {schuler_participatory_1993}
\bibfield{author}{\bibinfo{person}{Douglas Schuler} {and} \bibinfo{person}{Aki Namioka}.} \bibinfo{year}{1993}\natexlab{}.
\newblock \bibinfo{booktitle}{\emph{Participatory {Design}: {Principles} and {Practices}}}.
\newblock \bibinfo{publisher}{CRC Press}, \bibinfo{address}{Boca Raton, FL, USA}.
\newblock


\bibitem[Spiel et~al\mbox{.}(2019)]%
        {spiel_how_2019}
\bibfield{author}{\bibinfo{person}{Katta Spiel}, \bibinfo{person}{Oliver~L. Haimson}, {and} \bibinfo{person}{Danielle Lottridge}.} \bibinfo{year}{2019}\natexlab{}.
\newblock \showarticletitle{How to do better with gender on surveys: a guide for {HCI} researchers}.
\newblock \bibinfo{journal}{\emph{Interactions}} \bibinfo{volume}{26}, \bibinfo{number}{4} (\bibinfo{date}{June} \bibinfo{year}{2019}), \bibinfo{pages}{62--65}.
\newblock
\showISSN{1072-5520}
\urldef\tempurl%
\url{https://doi.org/10.1145/3338283}
\showDOI{\tempurl}


\bibitem[Sprouse(2011)]%
        {sprouse_validation_2011}
\bibfield{author}{\bibinfo{person}{Jon Sprouse}.} \bibinfo{year}{2011}\natexlab{}.
\newblock \showarticletitle{A validation of {Amazon} {Mechanical} {Turk} for the collection of acceptability judgments in linguistic theory}.
\newblock \bibinfo{journal}{\emph{Behavior Research Methods}} \bibinfo{volume}{43}, \bibinfo{number}{1} (\bibinfo{date}{March} \bibinfo{year}{2011}), \bibinfo{pages}{155--167}.
\newblock
\showISSN{1554-3528}
\urldef\tempurl%
\url{https://doi.org/10.3758/s13428-010-0039-7}
\showDOI{\tempurl}


\bibitem[Södergren(2024)]%
        {sodergrenEnsuringInclusivityWellbeing2024a}
\bibfield{author}{\bibinfo{person}{Antonia~Clasina Södergren}.} \bibinfo{year}{2024}\natexlab{}.
\newblock \showarticletitle{Ensuring {Inclusivity} and {Well}-being of {Children} {Call} for {Accuracy} in {Ethical} ({Design}) {Practices}: {Making} the {Interpersonal} {Aspects} of {Data} {Collection} {Explicit} and {Value}-centred by {Scaffolding} the {Quality} {Criterion} of '{Sincerity}' in {Human}-involved {Research}}. In \bibinfo{booktitle}{\emph{Proceedings of the 23rd {Annual} {ACM} {Interaction} {Design} and {Children} {Conference}}} \emph{(\bibinfo{series}{{IDC} '24})}. \bibinfo{publisher}{Association for Computing Machinery}, \bibinfo{address}{New York, NY, USA}, \bibinfo{pages}{856--868}.
\newblock
\showISBNx{9798400704420}
\urldef\tempurl%
\url{https://doi.org/10.1145/3628516.3659407}
\showDOI{\tempurl}


\bibitem[Sörries et~al\mbox{.}(2024)]%
        {sorriesAdvocatingValuesMeaningful2024a}
\bibfield{author}{\bibinfo{person}{Peter Sörries}, \bibinfo{person}{David Leimstädtner}, {and} \bibinfo{person}{Claudia Müller-Birn}.} \bibinfo{year}{2024}\natexlab{}.
\newblock \showarticletitle{Advocating {Values} through {Meaningful} {Participation}: {Introducing} a {Method} to {Elicit} and {Analyze} {Values} for {Enriching} {Data} {Donation} {Practices} in {Healthcare}}.
\newblock \bibinfo{journal}{\emph{Proceedings of the ACM on Human-Computer Interaction}} \bibinfo{volume}{8}, \bibinfo{number}{CSCW1} (\bibinfo{date}{April} \bibinfo{year}{2024}), \bibinfo{pages}{16:1--16:32}.
\newblock
\urldef\tempurl%
\url{https://doi.org/10.1145/3637293}
\showDOI{\tempurl}


\bibitem[Tazi et~al\mbox{.}(2024)]%
        {taziSoKAnalyzingPrivacy2024}
\bibfield{author}{\bibinfo{person}{Faiza Tazi}, \bibinfo{person}{Archana Nandakumar}, \bibinfo{person}{Josiah Dykstra}, \bibinfo{person}{Prashanth Rajivan}, {and} \bibinfo{person}{Sanchari Das}.} \bibinfo{year}{2024}\natexlab{}.
\newblock \showarticletitle{{SoK}: {Analyzing} {Privacy} and {Security} of {Healthcare} {Data} from the {User} {Perspective}}.
\newblock \bibinfo{journal}{\emph{ACM Transactions on Computing for Healthcare}} \bibinfo{volume}{5}, \bibinfo{number}{2} (\bibinfo{date}{April} \bibinfo{year}{2024}), \bibinfo{pages}{1--31}.
\newblock
\showISSN{2691-1957, 2637-8051}
\urldef\tempurl%
\url{https://doi.org/10.1145/3650116}
\showDOI{\tempurl}


\bibitem[Terpstra et~al\mbox{.}(2019)]%
        {terpstra_improving_2019}
\bibfield{author}{\bibinfo{person}{Arnout Terpstra}, \bibinfo{person}{Alexander Schouten}, \bibinfo{person}{Alwin de Rooij}, {and} \bibinfo{person}{Ronald Leenes}.} \bibinfo{year}{2019}\natexlab{}.
\newblock \showarticletitle{Improving {Privacy} {Choice} through {Design}: {How} {Designing} for {Reflection} {Could} {Support} {Privacy} {Self}-{Management}}. In \bibinfo{booktitle}{\emph{First {Monday}}}, Vol.~\bibinfo{volume}{24}. \bibinfo{publisher}{First Monday}, \bibinfo{pages}{1--13}.
\newblock
\urldef\tempurl%
\url{https://doi.org/10.5210/fm.v24i7.9358}
\showDOI{\tempurl}
\newblock
\shownote{ISSN: 1396-0466 Issue: 7}.


\bibitem[Tschider(2018)]%
        {tschiderConsentMythImproving2018}
\bibfield{author}{\bibinfo{person}{Charlotte Tschider}.} \bibinfo{year}{2018}\natexlab{}.
\newblock \showarticletitle{The {Consent} {Myth}: {Improving} {Choice} for {Patients} of the {Future}}.
\newblock \bibinfo{journal}{\emph{SSRN Electronic Journal}} \bibinfo{volume}{96 Washington University Law Review}, \bibinfo{number}{1505} (\bibinfo{year}{2018}).
\newblock
\showISSN{1556-5068}
\urldef\tempurl%
\url{https://doi.org/10.2139/ssrn.3171801}
\showDOI{\tempurl}


\bibitem[Tseng et~al\mbox{.}(2024)]%
        {tsengDataStewardshipClinical2024}
\bibfield{author}{\bibinfo{person}{Emily Tseng}, \bibinfo{person}{Rosanna Bellini}, \bibinfo{person}{Yeuk-Yu Lee}, \bibinfo{person}{Alana Ramjit}, \bibinfo{person}{Thomas Ristenpart}, {and} \bibinfo{person}{Nicola Dell}.} \bibinfo{year}{2024}\natexlab{}.
\newblock \showarticletitle{Data {Stewardship} in {Clinical} {Computer} {Security}: {Balancing} {Benefit} and {Burden} in {Participatory} {Systems}}.
\newblock \bibinfo{journal}{\emph{Proceedings of the ACM on Human-Computer Interaction}} \bibinfo{volume}{8}, \bibinfo{number}{CSCW1} (\bibinfo{date}{April} \bibinfo{year}{2024}), \bibinfo{pages}{39:1--39:29}.
\newblock
\urldef\tempurl%
\url{https://doi.org/10.1145/3637316}
\showDOI{\tempurl}


\bibitem[Utz et~al\mbox{.}(2019)]%
        {utz_informed_2019}
\bibfield{author}{\bibinfo{person}{Christine Utz}, \bibinfo{person}{Martin Degeling}, \bibinfo{person}{Sascha Fahl}, \bibinfo{person}{Florian Schaub}, {and} \bibinfo{person}{Thorsten Holz}.} \bibinfo{year}{2019}\natexlab{}.
\newblock \showarticletitle{({Un})informed {Consent}: {Studying} {GDPR} {Consent} {Notices} in the {Field}}. In \bibinfo{booktitle}{\emph{Proceedings of the 2019 {ACM} {SIGSAC} {Conference} on {Computer} and {Communications} {Security}}}. \bibinfo{publisher}{ACM}, \bibinfo{address}{London United Kingdom}, \bibinfo{pages}{973--990}.
\newblock
\showISBNx{978-1-4503-6747-9}
\urldef\tempurl%
\url{https://doi.org/10.1145/3319535.3354212}
\showDOI{\tempurl}


\bibitem[Waldman(2020)]%
        {waldman_cognitive_2020}
\bibfield{author}{\bibinfo{person}{Ari~Ezra Waldman}.} \bibinfo{year}{2020}\natexlab{}.
\newblock \showarticletitle{Cognitive biases, dark patterns, and the ‘privacy paradox’}.
\newblock \bibinfo{journal}{\emph{Current Opinion in Psychology}}  \bibinfo{volume}{31} (\bibinfo{date}{Feb.} \bibinfo{year}{2020}), \bibinfo{pages}{105--109}.
\newblock
\showISSN{2352250X}
\urldef\tempurl%
\url{https://doi.org/10.1016/j.copsyc.2019.08.025}
\showDOI{\tempurl}


\bibitem[Wang et~al\mbox{.}(2014)]%
        {wang_field_2014}
\bibfield{author}{\bibinfo{person}{Yang Wang}, \bibinfo{person}{Pedro~Giovanni Leon}, \bibinfo{person}{Alessandro Acquisti}, \bibinfo{person}{Lorrie~Faith Cranor}, \bibinfo{person}{Alain Forget}, {and} \bibinfo{person}{Norman Sadeh}.} \bibinfo{year}{2014}\natexlab{}.
\newblock \showarticletitle{A {Field} {Trial} of {Privacy} {Nudges} for {Facebook}}. In \bibinfo{booktitle}{\emph{Proceedings of the {SIGCHI} {Conference} on {Human} {Factors} in {Computing} {Systems}}} \emph{(\bibinfo{series}{{CHI} '14})}. \bibinfo{publisher}{Association for Computing Machinery}, \bibinfo{address}{New York, NY, USA}, \bibinfo{pages}{2367--2376}.
\newblock
\showISBNx{978-1-4503-2473-1}
\urldef\tempurl%
\url{https://doi.org/10.1145/2556288.2557413}
\showDOI{\tempurl}


\bibitem[Warberg et~al\mbox{.}(2019)]%
        {warberg_can_2019}
\bibfield{author}{\bibinfo{person}{Logan Warberg}, \bibinfo{person}{Alessandro Acquisti}, {and} \bibinfo{person}{Douglas Sicker}.} \bibinfo{year}{2019}\natexlab{}.
\newblock \showarticletitle{Can {Privacy} {Nudges} {Be} {Tailored} to {Individuals}' {Decision} {Making} and {Personality} {Traits}?}. In \bibinfo{booktitle}{\emph{Proceedings of the 18th {ACM} {Workshop} on {Privacy} in the {Electronic} {Society}}} \emph{(\bibinfo{series}{{WPES}'19})}. \bibinfo{publisher}{Association for Computing Machinery}, \bibinfo{address}{New York, NY, USA}, \bibinfo{pages}{175--197}.
\newblock
\showISBNx{978-1-4503-6830-8}
\urldef\tempurl%
\url{https://doi.org/10.1145/3338498.3358656}
\showDOI{\tempurl}


\bibitem[Webb and Tangney(2022)]%
        {webb_too_2022}
\bibfield{author}{\bibinfo{person}{Margaret~A. Webb} {and} \bibinfo{person}{June~P. Tangney}.} \bibinfo{year}{2022}\natexlab{}.
\newblock \showarticletitle{Too {Good} to {Be} {True}: {Bots} and {Bad} {Data} {From} {Mechanical} {Turk}}.
\newblock \bibinfo{journal}{\emph{Perspectives on Psychological Science}} (\bibinfo{date}{Nov.} \bibinfo{year}{2022}), \bibinfo{pages}{174569162211200}.
\newblock
\showISSN{1745-6916, 1745-6924}
\urldef\tempurl%
\url{https://doi.org/10.1177/17456916221120027}
\showDOI{\tempurl}


\bibitem[Wilcox et~al\mbox{.}(2023)]%
        {wilcoxAIConsentFutures2023}
\bibfield{author}{\bibinfo{person}{Lauren Wilcox}, \bibinfo{person}{Robin Brewer}, {and} \bibinfo{person}{Fernando Diaz}.} \bibinfo{year}{2023}\natexlab{}.
\newblock \showarticletitle{{AI} {Consent} {Futures}: {A} {Case} {Study} on {Voice} {Data} {Collection} with {Clinicians}}.
\newblock \bibinfo{journal}{\emph{Proceedings of the ACM on Human-Computer Interaction}} \bibinfo{volume}{7}, \bibinfo{number}{CSCW2} (\bibinfo{date}{Oct.} \bibinfo{year}{2023}), \bibinfo{pages}{316:1--316:30}.
\newblock
\urldef\tempurl%
\url{https://doi.org/10.1145/3610107}
\showDOI{\tempurl}


\bibitem[Witteman et~al\mbox{.}(2016)]%
        {witteman_design_2016}
\bibfield{author}{\bibinfo{person}{Holly~O. Witteman}, \bibinfo{person}{Laura~D. Scherer}, \bibinfo{person}{Teresa Gavaruzzi}, \bibinfo{person}{Arwen~H. Pieterse}, \bibinfo{person}{Andrea Fuhrel-Forbis}, \bibinfo{person}{Selma Chipenda~Dansokho}, \bibinfo{person}{Nicole Exe}, \bibinfo{person}{Valerie~C. Kahn}, \bibinfo{person}{Deb Feldman-Stewart}, \bibinfo{person}{Nananda~F. Col}, \bibinfo{person}{Alexis~F. Turgeon}, {and} \bibinfo{person}{Angela Fagerlin}.} \bibinfo{year}{2016}\natexlab{}.
\newblock \showarticletitle{Design {Features} of {Explicit} {Values} {Clarification} {Methods}: {A} {Systematic} {Review}}.
\newblock \bibinfo{journal}{\emph{Medical Decision Making}} \bibinfo{volume}{36}, \bibinfo{number}{4} (\bibinfo{date}{May} \bibinfo{year}{2016}), \bibinfo{pages}{453--471}.
\newblock
\showISSN{0272-989X, 1552-681X}
\urldef\tempurl%
\url{https://doi.org/10.1177/0272989X15626397}
\showDOI{\tempurl}


\bibitem[Wong and Mulligan(2019)]%
        {wong_bringing_2019}
\bibfield{author}{\bibinfo{person}{Richmond~Y. Wong} {and} \bibinfo{person}{Deirdre~K. Mulligan}.} \bibinfo{year}{2019}\natexlab{}.
\newblock \showarticletitle{Bringing {Design} to the {Privacy} {Table}: {Broadening} “{Design}” in “{Privacy} by {Design}” {Through} the {Lens} of {HCI}}. In \bibinfo{booktitle}{\emph{Proceedings of the 2019 {CHI} {Conference} on {Human} {Factors} in {Computing} {Systems}}}. \bibinfo{publisher}{Association for Computing Machinery}, \bibinfo{address}{New York, NY, USA}, \bibinfo{pages}{1--17}.
\newblock
\showISBNx{978-1-4503-5970-2}
\urldef\tempurl%
\url{https://doi.org/10.1145/3290605.3300492}
\showDOI{\tempurl}


\bibitem[{World Medical Association}(2013)]%
        {worldmedicalassociationWorldMedicalAssociation2013}
\bibfield{author}{\bibinfo{person}{{World Medical Association}}.} \bibinfo{year}{2013}\natexlab{}.
\newblock \showarticletitle{World {Medical} {Association} {Declaration} of {Helsinki}: ethical principles for medical research involving human subjects}.
\newblock \bibinfo{journal}{\emph{JAMA}} \bibinfo{volume}{310}, \bibinfo{number}{20} (\bibinfo{date}{Nov.} \bibinfo{year}{2013}), \bibinfo{pages}{2191--2194}.
\newblock
\showISSN{1538-3598}
\urldef\tempurl%
\url{https://doi.org/10.1001/jama.2013.281053}
\showDOI{\tempurl}


\bibitem[Yoo et~al\mbox{.}(2013)]%
        {yoo_value_2013}
\bibfield{author}{\bibinfo{person}{Daisy Yoo}, \bibinfo{person}{Alina Huldtgren}, \bibinfo{person}{Jill~Palzkill Woelfer}, \bibinfo{person}{David~G. Hendry}, {and} \bibinfo{person}{Batya Friedman}.} \bibinfo{year}{2013}\natexlab{}.
\newblock \showarticletitle{A {Value} {Sensitive} {Action}-{Reflection} {Model}: {Evolving} a {Co}-{Design} {Space} with {Stakeholder} and {Designer} {Prompts}}. In \bibinfo{booktitle}{\emph{Proceedings of the {SIGCHI} {Conference} on {Human} {Factors} in {Computing} {Systems}}}. \bibinfo{publisher}{Association for Computing Machinery}, \bibinfo{address}{New York, NY, USA}, \bibinfo{pages}{419--428}.
\newblock
\showISBNx{978-1-4503-1899-0}
\urldef\tempurl%
\url{https://doi.org/10.1145/2470654.2470715}
\showDOI{\tempurl}


\bibitem[Zenker et~al\mbox{.}(2022)]%
        {zenker_data_2021}
\bibfield{author}{\bibinfo{person}{Sven Zenker}, \bibinfo{person}{Daniel Strech}, \bibinfo{person}{Kristina Ihrig}, \bibinfo{person}{Roland Jahns}, \bibinfo{person}{Gabriele Müller}, \bibinfo{person}{Christoph Schickhardt}, \bibinfo{person}{Georg Schmidt}, \bibinfo{person}{Ronald Speer}, \bibinfo{person}{Eva Winkler}, \bibinfo{person}{Sebastian~Graf von Kielmansegg}, {and} \bibinfo{person}{Johannes Drepper}.} \bibinfo{year}{2022}\natexlab{}.
\newblock \bibinfo{booktitle}{\emph{Data protection-compliant broad consent for secondary use of health care data and human biosamples for (bio)medical research: {Towards} a new {German} national standard}}.
\newblock \bibinfo{type}{{T}echnical {R}eport}. \bibinfo{pages}{104096} pages.
\newblock
\showISSN{1532-0464}
\urldef\tempurl%
\url{https://doi.org/10.1016/j.jbi.2022.104096}
\showDOI{\tempurl}


\bibitem[Zhang et~al\mbox{.}(2021)]%
        {zhang_nudge_2021}
\bibfield{author}{\bibinfo{person}{Weiyu Zhang}, \bibinfo{person}{Tian Yang}, {and} \bibinfo{person}{Simon Tangi~Perrault}.} \bibinfo{year}{2021}\natexlab{}.
\newblock \showarticletitle{Nudge for {Reflection}: {More} {Than} {Just} a {Channel} to {Political} {Knowledge}}. In \bibinfo{booktitle}{\emph{Proceedings of the 2021 {CHI} {Conference} on {Human} {Factors} in {Computing} {Systems}}}. \bibinfo{publisher}{ACM}, \bibinfo{address}{Yokohama Japan}, \bibinfo{pages}{1--10}.
\newblock
\showISBNx{978-1-4503-8096-6}
\urldef\tempurl%
\url{https://doi.org/10.1145/3411764.3445274}
\showDOI{\tempurl}


\end{thebibliography}

\appendix

\section{Supplementary Information: Online Experiment}

\subsection{Scenario}
\label{sec:scenario}
We illustrated a highly descriptive scenario to familiarize participants and maintained terminology throughout the study to avoid misinterpretations. 
After being introduced to the company details, participants are asked to test new parts of the user interface of this health data platform.
An interactive, multi-page prototype of the hypothetical health data platform was created for this purpose.
Extending the prototype beyond the value-centered reflection interface follows the goal of setting up a user flow that is consequently interrupted by the \textit{Value-Centered consent interface} (see \autoref{sec:design}) as a design friction~\cite{terpstra_improving_2019}.
In addition, consistent interface design, including a company logo, was used throughout the interface to present a realistic scenario, following Warberg et al.~\cite{warberg_can_2019} suggestion that non-realistic scenarios may fail to capture risk perception when participants are asked to make \enquote{authentic} decisions.

\subsubsection{Scenario Text}
\label{sec:scenariotext}

The participants were presented with the following scenario text presenting a fictional company named ``\textit{HealthBridge}'':

\begin{description}
    \item 
    \textbf{What Is HealthBridge?}
    
    When you get first treated by a new medical professional, the availability of previous health records can greatly improve the quality of your treatment. Time that is used on gathering already known health information could be spent on medical treatment instead.
    To help with this, our company HealthBridge provides you with an easy and secure way to receive, store, and share your personal health data. This can include previous diagnoses, medical certifications, treatment results, and data collected through wearable health devices.
    We offer an electronic personal health data storage solution through an online platform, which allows you to have your health information ready whenever and wherever needed.
    Our vision is to keep patients fully in control of their data. You alone control what and how health data is stored in HealthBridge: you decide which data should be stored or deleted and which healthcare providers can access your data. Thereby, all data is stored in a secure and encrypted form.
    We also envision our platform as contributing to medical innovation and increased quality of healthcare overall by giving our users the option of making their data available for clinical research purposes.
\end{description}

\begin{description}
    \item 
    \textbf{What Is Your Task?}
    
    We are currently testing new user interface functionalities of the HealthBridge platform.
    Please imagine you have recently created a new HealthBridge account and use it to store medical treatment results and smartwatch health data.
    Please answer all questions truthfully, providing answers as you would in real life.
\end{description}

\subsection{``Disclosure Options'' Texts}
\label{sec:dataoptionstext}
The following text described the ``Disclosure Options'' within the Value-Centered Consent Interface:

%
\emph{You have the possibility to make certain data available for medical research at universities. This is optional and can help in driving healthcare innovation and gaining new medical insights. You can revoke your consent at any time.}

\begin{itemize}
    \item 
    \textbf{Personal Contact Data}
    
    If you agree, your personal contact data, specifically your e-mail address, is made available to university researchers, only for the purpose of informing you about current research projects.
    \item 
    \textbf{Health-Wearables Data}
    
    If you agree, your health data collected by wearables are made available to university researchers for the purpose of medical research, in an anonymized form. Wearable health devices like fitness trackers or smartwatches are valuable for data-driven medical research.
    \item 
    \textbf{Medical Diagnoses and Treatment Data}
    
    If you agree, your medical diagnoses and treatment data are made available to university researchers for the purpose of medical research, without including your name or other directly identifying information.
\end{itemize}

\subsection{Knowledge Questionnaire}
\label{sec:knowledgequest}
The following comprehension questions constituted the \knowledge measurement. Two questions are asked about per data option and two concerning the overall interface.
The answer options are binary \enquote{Yes, No}, with an additional \enquote{Don't know} option to avoid guessing. The resulting comprehension score is calculated by adding one point for each correct answer and subtracting one point for each incorrect answer. 
\begin{itemize}
    \item I had the option to allow for my data to be sold to commercial providers.
    \item I was able to choose if my home address is made available to university researchers.
    \item All data-sharing options concerned medical research conducted at universities.
    \item I was able to choose if my e-mail address is made available to university researchers.
    \item According to the \textit{Additional Data Options} I can revoke my consent at any time.
    \item Making my medical diagnosis and treatment data available would also include directly identifying information such as my name.
    \item I could agree to make my health-wearable data available for university researchers.
    \item I had the option to allow for my keyboard inputs and IP address to be collected.	
\end{itemize}

\subsection{Inclusion Criteria}
\label{sec:inclusioncriteria}
We received a total of 621 complete submissions, of which we excluded 414 for failing at least two attention checks. Following the recommendations for collecting quality data on Mturk~\cite{webb_too_2022, ramirez_state_2021}, we required 1,000 completed HITs with a greater than 95\% acceptance rate for workers to participate in our study. However, we had to reject 60.47\% of all MTurk submissions.
For this, a manual review of all participant data was conducted, where at least two rejection criteria, i.e., failed attention check or incoherent or irrelevant text in qualitative feedback, had to be violated to result in a rejection.
Popular recommendations for research on MTurk suggest that nonconforming responses should be expected to make up 15-30\% of the total sample collected~\cite{aguinis_mturk_2021}. However, these expectations are based on research conducted a decade earlier~\cite{sprouse_validation_2011}.
More recent studies provide a starkly contrasting picture: Dupuis et al.~\cite{dupuis_crowdsourcing_2022} report a noncompliance rate of over 85\% for qualitative responses in their 2022 paper, describing response patterns very similar to those in our study. Another report from March 2022 by Webb and Tangney~\cite{webb_too_2022} reports that only 2.6\% managed to meet the eligibility criteria. 
In our estimation, the availability of automated filling and text generation tools has greatly reduced the quality of response data on Mturk. Thus, our experience raises the question of whether the effort to identify quality contributions is worth using MTurk in future research.

\subsection{Table: Hypothesis 1 Tests}
\label{tab:h1table}

\begin{table}[ht]
\centering
\begin{tabular}{p{4cm}p{2cm}p{2.2cm}p{2.3cm}}
  \hline
 & Refl & ValueQuestRefl & ValueReflPrompt \\ 
  \hline
  \textit{ValueDiscrepancyScore} \newline of \textit{Selection 1} & M=0.614, \newline SD= 0.278 & M=0.548, \newline SD=0.304 & M=0.555, \newline SD= 0.290\\  
  \\
  \textit{ValueDiscrepancyScore} \newline of \textit{Selection 2} & M=0.629, \newline SD= 0.274 & M=0.565, \newline SD=0.301 & M=0.539, \newline SD=0.292\\
  \\
  Shapiro-Wilk test \newline for \textit{Selection 1} & W = 0.893, \newline p < 0.000 & W = 0.907, \newline p < 0.000 & W = 0.892, \newline p < 0.000 \\ 
  \\
  Shapiro-Wilk test \newline for \textit{Selection 2} & W = 0.892, \newline p < 0.000 & W = 0.906, \newline p < 0.000 & W = 0.904, \newline p < 0.000 \\ 
  \\
  Wilcoxon signed-Rank test & W = 3.5, \newline p = 0.889 & W = 2.5, \newline p = 0.932 & W = 8.5, \newline p = 0.135  \\ 
   \hline
\end{tabular}
\end{table}

\section{Supplementary Information: Expert Workshop}

\subsection{Activity Cards}
\label{sec:activitycards}

\begin{enumerate}
    \item \textbf{Direct Stakeholders}
    
    \emph{Direct stakeholders are individuals or groups of individuals who interact in a context and are directly affected by it.}
    
    From your perspective, write down individuals or groups of individuals who are directly involved in a health data donation regarding the broad consent.
    \item \textbf{Indirect Stakeholders}
    
    \emph{Indirect stakeholders are individuals or groups of individuals who do not interact directly in a context but may still be affected by it.}
    
    From your perspective, write down individuals or groups of individuals who are indirectly related to a health data donation regarding the broad consent.
    \item \textbf{Impact}
    
    \emph{Imagine that the broad consent is used at all German university hospitals.}
    
    From your perspective, what would be the associated effects for patients?
    \item \textbf{Values}
    
    \emph{Values can be defined as goals that are desirable, worthwhile, or valuable, that transcend a specific situation, and are generally considered applicable to social life.}
    
    From your expertise, what values are important in the context of medical data donation?
    \item \textbf{Reasoning}
    
    \emph{Select three values you have identified and that you consider essential for medical data donation.}
    
    Describe why and how these values should be considered in the broad consent.
\end{enumerate}

\subsection{Value Elicitation Questions for Patients} 
\label{sec:valuequestions}

Addressing \textit{``value''} terminology confusion brought about by our expert panel, we adapted phrasing similar to Friedman et al.'s~\cite{friedman_survey_2017} value definition to convey what values ought to be the subject of reflection without explicitly referring to \textit{``values.''}

\begin{enumerate}
    \item \textbf{Autonomy}
    \emph{What is important to you when you share your data?}
    \item \textbf{Non-maleficence}
    \emph{Do you have concerns about sharing your data? If so, what are they?}
    \item \textbf{Beneficence/Justice}
    \emph{What might be the benefit to you or to society of donating your data?}
\end{enumerate}

\section{Supplementary Information: User Interface Evaluations}

\subsection{User Instruction}
\label{sec:usertestintructions}

``Welcome. Thank you for agreeing to take part in this user interface evaluation. The aim of the user interface evaluation is to examine a newly developed application for data-sharing decisions. It is important to note that the application is being tested - not you. If something is not understandable or confusing, it is the application's fault - not yours. We would like to know what you are thinking while using the application to understand what needs improvement. For this purpose, we ask you to constantly speak your thoughts out loud when using the application. Just freely share whatever is on your mind. You do not have to try to explain anything to the conductor. Simply act as if you were alone in the room and talking to yourself.
Should you forget to speak for a longer period of time, I will remind you to continue doing so. Do you have any further questions?''
[Use a phrase such as “Remember to speak your thoughts out loud” when they are silent for a few seconds].''

\subsection{Scenario Text}
\label{sec:usertestscenario}
Imagine the following: \textit{``You are staying at a local university hospital as part of a treatment program. The hospital staff asks you to decide whether the medical data collected during your stay can be used for medical research. You are presented with an application that you can use to make this decision. In a quiet moment, when you have sufficient time to think about this possibly new topic, you launch the following application[...].''}

\subsection{Interview Questionnaire}
\label{sec:usertestinterview}

\textbf{Feedback on the Demonstrator (7 min)}
\begin{itemize}
    \item{What impression did the application leave on you now that you have used it?}
    \item{Was there something about the application that struck you as negative? Were there any sections that you found incomprehensible?}
    \item{Was there something about the application that struck you as positive?}
\end{itemize}

\textbf{Interface Elements and Reflection (10 min)}
\begin{itemize}
    \item{How do you decide whether to donate data? Which factors were decisive?}
    \item{Did entering your personal needs and values help you to think about medical data donation?}
    \item{The application provided you with information in different places and in different forms (e.g., text form, video, questions, info boxes). How did you perceive this information? Which ones were especially important to you, or which ones do you remember?}
    \item{Before you confirmed your decision regarding data donation, you were shown your personal needs and values. How did you perceive this?}
\end{itemize}

\textbf{Medical Data Donation (3 min):}
\begin{itemize}
\item{Are you familiar with the term medical data donation? If so, how do you know this term?}
\item{What is your general attitude toward medical data donation? Which aspects do you find good? What worries you about it?}
\end{itemize}

\textbf{Conclusion}
\begin{itemize}
\item{Is there anything else you would like to mention about the application or medical data donation?}
\end{itemize}

\subsection{Qualitative Analysis: Categories}
\label{sec:userstudycategories}

The valence of the user statements is indicated by ``(+)'' for a positive sentiment and ``(-)'' indicating negative sentiment.
\newline

\textbf{Individual Decision-Making Processes}
\begin{itemize}
    \item{Pre-determination: Immediate decision based on individual values\newline[P1,P2,P4,P7,P8,P9,P19,P12]}
    \item{Uncertainty: Single aspects compromise sharing intention as uncertainty remains unresolved  by the information provided (-)}, specifically regarding: 
    \SubItem{Long timeline of data donation [P2,P11]}
    \SubItem{Scope of data donation [P3]}
    \SubItem{Possibility of negative health insurance effects[P4,P5]}
    \SubItem{Request: Accommodate individual information needs through e.g. a hotline  [P7]}
    \item{Control: Sovereignty over one's data (+):}
    \SubItem{Feeling of being in Control (+) [P3,P7]}
    \SubItem{Request: More granularity in decision [P3]}
    \item{Privacy: Central consideration for decision outcome \newline [P1,P2,P4,P7,P8,P9,P10,P12]}
    \item{Motivation to Share:}    
    \SubItem{Medical Advancements  [P1,P2,P4,P7,P9,P10,P12]}
    \SubItem{Personal Medical Gain[P7,P9,P10,P11]}
    \SubItem{Altruism  [P1,P4,P6,P7,P8,P10,P11]}    
\end{itemize}

\textbf{Overall Interface Evaluation}
\begin{itemize}
    \item{Ease of use (+) [P4]}
    \item{Individual criticism of visual design elements (-) \newline [P4,P5,P7,P9]}
    \item{Sufficent Information Provision (+) \newline [P1,P2,P3,P5,P8]}
    \item{Language:}
    \SubItem{Concise descriptions (+) [P1,P3,P12]}
    \SubItem{Good understandability (+) [P1,P2,P3,P6,P7,P8,P9,P12]}
    \SubItem{Vocabulary: Certain terms not suitable for all user groups (-) [P1,P4]}
    \item{Multi-modal information presentation:}
    \SubItem{Explanatory video (+) [P1,P2,P3,P4,P5,P6,P8,P9,P10,P11,P12]}
    \SubItem{Provision of subtitles for video(+) [P5,P6]}
    \item{Structure: Multi-Layered Information Provision:} 
    \SubItem{Clear Structure (+) [P9]}
    \SubItem{Segmentation of information provision (+)[P1,P3,P5,P11]}
    \SubItem{Possibility to adjust information to personal needs (+)[P8]}
    \SubItem{Availability of supplemental information (+)[P3,P8]}
    \SubItem{Text Volume: Too long and/or repetitive (-) [P2,P7,P8,P9,P10]}
    \SubItem{Request: Adjustment of interface to individual preferences [P12]}
\end{itemize}

\textbf{Overall Interface Evaluation}
\begin{itemize}
    \item{Possibility of active reflection and review (+)  [P2,P3,P4,P5,P8,P11]}
    \item{Reflection Questions: Good understandability (+) [P2,P5,P6]}
    \item{Provides support against uncertainty (+)[P2]}
    \item{Reflection Prompt: Decision review provides control (+) [P1,P5,P7]}
    \item{Decison review: Perceived as unnecessary task (-)	 [P10]}
    \item{Reflection prompt: Mistaken as data collection (-) [P7]}
\end{itemize}

\end{document}